\RequirePackage{fix-cm}
\documentclass[smallextended]{svjour3}

\usepackage{adjustbox}
\usepackage{afterpage}
\usepackage{balance}
\usepackage{booktabs}
\usepackage{caption}
\usepackage{cite}
\usepackage{colortbl}
\usepackage{collectbox}
\usepackage{comment}
\usepackage{enumitem}
\usepackage{framed}
\usepackage{graphicx}
\usepackage{lscape}
\usepackage{lipsum}
\usepackage{listings}
\usepackage{longtable}
\usepackage{multirow}
\usepackage{tabularx}
\usepackage{txfonts}
\usepackage{url}
\usepackage{varwidth}
\usepackage{xspace}
\usepackage[dvipsnames]{xcolor}
\usepackage[tight,footnotesize]{subfigure}
\usepackage{tcolorbox}

\usepackage[T1]{fontenc}
\usepackage{array}      
\usepackage[strict]{changepage}
\usepackage{arydshln}
\usepackage{bm}

\renewcommand{\arraystretch}{1.3}

\tcbuselibrary{raster,skins}
\definecolor{light-gray}{gray}{0.85}

\definecolor{formalshade}{rgb}{0.95,0.96,0.96}
\definecolor{side}{rgb}{0.0,0.2,0.6}

\newenvironment{formal}{%
  \MakeFramed{\advance\hsize-\width\FrameRestore}%
  \noindent\hspace{-4.55pt}
  \begin{adjustwidth}{}{7pt}%
  \vspace{2pt}\vspace{2pt}%
}
{%
\vspace{2pt}\end{adjustwidth}\endMakeFramed%
}

\newcommand{\countobservations}{
    \def \countobservations{1}
}
\newcounter{observation}
\newenvironment{observation}[1]
{   \refstepcounter{observation}
    \noindent \textbf{Observation~\theobservation) \textit{#1}}
}

\countobservations

\newcommand{\countmess}{
    \def \countmess{1}
}
\newcounter{mes}
\countmess

\def\summarybox#1#2{
\medskip
\begin{tcolorbox}[enhanced,title=#1, attach boxed title to top left= {xshift=0mm, yshift*=-\tcboxedtitleheight/2}]
    #2
\end{tcolorbox}
}

\newlist{stepitemize}{itemize}{1}
\setlist[stepitemize,1]{leftmargin=1.26cm}

\definecolor{diffstart}{named}{Gray}
\definecolor{diffincl}{named}{Green}
\definecolor{diffrem}{named}{Purple}
\lstdefinelanguage{diff}{
basicstyle=\ttfamily\small,
float,
floatplacement=t,
frame=tb,
morecomment=[f][\color{diffstart}]{@@},
morecomment=[f][\color{diffincl}]{+},
morecomment=[f][\color{diffrem}]{-},
}

\lstdefinelanguage{difftab}{
basicstyle=\ttfamily\scriptsize,
breaklines=true,
morecomment=[f][\color{diffstart}]{@@},
morecomment=[f][\color{diffincl}]{+},
morecomment=[f][\color{diffrem}]{-},
}

\newcommand{\countimplications}{
    \def \countimplications{1}
}
\newcounter{implication}
\newenvironment{implication}[1]
{
    \refstepcounter{implication}
    \noindent \textbf{Implication~\theimplication ) \textit{#1}}
}

\countimplications

\newcommand{\red}[1]{\textcolor{black}{{#1}}}

\def\et{et\ al.\xspace}

\def\eg{e.g.}

\def\sec#1{Section~\ref{#1}}

\newlength\MAX  

\newsavebox{\fminipagebox}
\NewDocumentEnvironment{hassanbox}{m O{\fboxsep}}
 {\par\kern#2\noindent\begin{lrbox}{\fminipagebox}
  \begin{minipage}{#1}\ignorespaces}
 {\end{minipage}\end{lrbox}%
  \makebox[#1]{%
    \kern\dimexpr-\fboxsep-\fboxrule\relax
    \fbox{\usebox{\fminipagebox}}%
    \kern\dimexpr-\fboxsep-\fboxrule\relax
  }\par\kern#2
 }

\newcommand{\RQtwo}{To what extent does the natural language of the problem statement affect LLM code generation performance?}
\newcommand{\RQthree}{Can translating a problem statement into the best-performing language improve LLM code generation performance?}

\usepackage[hidelinks]{hyperref} 

\definecolor{ABlue}{HTML}{127bca}
\definecolor{LHScolor}{HTML}{555555}

\newcommand{\DOIbox}[1]{
\tcbsidebyside[
        bicolor,
        sidebyside,
        fontupper=\footnotesize\ttfamily\bfseries,
        fontlower=\footnotesize\sffamily\mdseries,
        nobeforeafter,
        shrink tight,
        extrude bottom by=1mm,
        sidebyside adapt=both,
        sidebyside gap=5pt,
        top=2pt,left=3pt,right=3pt,bottom=2pt,
        boxrule=0pt,
        rounded corners,
        coltext=white,
        colback=LHScolor,
        colbacklower=ABlue,
]{%
DOI
}{%
	\nolinkurl{#1}
	}%
	}

\begin{document}

\title{A Study on the Impact of Natural Language Differences\\ in Prompts on Automatic Code Generation Using LLMs}
\titlerunning{A Study on the Impact of Natural Language Differences\\ in Prompts on Automatic Code Generation Using LLMs}

\author{Haruka Tokumasu \and
        Masanari Kondo \and
        Alexander Serebrenik \and
        Dong Wang \and
        Kei Koyanagi \and
        Kotaro Noguchi \and
        Naoyasu Ubayashi \and
        Yasutaka Kamei
}

\institute{
    Haruka Tokumasu\at
    Principles of Software Languages group (POSL)\\
    Kyushu University, Japan\\
    \email{tokumasu@posl.ait.kyushu-u.ac.jp}\\\\
    Masanari Kondo\at
    Principles of Software Languages group (POSL)\\
    Kyushu University, Japan\\
    \email{kondo@ait.kyushu-u.ac.jp}\\\\
    Alexander Serebrenik\at
    Eindhoven University of Technology, The Netherlands\\
    \email{a.serebrenik@tue.nl}\\\\
    Dong Wang\at
    Tianjin University, China\\
    \email{dong\_w@tju.edu.cn}\\\\
    Kei Koyanagi\at
    Kyushu University, Japan\\
    \email{koyanagi@posl.ait.kyushu-u.ac.jp}\\\\
    Kotaro Noguchi\at
    Kyushu University, Japan\\
    \email{noguchi@posl.ait.kyushu-u.ac.jp}\\\\
    Naoyasu Ubayashi\at
    Waseda University, Tokyo, Japan\\
    \email{ubayashi@aoni.waseda.jp}\\\\
    Yasutaka Kamei\at
    Principles of Software Languages group (POSL)\\
    Kyushu University, Japan, and\\
    Inamori Research Institute for Science, Kyoto, Japan\\
    \email{kamei@ait.kyushu-u.ac.jp}\\\\
}

\authorrunning{Tokumasu~\et{}} 

\maketitle

\makeatletter
\def\ps@headings{%
\def\@oddfoot{\scriptsize \hfill \thepage }%
\def\@evenfoot{\scriptsize \thepage \hfill}}
\makeatother
\pagestyle{headings}

\begin{abstract}
\textbf{Context:}
Large Language Models (LLMs) have demonstrated remarkable performance in automatic code generation tasks, thereby encouraging new research in this area. Although numerous studies have explored LLM-based code generation, the impact of the natural language in input prompts remains unexplored (\textit{language bias}). As previous work has revealed natural language biases in LLMs, it is necessary to understand their impact on code generation tasks.

\noindent
\textbf{Objectives:}
This study aims to (1) quantify how the natural language of input prompts influences LLM-based code generation performance and (2) evaluate a mitigation strategy to reduce language bias in code generation.
This is a conceptual replication of our previous work (Koyanagi et al., MSR2024). Since the previous work focused on a single LLM (GitHub Copilot) and a single dataset (AtCoder), this study extends the previous work and explores a mitigation strategy for language bias. 

\noindent
\textbf{Methods:}
We assess code generation $\mbox{\sl Accuracy}$ on 756 AtCoder problems (difficulty levels A, B, C, and D), 100 LeetCode problems (easy, medium, and hard), and 288 BigCodeBench problems.
To quantify the language bias on code generation, each problem is presented in English, Japanese, and Chinese. We use seven LLMs (GPT-4o, o3-mini, DeepSeek-V3.2, Llama-3, Qwen2.5-Coder-14B, Qwen2.5-Coder-0.5B, and GitHub Copilot) and assess their performance in terms of $\mbox{\sl Accuracy}$ (the number of problems for which generated code passes all test cases).
Since a prior study indicates that translating prompts into a certain language, in which LLMs show better performance, can mitigate language bias, we compare $\mbox{\sl Accuracy}$ before and after translation to evaluate the effectiveness of translation as a mitigation strategy.

\noindent
\textbf{Results:}
We observed that the natural language of problem statements affects LLM-based code generation performance. Specifically, the languages officially supported by each dataset achieved the highest median $\mbox{\sl Accuracy}$. Also, translation improved Accuracy, but its effectiveness was not consistent across datasets and model types. While translation improved $\mbox{\sl Accuracy}$ for LeetCode and BigCodeBench, its effect was mixed for AtCoder. To understand this gap, we analyzed narrative aspects and problem statement length. We found that AtCoder contained a particularly high proportion of narrative-style problem statements and longer problem statements.

\noindent
\textbf{Conclusions:}
Natural language significantly affects LLM code generation accuracy.
Translation can mitigate language bias in some settings, but its effectiveness depends on the dataset and model type.
Furthermore, the narrative aspects and context length of input prompts are important factors related to language bias and the effectiveness of translation as a mitigation strategy.

\end{abstract}
\section{Introduction}\label{sec:introduction}
Large Language Models (LLMs) have been extensively studied for their potential applications in various software development tasks~\cite{pearce2022asleep,asare2023github,nguyen2022MSR,mastropaolo2023robustness,dakhel2023github,al2022readable,daniotti2026science,Bonan2024FSE,jimenez2024ICLR,liu2023repobench,jain2025livecodebench}. 
Among these, significant attention has been given to the domain of automatic code generation, where LLMs have demonstrated strong performance~\cite{nguyen2022MSR,mastropaolo2023robustness,dakhel2023github,al2022readable}. 
For instance, Nguyen et al.~\cite{nguyen2022MSR} evaluated the code generation performance of GitHub Copilot and found that it achieved a 57\% accuracy in Java code generation tasks.
More recently, large-scale empirical evidence suggests that LLM-based code generation
is already widely adopted in practice. Daniotti et al.~\cite{daniotti2026science} analyzed over 30 million
GitHub commits and found that approximately 29\% of Python functions are generated
by AI, highlighting the growing prevalence and real-world impact of LLMs in software
development.

While LLMs perform well in code generation tasks, previous studies also demonstrated the \textit{natural language biases} of LLMs. 
For example, Liu et al.~\cite{liu2024arxiv} found that LLMs produce the most accurate results for English inputs, but not for other languages, specifically low-resource languages. 
Language-related limitations also arise in practice: e.g., a survey of Spanish-speaking developers found that AI coding tools can struggle with non-English code elements and local naming conventions, contributing to inefficiency, mistrust, and rejection of generated suggestions~\cite{BottoTobar2025AI}.
To overcome these limitations, previous studies proposed various approaches, such as proposing multilingual LLMs~\cite{sun2024arxiv,liu2024arxiv,raihan-etal-2025-mhumaneval,li2025bridginglanguagegapenhancing,daniotti2026science}. 
However, the natural language biases of LLMs in code generation tasks had not been explored. 

To understand the natural language biases in code generation tasks, our previous work, Koyanagi et al.~\cite{koyanagi2023MSR}, conducted experiments using problems from AtCoder, a multilingual programming contest platform, as input to GitHub Copilot. The findings revealed that language bias exists in code generation tasks. For example, GitHub Copilot could generate accurate code even for problems presented in Japanese.
Despite this initial analysis, the scope of the study was limited, as it focused solely on AtCoder problems and evaluated only GitHub Copilot.

In this study, we build upon the work of Koyanagi et al.~\cite{koyanagi2023MSR}.
To improve the generalizability of the findings, we include an additional dataset that closely resembles AtCoder. 
Specifically, we select LeetCode, as it is a widely used programming contest platform that, similar to AtCoder, officially supports multiple natural languages for problem statements.
Furthermore, we extend the scope of our analysis by incorporating BigCodeBench, which represents specification-based code generation tasks.
Unlike AtCoder and LeetCode, which focus on competitive programming tasks, BigCodeBench consists of
code generation tasks with natural language descriptions embedded in code contexts. 
This allows us to examine whether the observed language biases persist beyond contest-style problems.
AtCoder and LeetCode are popular online programming contest platforms that officially support multiple natural languages for their problem statements, while BigCodeBench enables us to analyze specification-based code generation tasks.
Additionally, we add more recent LLMs, including GPT-4o, o3-mini, and DeepSeek-V3.2, thereby incorporating LLMs beyond GitHub Copilot into the evaluation framework. Furthermore, we include open-source models, such as Llama3 and Qwen2.5-Coder (0.5B and 14B), to enable a broader comparison across both proprietary and open-source LLMs.

Although Koyanagi et al.~\cite{koyanagi2023MSR} reported that the performance of code generation varies depending on the natural language, proposing a mitigation strategy to alleviate language bias is still an open question.
We deduce that translating problem statements into the language that yields the best code generation performance may improve the accuracy of code generation, as Liu et al.~\cite{liu2024arxiv} demonstrated that translating input prompts into English can enhance the performance of LLMs in natural language processing tasks.
Hence, as a mitigation strategy, we evaluate the effectiveness of translation in code generation tasks.

This manuscript aims to (1) provide a comprehensive understanding of the impact of natural language on code generation performance, and (2) explore the potential mitigation strategy to relieve the language bias in code generation tasks.
For this purpose, we conduct experiments and address the following two research questions (RQs):
\begin{itemize}[leftmargin=11mm]
  \item[\textbf{(RQ1)}] \textbf{\RQtwo{}}
  \\ 
  \emph{Motivation:}
  Koyanagi et al.~\cite{koyanagi2023MSR} demonstrated that the natural language of problem statements influences code generation performance. This RQ is a conceptual replication of their work, expanding the scope to include multiple LLMs and datasets. In this setting, we aim to measure the impact of the natural languages on code generation performance. 
  \\
  \emph{Results:}
  We define the language gap as the difference between the highest and lowest median $\mbox{\sl Accuracy}$ across natural languages for each dataset and model type. We observed language gaps across all three datasets, indicating that the natural language of problem statements affects LLM code generation performance. The size of these gaps varied depending on both the dataset and model type, ranging from 3.6\% to 6.6\% for open-source LLMs and from 2.7\% to 15.6\% for closed-source LLMs. In addition, the languages officially supported by each dataset frequently yielded the highest median $\mbox{\sl Accuracy}$.
  \item[\textbf{(RQ2)}] \textbf{\RQthree{}}
  \\
  \emph{Motivation:}
  Machine translation is widely used to overcome language barriers and Liu et al.~\cite{liu2024arxiv} demonstrated that translating input prompts into English can improve the performance of LLMs in natural language processing tasks. We investigated whether translating problem statements into the better-performing language could enhance code generation performance.
  \\
  \emph{Results:}
  The impact of translation varied depending on both the dataset and the type of LLM. Translation improved Accuracy for LeetCode and BigCodeBench, but its effect was mixed for AtCoder. For example, when translating AtCoder problems into English, open-source models showed performance improvements, with a median change of +3.2\%, whereas closed-source models showed performance degradation, with a median change of -5.3\%. These results indicate that translation can mitigate language bias in some settings, but it is not universally effective and depends on both the task characteristics and the underlying model.
\end{itemize}

Based on the results of the RQs, we pose an additional question: why do language bias and the impact of translating problem statements vary across different datasets?
To further investigate this issue, we conduct several complementary analyses.

First, to better understand the potential factors behind the observed differences across datasets,
we analyze the characteristics of problem statements, focusing on their narrative properties and length. 
In particular, since AtCoder problems tend to be more narrative-style and longer than LeetCode problems, we examine how these aspects differ between the two datasets.

To investigate the narrative aspect, we define a narrative-style problem statement using three criteria: (1) the presence of a concrete subject, (2) a contextualized objective, and (3) sentences that are not directly required for solving the task. Based on these criteria, we conduct a quantitative comparison between AtCoder and LeetCode problems. \red{We selected LeetCode as a representative dataset of non-narrative-style problems instead of BigCodeBench.}
We find that AtCoder problems are more frequently classified as narrative-style (74.0\%) than LeetCode problems (21.2\%). In addition, the median length of AtCoder problem statements is more than twice that of LeetCode, indicating that AtCoder problems generally contain richer contextual information. These findings suggest that differences in narrative structure and problem length may be related to the variation in code generation performance across datasets.

In addition to this analysis, we conduct further experiments to evaluate the generalizability of our findings. Specifically, we extend our analysis in two directions. First, we analyze code generation performance for C++ on AtCoder to examine whether the observed trends hold across different programming languages. Second, we investigate code generation performance in low-resource natural languages to assess how language bias manifests under limited-resource settings.

These complementary analyses allow us to both explore potential explanations for dataset
differences and evaluate whether our findings remain consistent across different programming and linguistic conditions.

The contributions of this study are as follows:
\begin{itemize}
  \item 
  We conduct a conceptual replication of Koyanagi et al.~\cite{koyanagi2023MSR} with a broader scope, including seven more recent LLMs (GPT-4o, o3-mini, DeepSeek-V3.2, Copilot, Llama3, and Qwen2.5-Coder 0.5B/14B) and three datasets (AtCoder, LeetCode, and BigCodeBench). 
  The results show that the impact of natural language on code generation performance varies across datasets and models, and that prompts written in officially supported languages tend to achieve higher performance.
  \item 
  We evaluate a strategy for mitigating the language bias in code generation tasks by translating problem statements into the language that yields the best code generation performance.
  Our results show that its effect depends on both the dataset and the model. Translation into officially supported languages improves performance on LeetCode and BigCodeBench, whereas it tends to reduce performance for closed-source models on AtCoder.
  \item 
  This is the first study to investigate the narrative and length aspects of input prompts in the context of LLM-based code generation.
  We define a narrative-style problem statement using three criteria: (1) concrete subject, (2) contextualized objective, and (3) irrelevant sentence, and conduct a quantitative analysis of the narrative aspect of problem statements.
  We clarify the impact of the narrative aspect on code generation performance, which varies depending on the context information length.
  \item 
  We provide a replication package to encourage further research in this area.\footnote{\url{https://doi.org/10.5281/zenodo.20155616}}
\end{itemize}

The structure of this paper is as follows.
Section~\ref{sec:related} describes the related work and the novelty of this study. 
Section~\ref{sec:setup} presents the study design.
Sections~\ref{sec:rq1} and \ref{sec:rq2} show the results of RQ1 and RQ2, respectively.
Section~\ref{sec:ad} presents additional analyses.
Section~\ref{sec:discussion} provides the implications to LLM-based code generation for multiple natural languages and discuss future research directions.
Section~\ref{sec:threats} explains the threats to validity and 
Section~\ref{sec:conclusion} concludes the study.

\section{Related Work}
\label{sec:related}
In this section, we discuss the related work and highlight the differences between our study and previous studies.

\subsection{Code Generation Benchmarks for LLMs}
LLMs show high performance in various software engineering tasks~\cite{Feng2024RE,Su2024ICSE,Ma2024ICSE}. One of the most successful areas is code generation~\cite{nijkamp2023arxiv,li2023arxiv,luo2023arxiv,wang2023arxiv,nguyen2022MSR,Jiang2024TOSEM,Li2025ACM,Ouyang2025TOSEM,Li2024ACM}.
Since researchers and practitioners propose various methods for LLM code generation (e.g., new LLMs, prompts, and workflows), it is essential to evaluate and compare their performance on common benchmarks.
Indeed, previous studies~\cite{Li2025ACM,Ouyang2025TOSEM,Li2024ACM} have evaluated the code generation capabilities on benchmarks, such as HumanEval~\cite{chen2021evaluatinglargelanguagemodels}, MBPP~\cite{austin2021programsynthesislargelanguage}, and MBCPP~\cite{Li2025ACM}. 
These benchmarks consist of various problems, test cases, and correct programs. Researchers and practitioners 
automate the evaluation process and compare their proposed methods with existing ones.
For example, Li et al.~\cite{Li2025ACM} proposed a prompt called SCoT for code generation and evaluated its effectiveness using five different LLMs.
The evaluation, conducted on HumanEval, MBPP, and MBCPP, demonstrated that SCoT outperformed other prompts in code generation.

Previous studies also evaluated the code generation accuracy on programming contests~\cite{nguyen2022MSR,billah2024ESEIW}.
Similar to basic code generation benchmarks (e.g., HumanEval), programming contests also consist of various problems, test cases, and correct programs. Additionally, programming contests cover multiple natural languages and programming languages since they are held worldwide and target a diverse programming community. Consequently, programming contests are suitable for evaluating the impact of natural language and programming language on code generation.
For example, Nguyen et al.~\cite{nguyen2022MSR} evaluated the code generation accuracy of GitHub Copilot across easy, medium, and hard problems on LeetCode. It consists of 132 problems in total and is written in Java, Python, C, and JavaScript. They found that GitHub Copilot perform well in Java, with 57\% accuracy, but only 27\% in JavaScript. This indicates that the accuracy of GitHub Copilot relies on the programming language.

Billah et al.~\cite{billah2024ESEIW} evaluated the performance of ChatGPT-3.5, Gemini 1.0 Pro, and Meta AI on 98 problems from LeetCode, 126 problems from Codeforces, and two certification tests from HackerRank. They found that ChatGPT-3.5 achieved 71.43\% accuracy on LeetCode, but only 26.98\% on Codeforces. While ChatGPT-3.5 consistently leads to better performance, all LLMs result in low accuracy on high difficulty problems. This result indicates that LLMs still have room for improvement in code generation even for programming contests.
In contrast, our study investigates the impact of natural languages in problem statements rather than only comparing performance across programming languages, platforms, or difficulty levels.

\subsection{Biases in LLMs}
While LLMs perform well in various tasks such as code generation, it is well known that their outputs can be biased. In code-related tasks, the performance of LLMs can depend on both the programming language and the natural language used in the input.

Previous studies investigated the programming language bias of LLMs~\cite{athiwaratkun2023multilingual,Zheng2023,peng2024humanevalxlmultilingualcodegeneration}.
For example, Zheng et al.~\cite{Zheng2023} proposed a multilingual model CodeGeeX pre-trained on 23 programming languages. This model was evaluated on the HumanEval-X benchmark, which consists of 164 code generation and code translation problems in C++, Java, JavaScript, Go, and Python. The authors compared CodeGeeX with GPT-J-6B, GPT-NeoX-20B, InCoder-6.7B, and CodeGen-Multi-6B/16B. The results showed that CodeGeeX achieved the best average code generation performance. Also, they found that the best-performing programming language for each model was different (e.g., Python for CodeGeeX vs. Java for CodeGen-Multi-16B), indicating that the performance of LLMs relies on the programming language in the training data.

Previous studies also investigated the natural language bias of LLMs~\cite{liu2024arxiv,li2024arxiv,sun2024arxiv,wang2024exploringmultilingualbiaslarge,katzy2025qualitativeinvestigationllmgeneratedmultilingual,li2025bridginglanguagegapenhancing}.
Liu et al.~\cite{liu2024arxiv} demonstrated the impact of natural language translation on natural language processing (NLP) tasks. LLMs they used are ChatGPT~\cite{openai2024gpt4technicalreport} and Llama-2-70B-Chat~\cite{touvron2023llama2openfoundation}. They used user requests from ShareGPT\footnote{https://sharegpt.com/}, a platform collecting real-world conversations in ChatGPT. They selected 10 different natural languages, sampling 100 requests for each language\footnote{Due to the shortage of requests, they only used 53 requests for Romanian, 98 for Ukrainian, and 53 for Norwegian}, and compared the performance on original language requests with translated requests to English. These experiments showed that the impact of translation depends on the target language and the LLM used. For example, ChatGPT showed an improvement for Vietnamese but a decrease for Norwegian. Llama-2-70B-Chat showed an improvement for Chinese, whereas ChatGPT showed a decrease.
While Liu et al. focused on translation effects in general NLP tasks, our study examines translation effects in LLM-based code generation, where outputs are evaluated using executable test cases.

Li et al.~\cite{li2024arxiv} proposed a metric called \emph{Language Ranker} ranking languages based on the performance of LLMs. This metric can evaluate the similarity between the LLM performance in English and other target languages. They found that high-resource languages (e.g., German) result in higher similarity with English, while low-resource languages (e.g., Romany) result in lower similarity. 
Their study shows that LLM performance can differ across natural languages and that such differences can be analyzed through representation similarity to English.

Sun~\et~\cite{sun2024arxiv}~introduced FuxiTranyu, a multilingual LLM trained on 43 natural languages and 16 programming languages. While previous multilingual LLMs suffered from outdated or imbalanced language distribution of training data and remained undisclosed, FuxiTranyu aims to address these challenges. To address the training data issue, this paper manually controlled the language distribution. Also, its model was disclosed to the public. 
They evaluated FuxiTranyu on discriminative tasks and generative tasks. FuxiTranyu was compared with Llama2 and Mistral as English-centric LLMs, and BLOOM and PolyLM as multilingual LLMs. As a result, FuxiTranyu was the best multilingual LLM, but it was still worse than English-centric LLMs. On the other hand, the instruction tuned one outperformed the English-centric LLM in translation and summarization tasks.
This showed that developing multilingual LLMs requires careful consideration of the language distribution in the training data. Also, even if we build a balanced dataset, the performance of multilingual LLMs may not exceed that of English-centric LLMs.

Li et al.~\cite{li2025bridginglanguagegapenhancing} investigated multilingual prompt-based code generation and analyzed the impact of different natural languages on code LLM performance. They evaluated LLMs such as CodeLLaMa and GPT-4 on a translated MBPP dataset across six languages, including English and five non-English languages (Chinese, Spanish, Japanese, Russian, and Hindi). They compared multiple strategies, including direct prompting, prompt translation, chain-of-thought prompting, and fine-tuning with bootstrapped multilingual data. To further improve multilingual performance, they proposed a zero-shot cross-lingual transfer method that incorporates a multilingual encoder (LASER) and a projection mechanism to align multilingual representations with the LLM input space. Their results showed that code generation performance varies significantly across languages and that the proposed method improves performance across multiple languages.

Chen et al.~\cite{chen2026nlperturbator} proposed NLPerturbator, an automated framework for generating realistic natural language perturbations in code generation prompts. They defined multiple categories of perturbations, including lexical, syntactic, and semantic variations such as typos, paraphrases, and grammatical changes, and constructed benchmark datasets with manually validated perturbations. They evaluated several code LLMs, including Codex, CodeGen, and InCoder, on standard benchmarks such as HumanEval and MBPP using English prompts. Their experiments showed that even minor perturbations consistently degrade code generation performance across models, indicating that code LLMs are highly sensitive to variations in natural language prompts.

Wang et al.~\cite{wang2024exploringmultilingualbiaslarge} investigated multilingual bias in LLMs from both natural language and programming language perspectives using nine different large-scale code generation models (LCMs). Their study targeted English and Chinese as natural languages and Python, Java, and C++ as programming languages. The evaluated LCMs included StarCoder, CodeLlama, and DeepSeekCoder. 
They also investigated whether translating prompts into English could improve code generation performance. The results showed that self-translation translation performed by the same LLM used for code generation,
increased the bias, while external translation tools reduced the bias. Additionally, instruction tuning improved the performance of LCMs and reduced multilingual bias. 

Complementing these model-centred evaluations, Botto Tobar et al.~\cite{BottoTobar2025AI} examined the experiences of Spanish-speaking developers using AI coding tools. They found practical difficulties with non-English code elements and local naming conventions, showing that multilingual limitations affect not only benchmark accuracy but also developers’ acceptance and trust in generated suggestions.

Li et al.~\cite{li2024improvingnaturallanguagecapability} focused on the natural language processing capabilities of LLMs and proposed a new method to enhance LLMs from the perspective of natural language processing. They used LLMs such as GPT-4, GPT-3.5-turbo, WizardCoder (1B, 3B, 7B, and 15B), and CodeLlama (7B and 13B). 
Specifically, they extracted key phrases from prompts at the word, phrase, and sentence levels and emphasized them in the prompts. They showed that this method improved code generation performance.

Katzy et al.~\cite{katzy2025qualitativeinvestigationllmgeneratedmultilingual} evaluated five state-of-the-art code generation models, including CodeGemma, CodeLlama, CodeQwen1.5, GraniteCode, and StarCoder2, across five natural languages: Chinese, Dutch, English, Greek, and Polish. They conducted an open-coding study, resulting in a taxonomy of 26 distinct error categories in model-generated code comments. The study revealed significant variations in language cohesion, informativeness, and syntax adherence across different natural languages. Moreover, the authors found that standard evaluation metrics like BLEU, ROUGE, METEOR, and even neural metrics such as BERTScore failed to reliably reflect the correctness of generated comments across languages. Notably, there was a substantial overlap between expert-rated correct and incorrect comments, calling into question the effectiveness of these metrics in assessing multilingual code-related tasks.

\subsection{Novelty of This Study}
LLMs rely on training data, and it has been reported that the natural language used can introduce biases in the generated outputs~\cite{wang2024exploringmultilingualbiaslarge,koyanagi2023MSR}.
A previous study has shown that translating the input language into English can improve performance in natural language processing tasks using LLMs~\cite{liu2024arxiv}. This is likely because LLMs are primarily trained on English data, making English inputs more compatible with their training distribution.

However, many studies on LLM-based code generation have concentrated on widely used benchmarks, such as HumanEval and MBPP. As a result, existing findings may be biased toward these benchmarks, and the impact of natural language on other types of code generation tasks, such as programming contest problems and specification-based code generation tasks, remains insufficiently understood. 
Moreover, existing studies often focus on English and Chinese, leaving limited evidence on how other natural languages, such as Japanese, affect code generation performance.

Our initial analysis~\cite{koyanagi2023MSR} evaluated the impact of three natural languages (Japanese, Chinese, and English) on the code generation $\mbox{\sl Accuracy}$ of GitHub Copilot. The results showed that the language bias exists in code generation tasks.
For example, GitHub Copilot performed best on Japanese problems. However, this study only targeted AtCoder (\textbf{\emph{benchmark limitation}}) and GitHub Copilot (\textbf{\emph{model limitation}}). 
To address these two limitations, in this article,
we extended the evaluation to include
LeetCode and BigCodeBench, and evaluated seven LLMs, including GPT-4o, o3-mini, DeepSeek-V3.2, GitHub Copilot, Llama3, and Qwen2.5-Coder.
Moreover, we investigated a mitigation strategy to relieve the language bias in code generation. Specifically, we evaluated the impact of translation on the code generation $\mbox{\sl Accuracy}$ of LLMs. 
This manuscript is motivated by the need to examine whether language bias observed in our initial analysis generalizes across different datasets and LLMs, and whether translation can mitigate such bias. Unlike previous studies that primarily focus on widely used benchmarks or a limited set of natural languages, our study provides a broader empirical analysis of language bias and translation effects across programming contest problems and specification-based code generation tasks.

\section{Study Design} \label{sec:setup}
This section describes the study design.
Specifically, we explain target natural languages, data collection, studied LLMs, and experimental workflow.

\subsection{Target Natural Languages}
In this study, we focus on English, Chinese, and Japanese. We selected these languages based on the following three criteria:
\begin{itemize}
  \item
  The target language must be a high-resource language~\cite{liu2024arxiv}. This requirement ensures that we can collect datasets of sufficient size for our study. In practice, widely used benchmarks and programming contests (e.g., HumanEval) are already provided in high-resource languages, which confirms the feasibility of this criterion. 
  \item
  The target language must either be spoken by at least one of the authors as a native language or be widely used in academic contexts (i.e., English). This allows us to ensure the quality of the input prompts.  
  For native languages, we can manually verify correctness; for English, we assume we can verify correctness to an acceptable degree.
  \item
  The target language must be among those used in Koyanagi~\et~\cite{koyanagi2023MSR}. This requirement ensures that we can replicate and extend our previous study.
\end{itemize}

\subsection{Data Collection}
\label{sec:setup:data-collection}
To answer the research questions, we targeted two online programming contest platforms and one specification-based code generation benchmark: AtCoder~\cite{AtCoder}, LeetCode~\cite{LeetCode}, and BigCodeBench~\cite{BigCodeBench}. 
We selected programming contest problems (AtCoder and LeetCode) since they provide well-structured problem statements in multiple natural languages, test cases, and multiple difficulty levels. These ensure that we can compare the impact of natural language differences on code generation tasks.
Also, these platforms are widely used for evaluating code generation models~\cite{koyanagi2023MSR,mathew2021crosslanguage,yan2023ASE,tian2023chatgptultimateprogrammingassistant,nguyen2022MSR,mayer2024arxiv,Billah_2024}.
\red{However, programming contest problems may not fully reflect real-world programming scenarios. 
Therefore, we additionally included BigCodeBench~\cite{BigCodeBench},
a specification-based code generation benchmark that provides code generation tasks beyond competitive programming settings.
In this section, we describe our data collection process.}

\smallskip{}
\noindent{}
\textbf{AtCoder:}
AtCoder is one of the most popular and largest programming contest platforms, targeting users ranging from beginners to experts. It has also been widely used in previous studies~\cite{koyanagi2023MSR,mathew2021crosslanguage,yan2023ASE}. 

We used problems from the AtCoder Beginner Contest (ABC), a weekly contest series.
As of October 9, 2024, the ABC hosted 374 contests and supported problems in Japanese and English.
A problem consists of a problem statement, constraints, input/output format, and sample input/output.

Each contest includes up to eight problems featuring difficulty levels: A, B, C, D, E, F, G, and Ex (H), and the difficulty of the problems is arranged alphabetically, with A being the easiest. 
For this study, we used the same problems as in our previous research~\cite{koyanagi2023MSR}.
It includes a total of 756 questions with difficulty levels A, B, C, and D retrieved from contest numbers 99 to 287. 
Hence, each difficulty level is represented by 189 questions.
Below, we provide four examples of AtCoder problems ranging from difficulty levels A to D. We extracted the main problem statement only. 

\smallskip
\noindent
The question with \textbf{A} level~\cite{AtCoder123_a}: 
\begin{formal}
In AtCoder city, there are five antennas standing in a straight line. They are called Antenna $A, B, C, D$ and $E$ from west to east, and their coordinates are $a, b, c, d$ and $e$, respectively.
Two antennas can communicate directly if the distance between them is $k$ or less, and they cannot if the distance is greater than $k$.
Determine if there exists a pair of antennas that cannot communicate directly.
Here, assume that the distance between two antennas at coordinates $p$ and $q (p < q)$ is $q - p$.
\end{formal}

\smallskip
\noindent
The question with \textbf{B} level~\cite{AtCoder123_b}:
\begin{formal}
The restaurant AtCoder serves the following five dishes:

ABC Don (rice bowl): takes $A$ minutes to serve.

ARC Curry: takes $B$ minutes to serve.

AGC Pasta: takes $C$ minutes to serve.

APC Ramen: takes $D$ minutes to serve.

ATC Hanbagu (hamburger patty): takes $E$ minutes to serve.

Here, the time to serve a dish is the time between when an order is placed and when the dish is delivered.
This restaurant has the following rules on orders:
An order can only be placed at a time that is a multiple of 10 (time 0, 10, 20, ...).
Only one dish can be ordered at a time.
No new order can be placed when an order is already placed and the dish is still not delivered, but a new order can be placed at the exact time when the dish is delivered.
E869120 arrives at this restaurant at time 0. He will order all five dishes. Find the earliest possible time for the last dish to be delivered.
Here, he can order the dishes in any order he likes, and he can place an order already at time 0.
\end{formal}

\smallskip
\noindent
The question with \textbf{C} level~\cite{AtCoder123_c}:
\begin{formal}
In 2028 and after a continuous growth, AtCoder Inc. finally built an empire with six cities (City 1, 2, 3, 4, 5, 6)!

\noindent
There are five means of transport in this empire:

Train: travels from City 1 to 2 in one minute. A train can occupy at most $A$ people.

Bus: travels from City 2 to 3 in one minute. A bus can occupy at most $B$ people.

Taxi: travels from City 3 to 4 in one minute. A taxi can occupy at most $C$ people.

Airplane: travels from City 4 to 5 in one minute. An airplane can occupy at most $D$ people.

Ship: travels from City 5 to 6 in one minute. A ship can occupy at most $E$ people.

For each of them, one vehicle leaves the city at each integer time (time 0, 1, 2, ...).
There is a group of $N$ people at City 1, and they all want to go to City 6.
At least how long does it take for all of them to reach there? 
You can ignore the time needed to transfer.  
\end{formal}

\smallskip
\noindent
The question with \textbf{D} level~\cite{AtCoder123_d}:
\begin{formal}
The Patisserie AtCoder sells cakes with number-shaped candles.
There are $X, Y$ and $Z$ kinds of cakes with 1-shaped, 2-shaped and 3-shaped candles, respectively.
Each cake has an integer value called deliciousness, as follows:

The deliciousness of the cakes with 1-shaped candles are $A_1, A_2, ..., A_X$.

The deliciousness of the cakes with 2-shaped candles are $B_1, B_2, ..., B_Y$.

The deliciousness of the cakes with 3-shaped candles are $C_1, C_2, ..., C_Z$.

Takahashi decides to buy three cakes, one for each of the three shapes of the candles, to celebrate ABC 123.
There are $X \times Y \times Z$ such ways to choose three cakes.
We will arrange these $X \times Y \times Z$ ways in descending order of the sum of the deliciousness of the cakes.
Print the sums of the deliciousness of the cakes for the $first, second, ..., K$-$th$ ways in this list.  
\end{formal}

\smallskip{}
\noindent{}
\textbf{LeetCode:}
LeetCode~\cite{LeetCode} is also a popular platform widely used for coding interview preparation. 
It officially supports English and Chinese.
It has also been used in previous studies~\cite{tian2023chatgptultimateprogrammingassistant,nguyen2022MSR,mayer2024arxiv,Billah_2024}.

\red{LeetCode consists of 3,313 problems,\footnote{Final-access: October 9, 2024} categorized into three difficulty levels: easy, medium, and hard.
In this study, we use a subset of 100 LeetCode problems for evaluation.
The dataset consists of 16 easy, 52 medium, and 32 hard problems.
We do not use the full set of 3,313 problems due to the substantial effort required to prepare test cases for each problem, as LeetCode does not always provide them. 
We prepared 15 non-duplicated test cases involving an input and an output for each problem with the assistance of ChatGPT. The generated test cases were validated one by one using the official LeetCode website.
The prompt used to generate the test cases is included in our replication package.
This test case preparation process requires substantial manual effort. Hence, we determined that using 100 problems would be a feasible scale for this study.}
A problem consists of a problem statement, sample input/output, and constraints.

Below, we provide three examples of LeetCode problems ranging from difficulty levels easy to hard. We extracted the main problem statement only. 

\smallskip
\noindent
The question with \textbf{Easy} level:
\begin{formal}
  Given an array of integers nums and an integer target, return indices of the two numbers such that they add up to target.
  You may assume that each input would have exactly one solution, and you may not use the same element twice.
  You can return the answer in any order.
\end{formal}  

\smallskip
\noindent
The question with \textbf{Medium} level:
\begin{formal}
  A super ugly number is a positive integer whose prime factors are in the array primes.
  Given an integer n and an array of integers primes, return the $n^{th}$ super ugly number.
  The $n^{th}$ super ugly number is guaranteed to fit in a 32-bit signed integer.
\end{formal}

\smallskip
\noindent
The question with \textbf{Hard} level:
\begin{formal}
  Given an $m \times n$ matrix matrix and an integer $k$, return the max sum of a rectangle in the matrix such that its sum is no larger than $k$.
  It is guaranteed that there will be a rectangle with a sum no larger than $k$.
\end{formal}

\smallskip{}
\noindent{}
\red{
\textbf{BigCodeBench:}
BigCodeBench~\cite{zhuo2024bigcodebench} is a benchmark that incorporates diverse real-world development scenarios.
It provides more context-rich code generation tasks beyond competitive programming settings.
}

\red{
Each task consists of a problem description and corresponding test cases, and the dataset contains 1,140 tasks in total.
It also provides an official evaluation framework and leaderboard~\cite{BigCodeBench}, enabling standardized comparison across models.
}

\red{
In this study, to ensure statistical reliability while controlling computational cost, we determine the sample size based on confidence interval estimation.
Specifically, assuming a 95\% confidence level and a 5\% margin of error, the required sample size is 288.
Accordingly, we randomly sample 288 tasks from the full dataset.
}

\red{Below, we present an example task from BigCodeBench.}
\begin{formal}
Replace certain values in a DataFrame with a dictionary mapping and calculate the Pearson correlation coefficient between each pair of columns.
\end{formal}

\smallskip{}
\noindent{}
\textbf{Translating Problems:}
AtCoder, LeetCode, and BigCodeBench do not fully cover the three languages. AtCoder lacks Chinese support, LeetCode lacks Japanese support, and BigCodeBench primarily provides problem descriptions in English.
To cover all three languages, we translated the English problems into Chinese for AtCoder, into Japanese for LeetCode, and into Chinese and Japanese for BigCodeBench.

\begin{figure*}[t!]
    \begin{center}
    \includegraphics[width=0.8\linewidth]{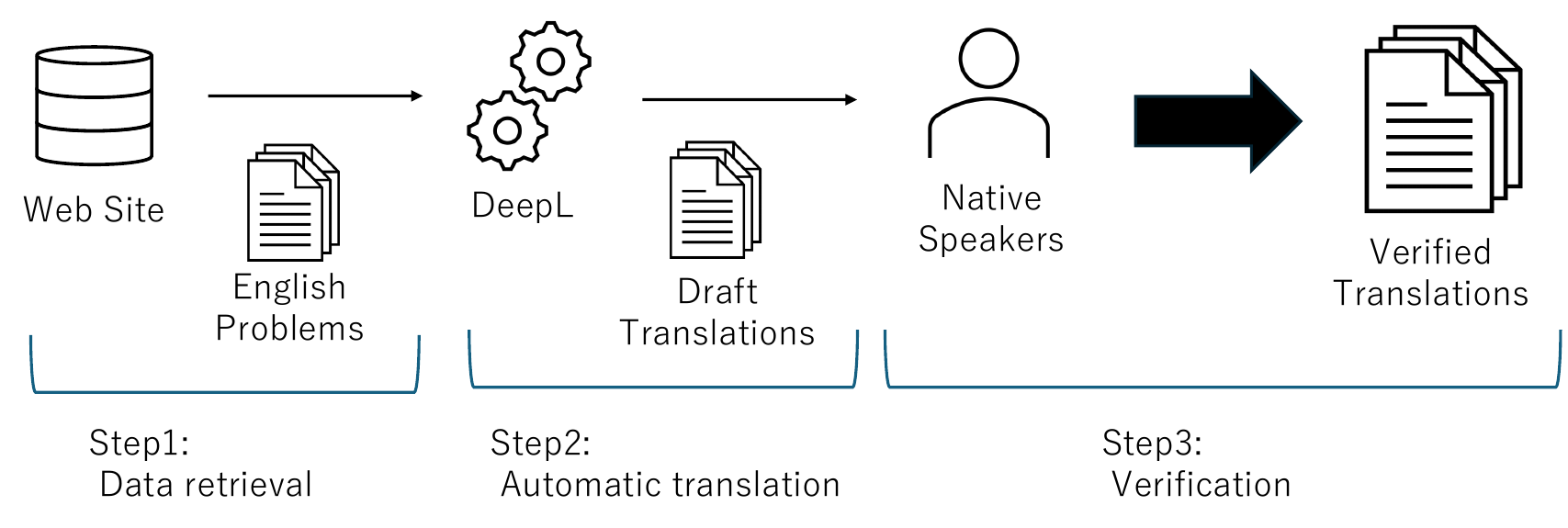}
    \caption{Overview of our translation process}
    \label{fig:trans_flow}
    \end{center}
\end{figure*}

Figure~\ref{fig:trans_flow} provides an overview of the translation process we follow in this study. Each step is explained below.
\begin{itemize}[leftmargin=11mm]
  \item[Step 1:] \textbf{(Automatic translation)} We utilized DeepL\footnote{https://www.deepl.com/} to automatically translate the English problems into Chinese for AtCoder, into Japanese for LeetCode, and into Chinese and Japanese for BigCodeBench.
  \item[Step 2:] \textbf{(Verification)} Translated problems were reviewed and refined by the authors of this paper, who are native speakers of Japanese and Chinese. 
\end{itemize}

Step 1 provides a draft translation. We selected English as the source language.
This is because (1) \red{all studied datasets provide problems or task descriptions in English}
and (2) English is the most common language used in the programming community\cite{guo2018chi,soosai2018does}.

DeepL is a well-known machine translation service and provides high-quality translations\cite{Linlin2024Procedia,Sebo2024PLOS}.
For instance, Li~\cite{Linlin2024Procedia} compared DeepL with other popular translation services such as Google Translate and Microsoft Translator, and found that DeepL achieved a semantic similarity score of 94.13 out of 100, outperforming the others in terms of accuracy, fluency, and naturalness. 
\red{However, the reported performance of DeepL does not necessarily guarantee translation quality for programming contest problems or for translations into Chinese and Japanese. 
Therefore, we used DeepL only to obtain draft translations and then manually reviewed and refined the translated problem statements in Step 2. 
This process allowed us to control the quality of translations used in our experiments.}
In fact, DeepL provides high-quality translations, but it is not perfect, such as ignoring some parts of the text and not translating them.

\red{To systematically identify and correct such issues, we developed a coding book for systematically identifying and correcting translation errors.
To build the coding book, we conducted an open coding process with a saturation criterion~\cite{hirao2019FSE}. Specifically, annotators reviewed the translated problems, identified translation errors, and developed codes.
To ensure the completeness of the coding book, we applied a saturation criterion.
Following prior studies~\cite{hirao2019FSE}, we continued the coding process until no new codes emerged over 50 consecutive samples.
Whenever a new code was identified, the counting was reset, and the process continued until saturation was reached.
This iterative process resulted in a stable and reliable coding book, which was then used to guide the manual verification and refinement of the translated problems.
The coding book is available in our replication package.\footnote{\url{https://github.com/posl/EMSE-2025-Multi-NL-CodeGeneration}}}

Note that this process generate problems that are not existed in the training data of LLMs. Hence, the evaluation with these problems is not biased by the training data and remedies the data leaking problem. 



\subsection{Studied LLMs}
Developers and researchers can use various LLMs to generate code from natural languages~\cite{liu2024TSE,du2024ICSE,fakih2024ICSE,sun2024ASE,huang2024ASE,chen2024ASE,feng2024ASE}.
Indeed, there exist many LLM services, such as ChatGPT~\cite{ChatGPT}, DeepSeek~\cite{DeepSeek-service}, and GitHub Copilot~\cite{Copilot}. Also, Hugging Face hosts various LLMs and provides a simple interface for anyone to use LLMs. Developers and researchers compare these LLMs and update the leaderboard of the best LLMs every day. Hence, the selection of the studied LLMs is important to ensure the validity of the study.

\red{In this study, we selected LLMs based on the following criteria, considering their practical usage in software development (as of May 2024):}
\begin{itemize}
  \item LLMs are widely used in the programming community.
  \item \red{LLMs cover both commercial (API-based) and open-source (locally deployable) models.}
  \item LLMs are developed in different countries.
\end{itemize}

\red{The first criterion aims to evaluate representative LLMs that are commonly adopted in practice~\cite{du2024ICSE,fakih2024ICSE,sun2024ASE,huang2024ASE,chen2024ASE,feng2024ASE}. 
The second criterion is introduced to reflect the diversity of real-world usage environments, including both API-based usage and local deployment. 
The third criterion is intended to analyze the impact of differences in training data distributions, particularly linguistic and cultural factors, on model performance. 
Prior work has shown that language-centric training data can influence the behavior of LLMs~\cite{sun2024arxiv}. It should be noted that we do not select an LLM developed in Japan, as there are no well-known LLMs in Japan that meet the above criteria.}

\red{Based on these criteria, we selected the following LLMs:}
\red{
\begin{itemize}
    \item \textbf{Closed-source models}: GPT-4o and o3-mini (OpenAI, United States), DeepSeek (DeepSeek,China), and GitHub Copilot (Microsoft, United States)
    \item \textbf{Open-source models}: Llama~3 (Meta, United States) and Qwen2.5-Coder (Alibaba, China)
\end{itemize}
}

\red{
By including both closed-source and open-source models, our study covers a broader range of practical usage scenarios. 
In particular, Llama~3 and Qwen2.5-Coder are locally deployable models and represent usage settings that do not rely on external APIs.}

Finally, both GPT and DeepSeek provide APIs to use. We can easily automate the code generation process using these APIs.
However, it is widely used in the programming community and provides a simple interface to use. Hence, we selected Copilot as one of the studied LLMs.

For GPT-4o, o3-mini, DeepSeek, Llama~3, and Qwen2.5-Coder, we can specify the version. We used gpt-4o-2024-05-13, o3-mini-2025-01-31, DeepSeek-V3.2, Llama-3.1-8B-Instruct, Qwen2.5-Coder-0.5B, and Qwen2.5-Coder-14B respectively.
Since Copilot does not provide version specification, we used the service with the default settings with Visual Studio Code (VSCode).
The code generation experiments using Copilot were conducted between May 24 and June 5, 2024.

\subsection{Machine Translation for Improving the Code Generation Accuracy}
As a mitigation strategy to relieve the language bias, we explore the use of machine translation to improve the accuracy of code generation.
We translate the problems from the language with the lowest $\mbox{\sl Accuracy}$ in code generation into the language with the highest $\mbox{\sl Accuracy}$ \red{using multiple machine translation tools, including DeepL, Google Translate, and GPT-4o.}
A previous study reported that translating the input language into English can improve performance~\cite{liu2024arxiv}. Based on this finding, we hypothesize that translation into a certain language (e.g., English) may lead to better code generation performance compared to other languages.

It should be noted that this translation process serves a different purpose than the one described in \sec{sec:setup:data-collection}.
\sec{sec:setup:data-collection} focuses on creating ground-truth data for each language with human verification to ensure semantic equivalence across languages. 
In contrast, this translation process aims solely to improve code generation accuracy in practical scenarios
by using machine translation tools without human verification.
Therefore, when translating problem statements into English, the resulting English prompts were sometimes machine-translated versions of non-English prompts rather than the original English statements.
By omitting human verification, we simulate a practical use case and
evaluate the impact of machine-translated prompts.
We then evaluate code generation performance on the translated problems and report the effectiveness of this approach.

\begin{figure*}[t!]
    \begin{center}
    \includegraphics[width=\linewidth]{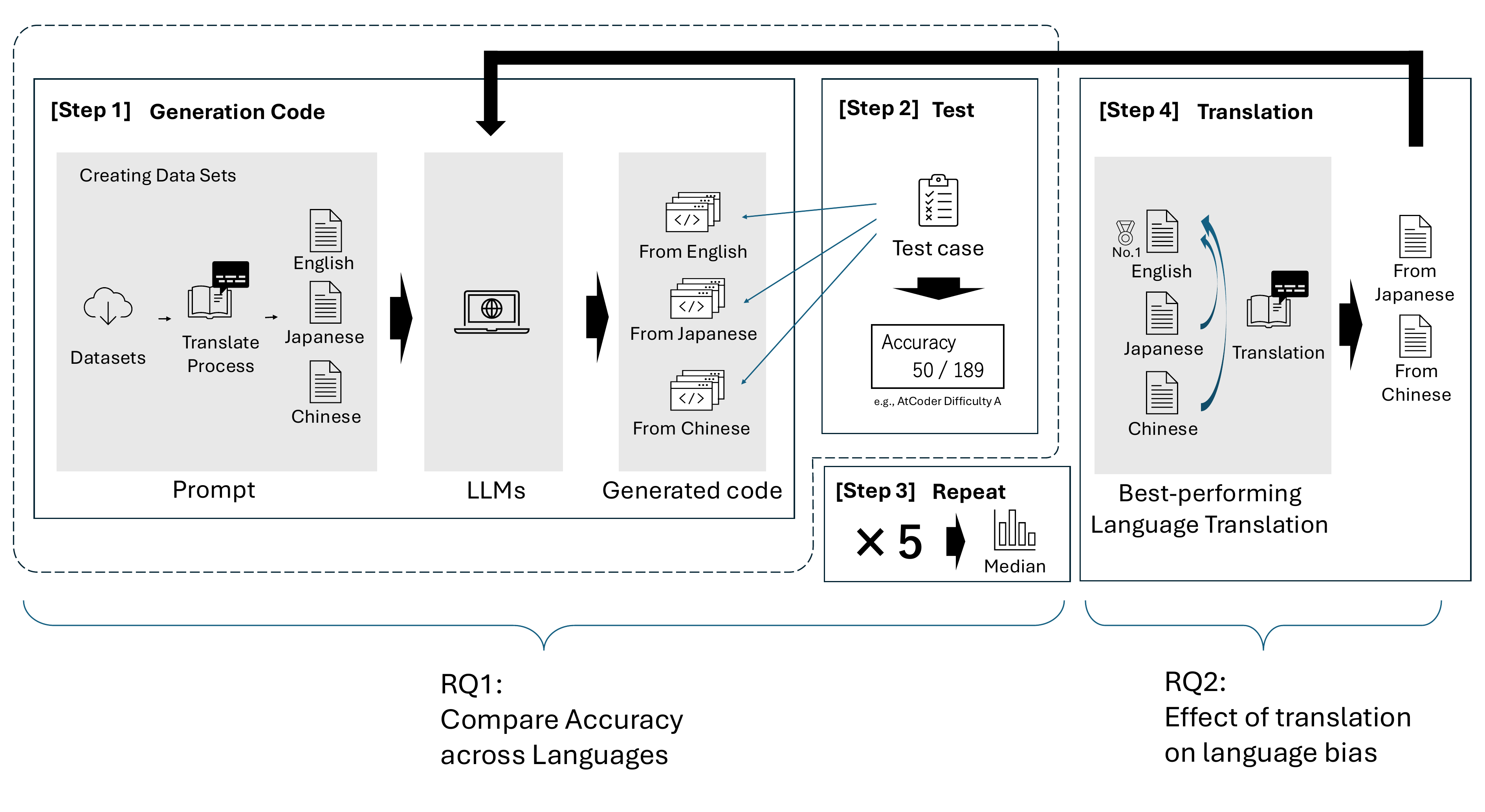}
    \caption{Overview of experimental workflow}
    \label{fig:overall}
    \end{center}
\end{figure*}

\subsection{Experimental Workflow}
Figure~\ref{fig:overall} illustrates the overall of our experimental workflow, which consists of four steps. 

\begin{figure*}[t!]
    \begin{center}
    \includegraphics[width=0.8\linewidth]{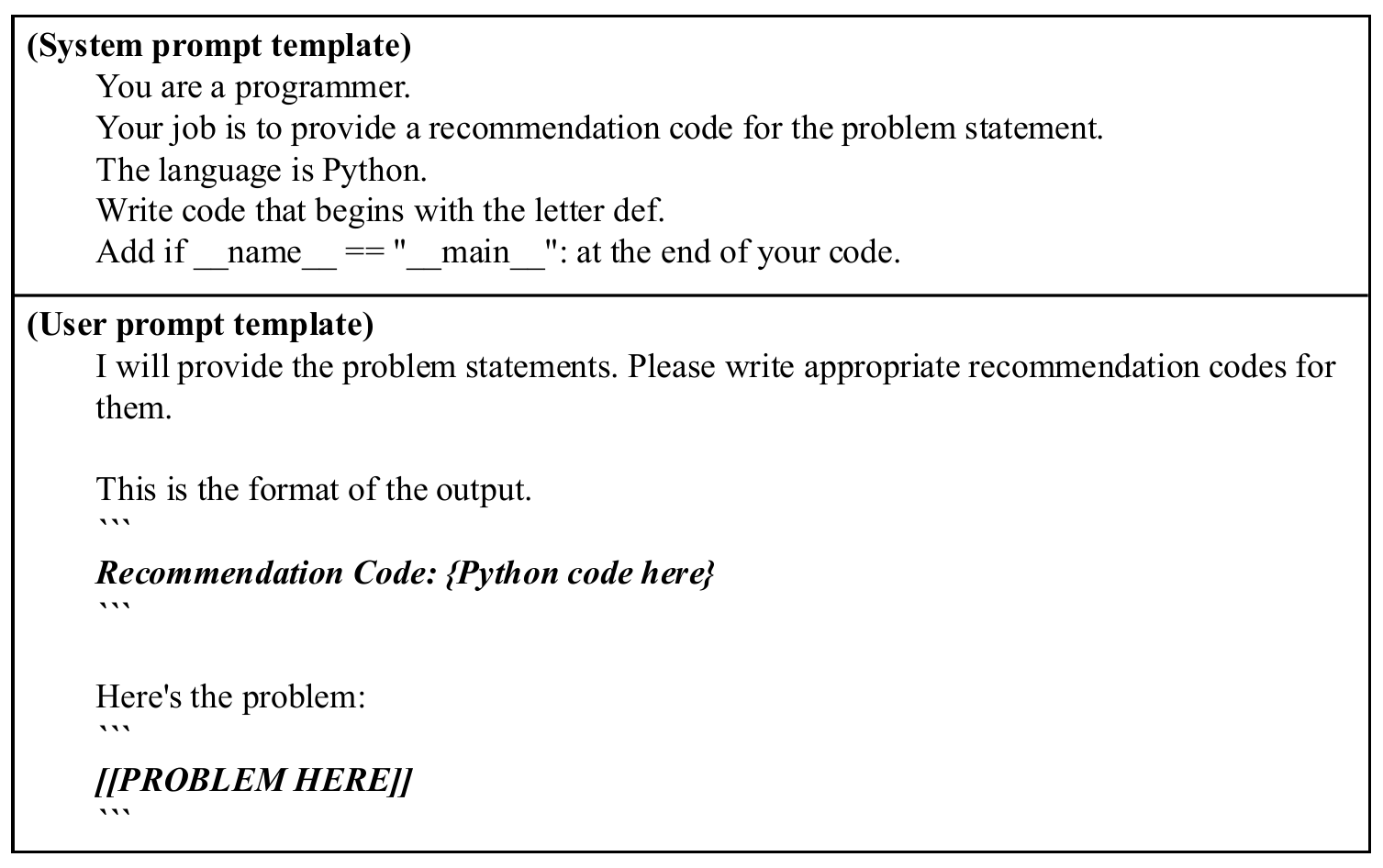}
    \caption{System and user prompts for GPT-4o, o3-mini, DeepSeek-V3.2, Llama3, and Qwen2.5-Coder}
    \label{fig:prompt-skeleton-gpt}
    \end{center}
\end{figure*}

\begin{figure*}[t!]
    \begin{center}
    \includegraphics[width=0.8\linewidth]{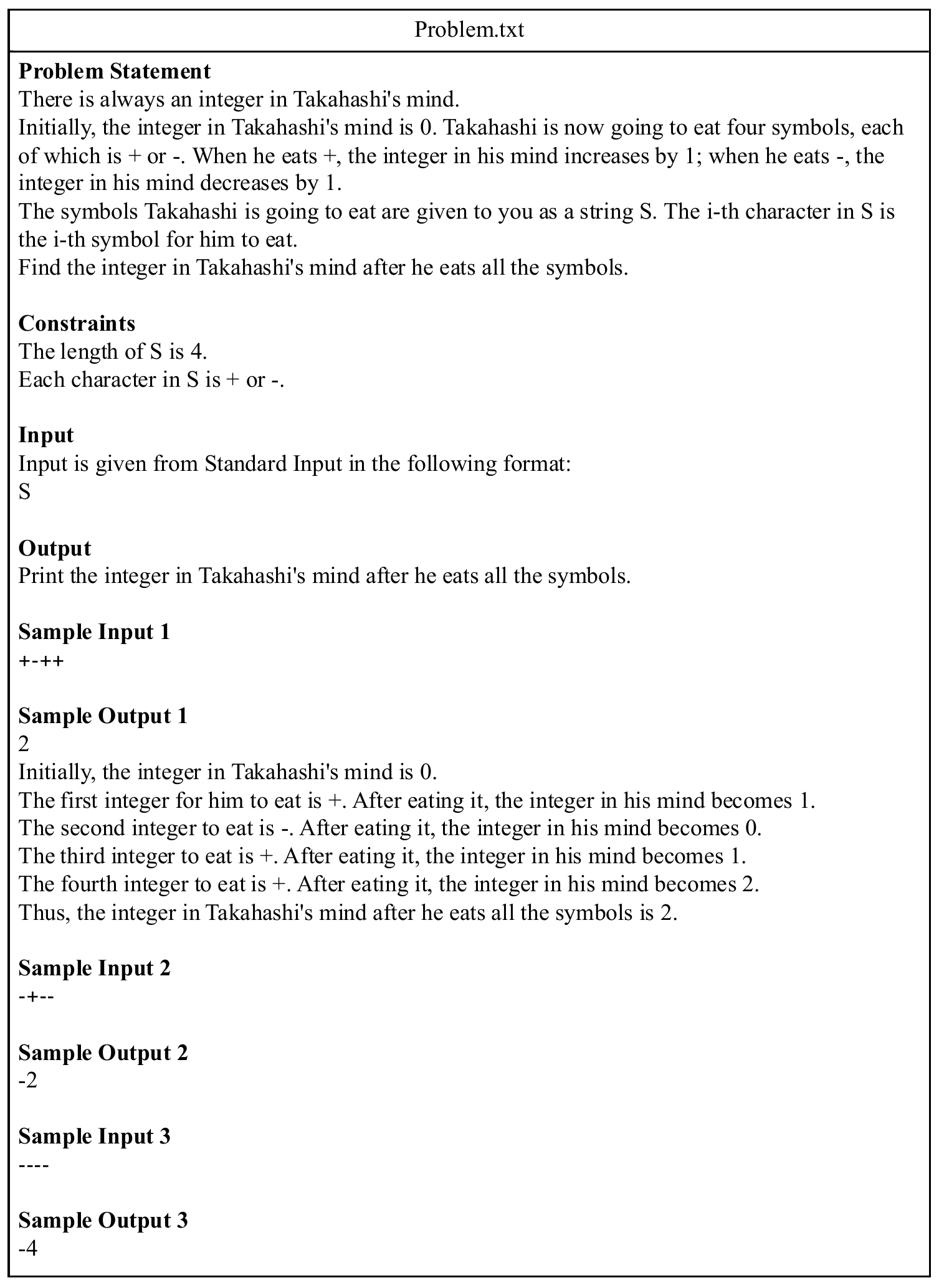}
    \caption{Problem Statement of AtCoder\cite{AtCoder101_a}}
    \label{fig:prompt-skeleton-GPT-problem}
    \end{center}
\end{figure*}

\smallskip{}
\noindent{}
\textbf{[Step 1: Generate code]}
In this study, we invoked LLMs to generate code written in Python. 
We first prepared prompt templates for each language. 
Figure~\ref{fig:prompt-skeleton-gpt} shows our English prompt template (we designed this prompt template before this study). 
We replaced the placeholder ``[[PROBLEM HERE]]'' with a concrete problem, such as an AtCoder problem. 
Specifically, the prompt embeds both the problem statement and supplementary information, such as input/output formatting details and their input/output examples. 
Figure~\ref{fig:prompt-skeleton-GPT-problem} illustrates an example of a question text that includes these elements.
Subsequently, code is generated by applying the constructed prompt to GPT-4o, o3-mini, DeepSeek-V3.2, Llama~3, and Qwen2.5-Coder models.

\begin{figure*}[t!]
    \begin{center}
    \includegraphics[width=0.8\linewidth]{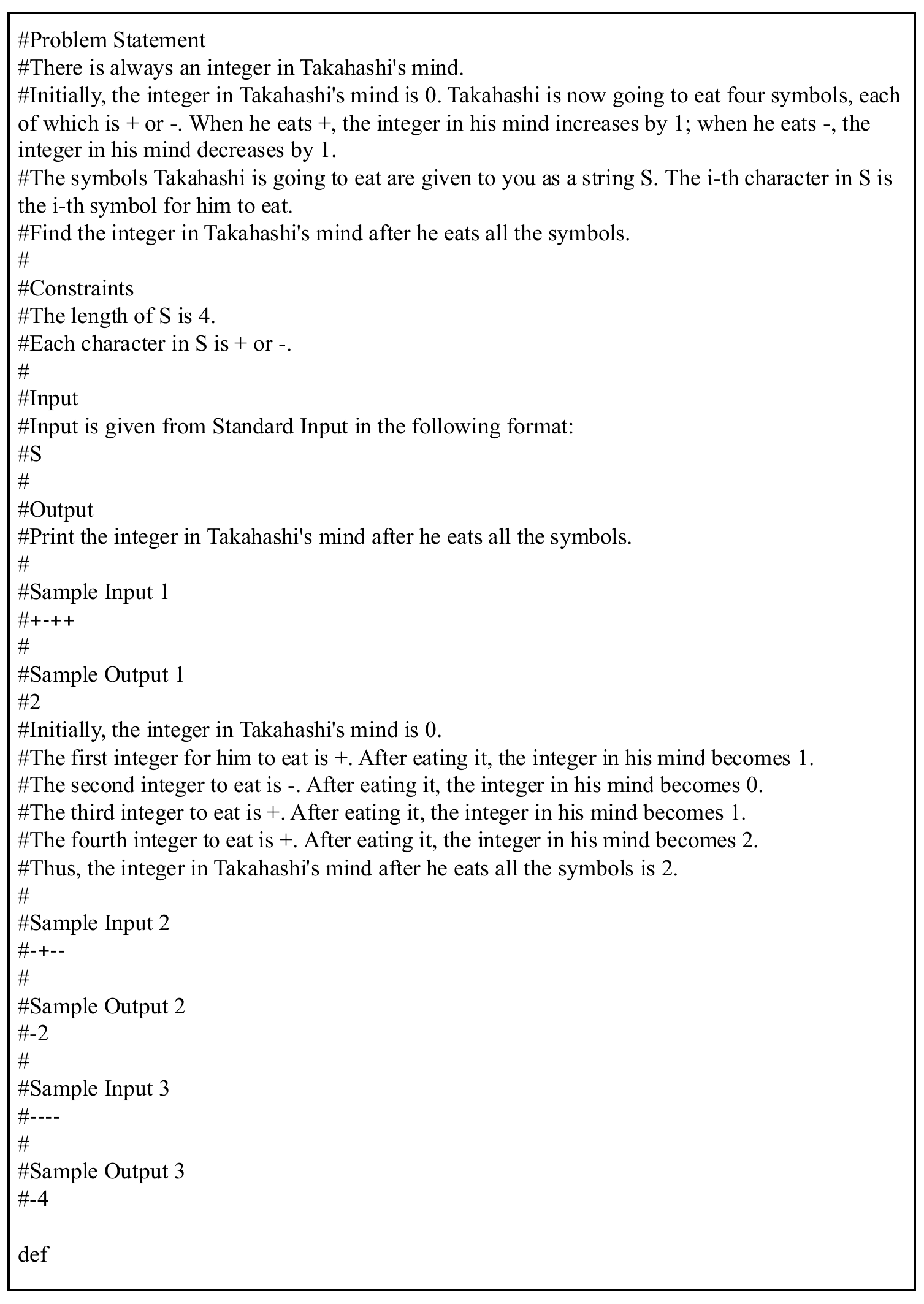}
    \caption{Code file used as input for Copilot\cite{AtCoder101_a}}
    \label{fig:prompt-skeleton-copilot}
    \end{center}
\end{figure*}

Copilot does not have APIs. Instead, it is integrated into VSCode and suggests code directly in a target source code file. 
To accommodate this interface, we prepared a source code file that includes the problem statement (as shown in Figure~\ref{fig:prompt-skeleton-GPT-problem}) along with supplementary information, such as input/output format descriptions, embedded as comments.  
Additionally, a def statement was appended to the end of the file to guide Copilot toward suggesting Python function implementations.
Figure~\ref{fig:prompt-skeleton-copilot} shows an example of a file.\footnote{All prompt templates in other natural languages can be found in our replication package.}
Copilot was activated following the comment section, and the code was subsequently generated based on the preceding context.

GPT-4o and DeepSeek-V3.2 occasionally generate two code snippets for a single problem. In such cases, both models explicitly indicate which snippet is considered more time-efficient within the output text. Since time efficiency is an important factor in programming contests, we selected the code identified as the more efficient one. Due to the nature of GitHub Copilot, it typically generates multiple code snippets. In these cases, we only used the first proposed code, as it is reasonable to assume that developers would typically choose the first option~\cite{ralph2016ease}.

For each combination of language and difficulty level, each LLM generates 189 code snippets
for AtCoder. For LeetCode, the number of generated code snippets
varies by difficulty level: 16 code snippets for easy, 52 for medium, and 32 for hard, totaling 100 code snippets for each language.
For BigCodeBench, each LLM generates code snippets for 288 tasks for each language.
We invoked GPT-4o from July 12 to July 27, 2024, o3-mini from Mar 6 to Mar 13, 2025, DeepSeek-V3.2 from Feb 6 to Apr 8, 2026, and Copilot from May 24 to June 5, 2024.

\smallskip{}
\noindent{}
\textbf{[Step 2: Distinguish correct code snippets]}
AtCoder and LeetCode provide test cases for each problem, and BigCodeBench provides an official evaluation framework with corresponding test cases.
We executed these test cases and distinguished whether the generated code passes all test cases. 
We marked the code as ``Pass'' if it passed all test cases and ``Fail'' if it failed at least one test case.  
Since we aim to evaluate the impact of the natural language differences on code generation, we calculated the percentage of code marked as ``Pass'' as $\mbox{\sl Accuracy}$ for each language and problem combination.
For example, if 50 of the AtCoder A problems in Japanese are marked as ``Pass,'' the $\mbox{\sl Accuracy}$
is calculated as $(50/189) \times 100 \approx 25.3\%$.

\smallskip{}
\noindent{}
\textbf{[Step 3: Repeat Steps 1 and 2 five times]}
Since LLMs include randomness, we repeated Step 1 and Step 2 five times. This process yields five $\mbox{\sl Accuracy}$ values for each language and problem combination. 
To mitigate the effect of randomness in the generated code, we calculated the median value of $\mbox{\sl Accuracy}$
and used it to evaluate the impact of natural language differences on code generation.

\smallskip{}
\noindent{}
\textbf{[Step 4: Translate prompts]}
We first translate the problem statements in the language with the lowest $\mbox{\sl Accuracy}$ into the language with the highest $\mbox{\sl Accuracy}$ 
using DeepL, Google Translate, and GPT-4o.
The lowest $\mbox{\sl Accuracy}$ language and the highest $\mbox{\sl Accuracy}$ language are determined by the results of Step 3.
Translated problems were then used as input for code generation. We invoked LLMs to generate code from translated problems and re-evaluated $\mbox{\sl Accuracy}$ following the procedures outlined in Steps 1 through 3.


\section{Influence of Problem Statement Language}
\label{sec:rq1}
In this section we address RQ1: \RQtwo{}


\subsection{Approach}
We analyze the $\mbox{\sl Accuracy}$. 
Also, to assess code quality in addition to functional correctness, we measure the number of coding rule violations with Pylint~\cite{pylint} and calculate cyclomatic complexity for each output.
Pylint is a Python linter that identifies stylistic and syntactic issues (e.g., naming conventions, unused imports, and missing docstrings). 
These analyses (functional correctness, coding rule violations, and cyclomatic complexity) allow us to examine not only whether the models solve the problems, but also whether these aspects of the generated code differ across natural languages.

\begin{table}[p]
\caption{Median $\mbox{\sl Accuracy}$ for each natural language and LLM}
\centering
\label{tab:all}
\scalebox{0.8}{
\begin{tabular}{lllrrr}
\hline
\multicolumn{1}{c}{Problems} & \multicolumn{1}{c}{Difficulty} & \multicolumn{1}{c}{Model} & \multicolumn{1}{c}{English} & \multicolumn{1}{c}{Japanese} & \multicolumn{1}{c}{Chinese} \\ \hline
\multirow{24}{*}{AtCoder} & \multirow{6}{*}{A} & 4o & 92.6\% & 92.6\% & \underline{\textbf{94.2\%}} \\
 &  & o3 & \underline{94.7\%} & \underline{\textbf{95.8\%}} & \underline{94.2\%} \\
 &  & deep & \underline{94.7\%} & \textbf{95.2\%} & \underline{94.2\%} \\
 &  & Copilot & 30.7\%  & \textbf{47.6\%}   & 34.9\%  \\ \cdashline{3-6}
 &  & llama3 & \textbf{67.2\%} & 51.3\% & 59.3\% \\
 &  & qwen14b & \textbf{87.8\%} & 76.7\% & 83.6\% \\
 &  & qwen05b & \textbf{9.0\%} & 2.1\% & 4.8\% \\
\cline{2-6}
 & \multirow{6}{*}{B} & 4o & 88.4\% & \textbf{91.0\%} & 87.3\% \\
 &  & o3 & 85.2\% & \textbf{90.5\%} & 87.8\% \\
 &  & deep & \underline{92.6\%} & \underline{\textbf{93.1\%}} & \underline{91.5\%} \\
 &  & Copilot & 31.2\%  & \textbf{38.1\%}   & 26.5\%  \\\cdashline{3-6}
 &  & llama3 & \textbf{46.6\%} & 33.9\% & 39.2\% \\
 &  & qwen14b & \textbf{76.2\%} & 49.7\% & 69.3\% \\
 &  & qwen05b & \textbf{2.1\%} & 0.0\% & 1.1\% \\
\cline{2-6}
 & \multirow{6}{*}{C} & 4o & \textbf{77.8\%} & 74.6\% & 73.5\% \\
 &  & o3 & 67.7\% & \underline{\textbf{79.9\%}} & 71.4\% \\
 &  & deep & \underline{79.9\%} & 78.8\% & \underline{\textbf{80.4\%}} \\
 &  & Copilot & \textbf{25.9\%}  & 24.9\%   & 18.0\%  \\\cdashline{3-6}
 &  & llama3 & \textbf{10.1\%} & 9.5\% & 8.5\% \\
 &  & qwen14b & \textbf{36.5\%} & 19.0\% & 33.3\% \\
 &  & qwen05b & \textbf{0.0\%} & \textbf{0.0\%} & \textbf{0.0\%} \\
\cline{2-6}
 & \multirow{6}{*}{D} & 4o & \textbf{47.1\%} & \textbf{47.1\%} & 45.5\% \\
 &  & o3 & 50.3\% & \underline{\textbf{63.5\%}} & 52.4\% \\
 &  & deep & \underline{52.9\%} & 55.0\% & \underline{\textbf{58.2\%}} \\
 &  & Copilot & \textbf{9.0\%}   & 7.4\%    & 7.40\%  \\\cdashline{3-6}
 &  & llama3 & \textbf{1.6\%} & 0.0\% & \textbf{1.6\%} \\
 &  & qwen14b & \textbf{11.1\%} & 4.8\% & 9.5\% \\
 &  & qwen05b & \textbf{0.0\%} & \textbf{0.0\%} & \textbf{0.0\%} \\
\hline
\multirow{18}{*}{LeetCode} & \multirow{6}{*}{Easy} & 4o & 75.0\% & \underline{\textbf{87.5\%}} & 81.2\% \\
 &  & o3 & \underline{\textbf{87.5\%}} & \underline{\textbf{87.5\%}} & \underline{\textbf{87.5\%}} \\
 &  & deep & \underline{\textbf{87.5\%}} & 68.8\% & \underline{\textbf{87.5\%}} \\\cdashline{3-6}
 &  & llama3 & 56.2\% & 56.2\% & \textbf{62.5\%} \\
 &  & qwen14b & 75.0\% & 62.5\% & \textbf{81.2\%} \\
 &  & qwen05b & \textbf{12.5\%} & 6.2\% & \textbf{12.5\%} \\
\cline{2-6}
 & \multirow{6}{*}{Medium} & 4o & \textbf{75.0\%} & 66.7\% & 73.1\% \\
 &  & o3 & 73.1\% & 58.8\% & \textbf{75.0\%} \\
 &  & deep & \underline{82.7\%} & \underline{68.6\%} & \underline{\textbf{86.5\%}} \\\cdashline{3-6}
 &  & llama3 & \textbf{38.5\%} & 25.5\% & \textbf{38.5\%} \\
 &  & qwen14b & 69.2\% & 66.7\% & \textbf{73.1\%} \\
 &  & qwen05b & \textbf{5.8\%} & 2.0\% & 1.9\% \\
\cline{2-6}
 & \multirow{6}{*}{Hard} & 4o & \textbf{84.4\%} & \underline{81.2\%} & 81.2\% \\
 &  & o3 & \textbf{84.4\%} & 62.5\% & \underline{\textbf{84.4\%}} \\
 &  & deep & \underline{\textbf{87.5\%}} & 71.9\% & 81.2\% \\\cdashline{3-6}
 &  & llama3 & \textbf{28.1\%} & 21.9\% & 21.9\% \\
 &  & qwen14b & \textbf{68.8\%} & 59.4\% & 59.4\% \\
 &  & qwen05b & 0.0\% & 0.0\% & \textbf{3.1\%} \\
\hline
\multirow{6}{*}{BigCodeBench} & \multirow{6}{*}{All} & 4o & 28.8\% & \underline{\textbf{57.3\%}} & 20.1\% \\
 &  & o3 & \textbf{62.2\%} & 56.6\% & 50.7\% \\
 &  & deep & \underline{\textbf{63.4\%}} & 56.9\% & \underline{55.6\%} \\\cdashline{3-6}
 &  & llama3 & \textbf{41.7\%} & 35.3\% & 36.7\% \\
 &  & qwen14b & \textbf{58.3\%} & 54.7\% & 54.7\% \\
 &  & qwen05b & \textbf{11.1\%} & 8.3\% & 9.4\% \\
\hline
\end{tabular}}
\end{table}
\begin{table}
    \caption{Count the highest $\mbox{\sl Accuracy}$ across the studied LLMs for each difficulty level and language combination (the number of underlined values in Table~\ref{tab:all})}
    \label{tab:highest_accuracy}
    \centering
    \scalebox{0.8}{
    \begin{tabular}{llrrrrrr}
    \toprule
    \multicolumn{1}{c}{Problem} & \multicolumn{1}{c}{Difficulty} & \multicolumn{1}{c}{GPT-4o} & \multicolumn{1}{c}{o3-mini} & \multicolumn{1}{c}{DeepSeek} & \multicolumn{1}{c}{Llama3} & \multicolumn{1}{c}{Qwen14B} & \multicolumn{1}{c}{Qwen0.5B} \\
    \midrule
    \multirow{4}{*}{AtCoder}
    & A & 1 & 3 & 2 & 0 & 0 & 0 \\
    & B & 0 & 0 & 3 & 0 & 0 & 0 \\
    & C & 0 & 1 & 2 & 0 & 0 & 0 \\
    & D & 0 & 1 & 2 & 0 & 0 & 0 \\
    \midrule
    \multirow{3}{*}{LeetCode}
    & Easy   & 1 & 3 & 2 & 0 & 0 & 0 \\
    & Medium & 0 & 0 & 3 & 0 & 0 & 0 \\
    & Hard   & 1 & 1 & 1 & 0 & 0 & 0 \\
    \midrule
    \multirow{1}{*}{BigCodeBench}
    & All & 1 & 0 & 2 & 0 & 0 & 0 \\
    \midrule
    Total &  & 4 & 9 & 17 & 0 & 0 & 0 \\
    \bottomrule
    \end{tabular}}
\end{table}

\begin{table}
\caption{Number of LLMs achieving the highest $\mbox{\sl Accuracy}$ for each language}
\label{tab:highest_accuracy_language}
\centering
\setlength{\tabcolsep}{7pt}
\renewcommand{\arraystretch}{1.15}
\begin{tabular}{llrrr}
\hline
\multicolumn{1}{c}{Problem} & \multicolumn{1}{c}{Difficulty} & \multicolumn{1}{c}{English} & \multicolumn{1}{c}{Japanese} & \multicolumn{1}{c}{Chinese} \\ \hline

\multirow{4}{*}{AtCoder}
& A & \textbf{\underline{3}} & \underline{2} & 1 \\
& B & \textbf{\underline{3}} & \textbf{\underline{3}} & 0 \\
& C & \textbf{\underline{4}} & \underline{3} & 3 \\
& D & \textbf{\underline{4}} & \underline{3} & 3 \\ \hdashline
Total (AtCoder)
&  & \textbf{\underline{14}} & \underline{11} & 7 \\ \hline

\multirow{3}{*}{LeetCode}
& Easy   & \underline{3} & 2 & \textbf{\underline{5}} \\
& Medium & \underline{3} & 0 & \textbf{\underline{4}} \\
& Hard   & \textbf{\underline{5}} & 0 & \underline{2} \\ \hdashline
Total (LeetCode)
&  & \textbf{\underline{11}} & 2 & \textbf{\underline{11}} \\ \hline

\multirow{1}{*}{BigCodeBench}
& All & \textbf{\underline{5}} & 1 & 0 \\ \hdashline
Total (BigCodeBench)
&  & \textbf{\underline{5}} & 1 & 0 \\ \hline

\textbf{Total (All)}
&  & \textbf{32} & 15 & 17 \\ \hline

\end{tabular}
\end{table}

\subsection{Results}
\observation{DeepSeek-V3.2 showed better code generation performance compared to other LLMs.}
Table~\ref{tab:all} shows the median $\mbox{\sl Accuracy}$ for each LLM across different difficulty levels and languages. 
Values typeset in boldface indicate the language that achieved the highest $\mbox{\sl Accuracy}$ among the three natural languages for each model and difficulty level.
Underlined values indicate the LLMs that achieved the highest $\mbox{\sl Accuracy}$ among the studied LLMs for each difficulty level and language combination.
When multiple LLMs achieved the same highest $\mbox{\sl Accuracy}$, all corresponding values were underlined.

Table~\ref{tab:highest_accuracy} presents the count of underlined values in Table~\ref{tab:all}. For each language and difficulty level in Table~\ref{tab:all}, the LLM with the best performance is counted. The maximum value in each cell is three, indicating that the LLM achieved the highest $\mbox{\sl Accuracy}$ in all three languages for that particular difficulty level.
\red{For example, at difficulty level D in AtCoder, o3-mini achieved the highest $\mbox{\sl Accuracy}$ in Japanese, while DeepSeek-V3.2 achieved the highest $\mbox{\sl Accuracy}$ in both English and Chinese. 
As a result, Table 2 shows that DeepSeek-V3.2 has the largest total count (17), followed by o3-mini (9) and GPT-4o (4), while the open-source models (Llama3, Qwen2.5-Coder-14B, and Qwen2.5-Coder-0.5B) do not achieve the highest $\mbox{\sl Accuracy}$ in any of these combinations.
}

\red{DeepSeek-V3.2 achieved the highest median $\mbox{\sl Accuracy}$ in 17 out of the 24 possible combinations (12 combinations from four difficulty levels and three languages in AtCoder, 9 combinations from three difficulty levels and three languages in LeetCode, and 3 combinations from one difficulty level and three languages in BigCodeBench). }
\red{However, as shown in Table~\ref{tab:all}, when focusing on closed-source models, Copilot exhibits substantially lower performance than the others. Specifically, Copilot has a much lower median $\mbox{\sl Accuracy}$ in AtCoder (25.1\%) compared to GPT-4o (76.0\%).
Moreover, Copilot does not provide API-based access and requires manual interaction for code generation, making it difficult to evaluate under consistent and reproducible conditions. In addition, its underlying model and version are not clearly specified, which further limits the validity of systematic comparisons.
Hence, we exclude Copilot from all subsequent results (i.e., the following Observations and Implications).}


\observation{Officially supported languages result in higher $\mbox{\sl Accuracy}$.}
Table~\ref{tab:highest_accuracy_language} presents the number of LLMs that achieved the highest $\mbox{\sl Accuracy}$ for each language.
The underlined values indicate the cases where the languages officially supported by the corresponding datasets are used; the bold values indicate the highest Accuracy for each difficulty level and dataset.
The maximum count is six (GPT-4o, o3-mini, DeepSeek-V3.2, llama-3, qwen2.5-coder-14b, qwen2.5-coder-0.5b), indicating that all six LLMs achieved the highest $\mbox{\sl Accuracy}$ in this language.

\red{In AtCoder, the highest total (AtCoder), which is the total count of the number of LLMs that achieved the highest $\mbox{\sl Accuracy}$ across all difficulty levels, is 14 for English, followed by Japanese, both of which are officially supported languages. In contrast, Chinese, which is not officially supported, achieved a lower total of 7.
In LeetCode, the officially supported languages (English and Chinese) resulted in slightly higher counts of highest $\mbox{\sl Accuracy}$ compared to Japanese (11 for English, 11 for Chinese, and 2 for Japanese), which is not officially supported.
In BigCodeBench, the officially supported language (English) achieved the highest total of 5, while the non-officially supported languages (Japanese and Chinese) achieved a total of 0 or 1.}

\begin{table}[t!]
\caption{Combination of language and LLM that showed the highest $\mbox{\sl Accuracy}$}
\centering
\label{tab:combination}
\setlength{\tabcolsep}{6pt}
\renewcommand{\arraystretch}{1.15}
\begin{tabular}{lrrr}
\hline
\multicolumn{1}{c}{Model} & \multicolumn{1}{c}{English} & \multicolumn{1}{c}{Japanese} & \multicolumn{1}{c}{Chinese} \\ \hline

GPT-4o 
& 0 & 1 & 0 \\

o3-mini     
& 1 & \textbf{\underline{3}} & 1 \\

DeepSeek-V3.2 
& \textbf{\underline{3}} & 1 & \textbf{\underline{3}} \\

Llama3      
& 0 & 0 & 0 \\

Qwen14B     
& 0 & 0 & 0 \\

Qwen0.5B    
& 0 & 0 & 0 \\

\hline
\end{tabular}
\end{table}

\observation{Japanese tended to yield higher $\mbox{\sl Accuracy}$ when paired with o3-mini, whereas English and Chinese tended to yield higher $\mbox{\sl Accuracy}$ when paired with DeepSeek-V3.2.}
\red{Table~\ref{tab:combination} shows the frequency with which each LLM and language combination achieved the highest $\mbox{\sl Accuracy}$.
The maximum count is eight (four difficulty levels in AtCoder, three difficulty levels in LeetCode, and one difficulty level in BigCodeBench).
Among them, the combination of o3-mini and Japanese achieved the highest $\mbox{\sl Accuracy}$ in three out of eight cases, accounting for 37.5\%.
Similarly, English and Chinese achieved the highest $\mbox{\sl Accuracy}$ with DeepSeek-V3.2 in three out of eight cases, respectively, accounting for 37.5\%.
These results suggest that, in this study, Japanese tended to yield higher $\mbox{\sl Accuracy}$ when paired with o3-mini, whereas English and Chinese tended to yield higher $\mbox{\sl Accuracy}$ when paired with DeepSeek-V3.2.
}

\begin{table}[t]
\centering
\caption{Median $\mbox{\sl Accuracy}$ differences between the highest- and lowest-performing languages for closed-source LLMs (E: English, J: Japanese, C: Chinese)}
\label{tab:lang_gap_closed}
\setlength{\tabcolsep}{6pt}
\renewcommand{\arraystretch}{1.15}
\begin{tabular}{llrrr}
\hline
\multicolumn{1}{c}{Problems} & \multicolumn{1}{c}{Difficulty} & \multicolumn{1}{c}{GPT-4o} & \multicolumn{1}{c}{o3-mini} & \multicolumn{1}{c}{DeepSeek-V3.2} \\
\hline

& A & 1.6\%(C-E) & 1.6\%(J-C) & 1.0\%(J-E) \\
& B & 3.7\%(J-C) & 5.3\%(J-E) & 1.6\%(J-C) \\
& C & 4.3\%(E-C) & 12.2\%(J-E) & 1.6\%(C-E) \\
\multirow{-4}{*}{AtCoder}
& D & 1.6\%(E-C) & 13.2\%(J-E) & 5.3\%(C-E) \\\cline{1-5}

& Easy   & 12.5\%(J-E) & 0.0\%(All) & 18.7\%(E-J) \\
LeetCode
& Medium & 8.3\%(E-J) & 16.2\%(C-J) & 17.9\%(C-J) \\
& Hard   & 3.2\%(E-J) & 21.9\%(E-J) & 15.6\%(E-J) \\\cline{1-5}

\multirow{1}{*}{BigCodeBench}
& All & 37.2\%(J-C) & 11.5\%(E-C) & 7.8\%(E-C) \\
\hline

\end{tabular}
\end{table}
\begin{table}[t]
\centering
\caption{Median $\mbox{\sl Accuracy}$ differences between the highest- and lowest-performing languages for open-source LLMs (E: English, J: Japanese, C: Chinese)}
\label{tab:lang_gap_local}
\setlength{\tabcolsep}{6pt}
\renewcommand{\arraystretch}{1.15}
\begin{tabular}{llrrr}
\hline
\multicolumn{1}{c}{Problems} & \multicolumn{1}{c}{Difficulty} & \multicolumn{1}{c}{Llama3} & \multicolumn{1}{c}{Qwen14B} & \multicolumn{1}{c}{Qwen0.5B} \\
\hline

& A & 15.9\%(E-J) & 11.1\%(E-J) & 6.9\%(E-J) \\
& B & 12.7\%(E-J) & 26.5\%(E-J) & 2.1\%(E-J) \\
& C & 1.6\%(E-C) & 17.5\%(E-J) & 0.0\%(E-J) \\
\multirow{-4}{*}{AtCoder}
& D & 1.6\%(E-J) & 6.3\%(E-J) & 0.0\%(E-J) \\\cline{1-5}

& Easy   & 6.3\%(C-E) & 18.7\%(C-J) & 6.3\%(E-J) \\
LeetCode
& Medium & 13.0\%(E-J) & 6.4\%(C-J) & 3.9\%(E-C) \\
& Hard   & 6.2\%(E-J) & 9.4\%(E-J) & 3.1\%(C-E) \\\cline{1-5}

\multirow{1}{*}{BigCodeBench}
& All & 6.4\%(E-J) & 3.6\%(E-J) & 2.8\%(E-J) \\
\hline

\end{tabular}
\end{table}

\observation{The magnitude of language bias differs across datasets and model types.}
\label{ob:lang_gap}
Table~\ref{tab:lang_gap_closed} and Table~\ref{tab:lang_gap_local} show the differences in $\mbox{\sl Accuracy}$ between the highest- and lowest-performing languages for each model. For open-source LLMs, the median language gap is 6.6\% for AtCoder, 6.3\% for LeetCode, and 3.6\% for BigCodeBench. In contrast, for closed-source LLMs, the median language gap is 2.7\% for AtCoder, 15.6\% for LeetCode, and 11.5\% for BigCodeBench. These results indicate that the magnitude of natural language bias varies depending on both the dataset and the model type: the bias is largest in AtCoder for open-source LLMs, whereas it is larger in LeetCode and BigCodeBench for closed-source LLMs.

\observation{Open-source models show relatively better performance in English, although their overall performance remains lower than that of closed-source models.}
\red{In Table~\ref{tab:all}, bold values indicate the language that achieves the highest performance compared to the others for each model and difficulty level.
From Table~\ref{tab:all}, it can be observed that open-source models (Llama3, Qwen14B, and Qwen0.5B) tend to achieve higher Accuracy in English compared to other languages.
On the other hand, these models generally exhibit lower overall performance than closed-source models, and they rarely achieve the highest Accuracy, as shown in Table~\ref{tab:highest_accuracy}.
Overall, although open-source models underperform in terms of absolute performance, they show relatively better performance in English.}



\observation{DeepSeek-V3.2 performs competitively with Chinese prompts.}
Table~\ref{tab:combination} shows the number of cases in which each combination of LLM and natural language achieved the highest $\mbox{\sl Accuracy}$. 
DeepSeek-V3.2 achieved the highest $\mbox{\sl Accuracy}$ most frequently among the studied LLMs, with three cases for English, one case for Japanese, and three cases for Chinese. 
In particular, although Chinese is not an officially supported language in AtCoder, Table~\ref{tab:all} shows that DeepSeek-V3.2 achieved comparable $\mbox{\sl Accuracy}$ with Chinese prompts to that with English prompts in AtCoder. 
For example, Chinese prompts achieved the highest $\mbox{\sl Accuracy}$ for difficulty levels C and D, while their $\mbox{\sl Accuracy}$ was close to English prompts for difficulty levels A and B. 
These results suggest that DeepSeek-V3.2 can perform competitively with Chinese prompts, even when Chinese is not an official language of the dataset.

\begin{table}[p]
\caption{Number of warnings of Pylint per line for each LLM and language}
\centering
\label{tab:pylint-all}
\scalebox{0.8}{
\begin{tabular}{lllrrrr}
\hline
\multicolumn{1}{c}{Problems} & \multicolumn{1}{c}{Difficulty} & \multicolumn{1}{c}{Model} & \multicolumn{1}{c}{English} & \multicolumn{1}{c}{Japanese} & \multicolumn{1}{c}{Chinese} & \multicolumn{1}{c}{Max-Min Diff} \\
\hline
\multirow{24}{*}{AtCoder} & \multirow{6}{*}{A} & GPT-4o & 0.237 & 0.211 & 0.217 & 0.025 \\
 &  & o3-mini & 0.000 & 0.000 & 0.000 & 0.000 \\
 &  & DeepSeek & 0.000 & 0.000 & 0.023 & 0.023 \\
 &  & Llama3 & 0.000 & 0.000 & 0.000 & 0.000 \\
 &  & Qwen2.5-Coder-0.5B & 0.101 & 0.000 & 0.067 & 0.101 \\
 &  & Qwen2.5-Coder-14B & 0.167 & 0.167 & 0.167 & 0.000 \\
\cline{2-7}
 & \multirow{6}{*}{B} & GPT-4o & 0.217 & 0.185 & 0.200 & 0.032 \\
 &  & o3-mini & 0.010 & 0.013 & 0.007 & 0.006 \\
 &  & DeepSeek & 0.046 & 0.037 & 0.042 & 0.009 \\
 &  & Llama3 & 0.029 & 0.043 & 0.036 & 0.014 \\
 &  & Qwen2.5-Coder-0.5B & 0.110 & 0.049 & 0.095 & 0.061 \\
 &  & Qwen2.5-Coder-14B & 0.182 & 0.150 & 0.182 & 0.032 \\
\cline{2-7}
 & \multirow{6}{*}{C} & GPT-4o & 0.174 & 0.167 & 0.171 & 0.007 \\
 &  & o3-mini & 0.010 & 0.015 & 0.008 & 0.007 \\
 &  & DeepSeek & 0.028 & 0.031 & 0.033 & 0.005 \\
 &  & Llama3 & 0.028 & 0.045 & 0.027 & 0.018 \\
 &  & Qwen2.5-Coder-0.5B & 0.115 & 0.056 & 0.100 & 0.060 \\
 &  & Qwen2.5-Coder-14B & 0.138 & 0.125 & 0.143 & 0.018 \\
\cline{2-7}
 & \multirow{6}{*}{D} & GPT-4o & 0.159 & 0.151 & 0.154 & 0.008 \\
 &  & o3-mini & 0.010 & 0.014 & 0.009 & 0.005 \\
 &  & DeepSeek & 0.024 & 0.024 & 0.025 & 0.001 \\
 &  & Llama3 & 0.018 & 0.043 & 0.025 & 0.026 \\
 &  & Qwen2.5-Coder-0.5B & 0.120 & 0.057 & 0.100 & 0.063 \\
 &  & Qwen2.5-Coder-14B & 0.129 & 0.091 & 0.130 & 0.040 \\
\hline
\multirow{18}{*}{LeetCode} & \multirow{6}{*}{Easy} & GPT-4o & 0.048 & 0.050 & 0.048 & 0.002 \\
 &  & o3-mini & 0.053 & 0.045 & 0.050 & 0.008 \\
 &  & DeepSeek & 0.056 & 0.031 & 0.040 & 0.025 \\
 &  & Llama3 & 0.046 & 0.060 & 0.044 & 0.015 \\
 &  & qwen14b & 0.051 & 0.050 & 0.062 & 0.012 \\
 &  & qwen05b & 0.043 & 0.036 & 0.039 & 0.007 \\
\cline{2-7}
 & \multirow{6}{*}{Medium} & GPT-4o & 0.047 & 0.053 & 0.048 & 0.006 \\
 &  & o3-mini & 0.043 & 0.039 & 0.041 & 0.004 \\
 &  & DeepSeek & 0.050 & 0.035 & 0.045 & 0.014 \\
 &  & Llama3 & 0.053 & 0.067 & 0.061 & 0.014 \\
 &  & qwen14b & 0.056 & 0.054 & 0.056 & 0.001 \\
 &  & qwen05b & 0.057 & 0.061 & 0.057 & 0.004 \\
\cline{2-7}
 & \multirow{6}{*}{Hard} & GPT-4o & 0.057 & 0.068 & 0.066 & 0.011 \\
 &  & o3-mini & 0.039 & 0.041 & 0.037 & 0.003 \\
 &  & DeepSeek & 0.049 & 0.045 & 0.046 & 0.004 \\
 &  & Llama3 & 0.059 & 0.079 & 0.068 & 0.021 \\
 &  & qwen14b & 0.054 & 0.070 & 0.055 & 0.016 \\
 &  & qwen05b & 0.077 & 0.095 & 0.062 & 0.033 \\
\hline
\multirow{6}{*}{BigCodeBench} & \multirow{6}{*}{All} & GPT-4o & 0.000 & 0.017 & 0.000 & 0.017 \\
 &  & o3-mini & 0.017 & 0.020 & 0.028 & 0.011 \\
 &  & DeepSeek & 0.017 & 0.016 & 0.017 & 0.001 \\
 &  & Llama3 & 0.021 & 0.018 & 0.023 & 0.005 \\
 &  & qwen14b & 0.018 & 0.029 & 0.024 & 0.010 \\
 &  & qwen05b & 0.021 & 0.025 & 0.023 & 0.004 \\
\hline
\end{tabular}}
\end{table}
\begin{table}[p]
\caption{Median LOC of generated code for each LLM and language}
\centering
\label{tab:loc-all}
\scalebox{0.8}{
\begin{tabular}{lllrrr}
\hline
\multicolumn{1}{c}{Problems} & \multicolumn{1}{c}{Difficulty} & \multicolumn{1}{c}{Model} & \multicolumn{1}{c}{English} & \multicolumn{1}{c}{Japanese} & \multicolumn{1}{c}{Chinese} \\
\hline
\multirow{24}{*}{AtCoder} & \multirow{6}{*}{A} & GPT-4o & 11.2 & 12.0 & 12.0 \\
 &  & o3-mini & 12.0 & 11.2 & 12.4 \\
 &  & DeepSeek & 10.0 & 11.0 & 13.0 \\
 &  & Llama3 & 16.0 & 9.0 & 16.0 \\
 &  & Qwen2.5-Coder-0.5B & 17.0 & 9.0 & 16.0 \\
 &  & Qwen2.5-Coder-14B & 13.0 & 10.0 & 11.0 \\
\cline{2-6}
 & \multirow{6}{*}{B} & GPT-4o & 17.8 & 18.4 & 18.0 \\
 &  & o3-mini & 18.2 & 17.8 & 19.4 \\
 &  & DeepSeek & 19.0 & 20.0 & 24.0 \\
 &  & Llama3 & 24.0 & 14.0 & 21.0 \\
 &  & Qwen2.5-Coder-0.5B & 20.0 & 13.0 & 18.0 \\
 &  & Qwen2.5-Coder-14B & 21.0 & 15.0 & 19.0 \\
\cline{2-6}
 & \multirow{6}{*}{C} & GPT-4o & 22.2 & 23.2 & 22.0 \\
 &  & o3-mini & 25.0 & 25.6 & 26.2 \\
 &  & DeepSeek & 29.6 & 27.8 & 31.0 \\
 &  & Llama3 & 25.0 & 17.0 & 22.0 \\
 &  & Qwen2.5-Coder-0.5B & 22.8 & 15.0 & 20.0 \\
 &  & Qwen2.5-Coder-14B & 27.0 & 15.0 & 26.0 \\
\cline{2-6}
 & \multirow{6}{*}{D} & GPT-4o & 28.0 & 28.6 & 29.4 \\
 &  & o3-mini & 31.4 & 31.2 & 32.4 \\
 &  & DeepSeek & 41.6 & 38.0 & 41.6 \\
 &  & Llama3 & 28.0 & 21.0 & 25.0 \\
 &  & Qwen2.5-Coder-0.5B & 23.4 & 15.0 & 22.0 \\
 &  & Qwen2.5-Coder-14B & 33.0 & 17.0 & 29.0 \\
\hline
\multirow{18}{*}{LeetCode} & \multirow{6}{*}{Easy} & GPT-4o & 21.5 & 22.0 & 21.8 \\
 &  & o3-mini & 23.1 & 21.5 & 22.7 \\
 &  & DeepSeek & 18.0 & 24.4 & 25.1 \\
 &  & Llama3 & 23.5 & 17.0 & 23.5 \\
 &  & qwen14b & 19.5 & 20.0 & 17.0 \\
 &  & qwen05b & 24.5 & 23.0 & 26.0 \\
\cline{2-6}
 & \multirow{6}{*}{Medium} & GPT-4o & 29.0 & 29.0 & 30.0 \\
 &  & o3-mini & 30.8 & 34.7 & 33.4 \\
 &  & DeepSeek & 29.0 & 33.1 & 33.8 \\
 &  & Llama3 & 28.0 & 24.0 & 25.0 \\
 &  & qwen14b & 29.0 & 28.0 & 25.0 \\
 &  & qwen05b & 29.0 & 25.0 & 25.0 \\
\cline{2-6}
 & \multirow{6}{*}{Hard} & GPT-4o & 31.1 & 31.9 & 28.9 \\
 &  & o3-mini & 38.8 & 44.4 & 38.2 \\
 &  & DeepSeek & 37.2 & 40.8 & 42.2 \\
 &  & Llama3 & 29.5 & 26.0 & 26.0 \\
 &  & qwen14b & 33.5 & 36.0 & 28.3 \\
 &  & qwen05b & 29.0 & 27.0 & 26.0 \\
\hline
\multirow{6}{*}{BigCodeBench} & \multirow{6}{*}{All} & GPT-4o & 15.6 & 38.3 & 12.0 \\
 &  & o3-mini & 23.0 & 19.1 & 14.6 \\
 &  & DeepSeek & 45.0 & 46.0 & 46.8 \\
 &  & Llama3 & 38.0 & 44.2 & 35.4 \\
 &  & qwen14b & 29.5 & 20.0 & 22.0 \\
 &  & qwen05b & 46.0 & 45.0 & 46.0 \\
\hline
\end{tabular}}
\end{table}
\begin{table}[p]
\caption{Median average cyclomatic complexity for each natural language and LLM}
\centering
\label{tab:complexity-all}
\scalebox{0.8}{
\begin{tabular}{lllrrr}
\hline
\multicolumn{1}{c}{Problems} & \multicolumn{1}{c}{Difficulty} & \multicolumn{1}{c}{Model} & \multicolumn{1}{c}{English} & \multicolumn{1}{c}{Japanese} & \multicolumn{1}{c}{Chinese} \\ \hline
\multirow{24}{*}{AtCoder} & \multirow{6}{*}{A} & 4o & \textbf{2.0} & \textbf{2.0} & \textbf{2.0} \\
 &  & o3 & \textbf{2.0} & \textbf{2.0} & \textbf{2.0} \\
 &  & deep & \textbf{2.0} & \textbf{2.0} & \textbf{2.0} \\
\cdashline{3-6}
 &  & llama3 & \textbf{2.0} & 1.0 & \textbf{2.0} \\
 &  & qwen14b & \textbf{2.0} & \textbf{2.0} & \textbf{2.0} \\
 &  & qwen05b & 2.0 & 1.0 & \textbf{3.0} \\
\cline{2-6}
 & \multirow{6}{*}{B} & 4o & \textbf{3.0} & \textbf{3.0} & \textbf{3.0} \\
 &  & o3 & \textbf{4.0} & 3.0 & \textbf{4.0} \\
 &  & deep & \textbf{3.0} & \textbf{3.0} & \textbf{3.0} \\
\cdashline{3-6}
 &  & llama3 & \textbf{3.0} & \textbf{3.0} & \textbf{3.0} \\
 &  & qwen14b & \textbf{3.0} & 2.0 & \textbf{3.0} \\
 &  & qwen05b & \textbf{3.0} & 2.0 & \textbf{3.0} \\
\cline{2-6}
 & \multirow{6}{*}{C} & 4o & \textbf{4.0} & \textbf{4.0} & \textbf{4.0} \\
 &  & o3 & \textbf{5.0} & 4.5 & \textbf{5.0} \\
 &  & deep & \textbf{4.0} & \textbf{4.0} & \textbf{4.0} \\
\cdashline{3-6}
 &  & llama3 & \textbf{4.0} & \textbf{4.0} & 3.5 \\
 &  & qwen14b & \textbf{4.0} & 2.0 & \textbf{4.0} \\
 &  & qwen05b & 3.0 & 3.0 & \textbf{4.0} \\
\cline{2-6}
 & \multirow{6}{*}{D} & 4o & 4.0 & 4.0 & \textbf{5.0} \\
 &  & o3 & \textbf{6.0} & \textbf{6.0} & \textbf{6.0} \\
 &  & deep & \textbf{6.0} & 5.0 & \textbf{6.0} \\
\cdashline{3-6}
 &  & llama3 & \textbf{4.8} & 4.0 & 4.5 \\
 &  & qwen14b & \textbf{5.0} & 2.0 & \textbf{5.0} \\
 &  & qwen05b & \textbf{4.0} & 3.0 & \textbf{4.0} \\
\hline
\multirow{18}{*}{LeetCode} & \multirow{6}{*}{Easy} & 4o & 5.0 & \textbf{5.5} & 5.0 \\
 &  & o3 & \textbf{5.5} & \textbf{5.5} & \textbf{5.5} \\
 &  & deep & \textbf{5.5} & \textbf{5.5} & \textbf{5.5} \\
\cdashline{3-6}
 &  & llama3 & \textbf{6.0} & 4.5 & 5.5 \\
 &  & qwen14b & \textbf{5.0} & \textbf{5.0} & 4.5 \\
 &  & qwen05b & \textbf{5.5} & \textbf{5.5} & 5.0 \\
\cline{2-6}
 & \multirow{6}{*}{Medium} & 4o & \textbf{5.5} & \textbf{5.5} & \textbf{5.5} \\
 &  & o3 & \textbf{5.5} & \textbf{5.5} & \textbf{5.5} \\
 &  & deep & \textbf{6.5} & \textbf{6.5} & \textbf{6.5} \\
\cdashline{3-6}
 &  & llama3 & \textbf{5.5} & \textbf{5.5} & \textbf{5.5} \\
 &  & qwen14b & \textbf{5.5} & \textbf{5.5} & \textbf{5.5} \\
 &  & qwen05b & 4.5 & \textbf{5.5} & \textbf{5.5} \\
\cline{2-6}
 & \multirow{6}{*}{Hard} & 4o & \textbf{7.5} & 6.5 & 6.5 \\
 &  & o3 & \textbf{6.5} & \textbf{6.5} & \textbf{6.5} \\
 &  & deep & \textbf{7.5} & \textbf{7.5} & \textbf{7.5} \\
\cdashline{3-6}
 &  & llama3 & \textbf{7.0} & 5.5 & 5.5 \\
 &  & qwen14b & 6.5 & \textbf{7.0} & 6.5 \\
 &  & qwen05b & 4.5 & \textbf{7.5} & 5.5 \\
\hline
\multirow{6}{*}{BigCodeBench} & \multirow{6}{*}{All} & 4o & \textbf{3.0} & \textbf{3.0} & 2.0 \\
 &  & o3 & \textbf{3.0} & \textbf{3.0} & \textbf{3.0} \\
 &  & deep & \textbf{3.0} & \textbf{3.0} & \textbf{3.0} \\
\cdashline{3-6}
 &  & llama3 & \textbf{3.0} & \textbf{3.0} & 2.0 \\
 &  & qwen14b & \textbf{3.0} & \textbf{3.0} & \textbf{3.0} \\
 &  & qwen05b & \textbf{2.0} & \textbf{2.0} & \textbf{2.0} \\
\hline
\end{tabular}}
\end{table}

\observation{The number of warnings per line is consistent across languages.}
Table~\ref{tab:pylint-all} shows the number of warnings per line and their difference between the maximum and minimum values across the three languages. The values are all less than 0.101, indicating that the number of warnings per line is consistent across languages. This suggests that the generated code adheres to common coding rules regardless of the natural language used in the prompt. 

One possible factor that biases the result is the difference of the lines of code (LOC) of generated code across languages. If LOC varies across languages, the fact that the number of warnings per line is consistent is not sufficient to lead to this conclusion.
Table~\ref{tab:loc-all} shows the median LOC of generated code. We observe that the LOC is similar across languages. Hence, the bias caused by LOC is negligible.

\observation{Cyclomatic complexity is primarily influenced by problem difficulty and model capacity, while the impact of natural language is limited.}
\red{Table~\ref{tab:complexity-all} shows the cyclomatic complexity of the generated code. The values are higher for more difficult problems and larger models, while the differences across languages are relatively small. This suggests that the complexity of the generated code is primarily influenced by problem difficulty and model capacity, while the impact of natural language is limited.}

\summarybox{Summary of RQ1}{
    In general, languages officially supported by the corresponding datasets or platforms tended to result in higher $\mbox{\sl Accuracy}$.
    This indicates that users should carefully select the natural language based on the target data source.
    In our study, Japanese tended to yield higher $\mbox{\sl Accuracy}$ when paired with o3-mini, whereas English and Chinese tended to yield higher $\mbox{\sl Accuracy}$ when paired with DeepSeek-V3.2.
    Additionally, the generated code adheres to common coding rules regardless of the natural language used in the prompt.
    }


\section{Translation for Performance Improvement}
\label{sec:rq2}
In this section we address
RQ2: \RQthree{}

\subsection{Approach}
\red{RQ1 indicated that the languages officially supported by each dataset tend to achieve higher performance. Specifically, English and Japanese showed higher performance for AtCoder, English and Chinese for LeetCode, and English for BigCodeBench.\\
On the other hand, the results did not show clear language bias across natural languages, models, and difficulty levels.
Based on these observations, and because code generation Accuracy is likely to depend on the training data of each LLM, we selected the target languages at the dataset level rather than separately for each model or difficulty level.
Specifically, we designed our experiments to translate prompts into the officially supported languages for each dataset.
}
\red{For AtCoder problems, we conducted experiments using (i) versions translated from Japanese and Chinese into English, and (ii) versions translated from English and Chinese into Japanese.}
\red{For LeetCode problems, we conducted experiments using (i) versions translated from Japanese and Chinese into English, and (ii) versions translated from English and Japanese into Chinese.}
\red{For BigCodeBench problems, we conducted experiments by translating inputs from other languages into English.}






\begin{table}[t]
\caption{Median $\mbox{\sl Accuracy}$ of code generation after translating problem statements into English and $\mbox{\sl Accuracy}$ difference from the original language}
\centering
\label{tab:translation-to-EN}
\scalebox{0.6}{
\begin{tabular}{lllrrrrrr}
\hline
\multicolumn{1}{c}{Problems} & \multicolumn{1}{c}{Difficulty} & \multicolumn{1}{c}{Model} & \multicolumn{2}{c}{Google} & \multicolumn{2}{c}{DeepL} & \multicolumn{2}{c}{GPT} \\
 &  &  & \multicolumn{1}{c}{From Japanese} & \multicolumn{1}{c}{From Chinese} & \multicolumn{1}{c}{From Japanese} & \multicolumn{1}{c}{From Chinese} & \multicolumn{1}{c}{From Japanese} & \multicolumn{1}{c}{From Chinese} \\ \hline
\multirow{24}{*}{AtCoder} & \multirow{6}{*}{A} & GPT-4o & 90.5\%$^{\textcolor{red}{-2.1}}$ & 91.0\%$^{\textcolor{red}{-3.2}}$ & 88.4\%$^{\textcolor{red}{-4.2}}$ & 93.1\%$^{\textcolor{red}{-1.1}}$ & 93.7\%$^{\textcolor{teal}{+1.1}}$ & 95.0\%$^{\textcolor{teal}{+0.8}}$ \\
 &  & o3-mini & 91.5\%$^{\textcolor{red}{-4.3}}$ & 89.9\%$^{\textcolor{red}{-4.3}}$ & 90.5\%$^{\textcolor{red}{-5.3}}$ & 92.1\%$^{\textcolor{red}{-2.1}}$ & 93.1\%$^{\textcolor{red}{-2.7}}$ & 91.5\%$^{\textcolor{red}{-2.7}}$ \\
 &  & DeepSeek-V3.2 & 89.9\%$^{\textcolor{red}{-5.3}}$ & 90.5\%$^{\textcolor{red}{-3.7}}$ & 85.7\%$^{\textcolor{red}{-9.5}}$ & 90.5\%$^{\textcolor{red}{-3.7}}$ & 90.5\%$^{\textcolor{red}{-4.7}}$ & 89.9\%$^{\textcolor{red}{-4.3}}$ \\
\cdashline{3-9}
 &  & Llama-3 & 57.7\%$^{\textcolor{teal}{+6.4}}$ & 56.1\%$^{\textcolor{red}{-3.2}}$ & 56.1\%$^{\textcolor{teal}{+4.8}}$ & 58.7\%$^{\textcolor{red}{-0.6}}$ & 66.7\%$^{\textcolor{teal}{+15.4}}$ & 61.4\%$^{\textcolor{teal}{+2.1}}$ \\
 &  & Qwen2.5-Coder-14B & 85.7\%$^{\textcolor{teal}{+9.0}}$ & 85.2\%$^{\textcolor{teal}{+1.6}}$ & 82.0\%$^{\textcolor{teal}{+5.3}}$ & 84.1\%$^{\textcolor{teal}{+0.5}}$ & 86.8\%$^{\textcolor{teal}{+10.1}}$ & 87.3\%$^{\textcolor{teal}{+3.7}}$ \\
 &  & Qwen2.5-Coder-0.5B & 5.8\%$^{\textcolor{teal}{+3.7}}$ & 11.1\%$^{\textcolor{teal}{+6.3}}$ & 5.8\%$^{\textcolor{teal}{+3.7}}$ & 9.0\%$^{\textcolor{teal}{+4.2}}$ & 5.3\%$^{\textcolor{teal}{+3.2}}$ & 13.8\%$^{\textcolor{teal}{+9.0}}$ \\
\cline{2-9}
 & \multirow{6}{*}{B} & GPT-4o & 86.8\%$^{\textcolor{red}{-4.2}}$ & 86.8\%$^{\textcolor{red}{-0.5}}$ & 84.1\%$^{\textcolor{red}{-6.9}}$ & 85.7\%$^{\textcolor{red}{-1.6}}$ & 88.4\%$^{\textcolor{red}{-2.6}}$ & 86.4\%$^{\textcolor{red}{-0.9}}$ \\
 &  & o3-mini & 85.2\%$^{\textcolor{red}{-5.3}}$ & 84.1\%$^{\textcolor{red}{-3.7}}$ & 81.0\%$^{\textcolor{red}{-9.5}}$ & 82.5\%$^{\textcolor{red}{-5.3}}$ & 85.2\%$^{\textcolor{red}{-5.3}}$ & 86.9\%$^{\textcolor{red}{-0.9}}$ \\
 &  & DeepSeek-V3.2 & 83.6\%$^{\textcolor{red}{-9.5}}$ & 82.0\%$^{\textcolor{red}{-9.5}}$ & 79.4\%$^{\textcolor{red}{-13.7}}$ & 77.8\%$^{\textcolor{red}{-13.7}}$ & 81.0\%$^{\textcolor{red}{-12.1}}$ & 77.8\%$^{\textcolor{red}{-13.7}}$ \\
\cdashline{3-9}
 &  & Llama-3 & 46.0\%$^{\textcolor{teal}{+12.1}}$ & 41.3\%$^{\textcolor{teal}{+2.1}}$ & 39.2\%$^{\textcolor{teal}{+5.3}}$ & 39.2\%$^{\textcolor{red}{-0.0}}$ & 43.9\%$^{\textcolor{teal}{+10.0}}$ & 45.5\%$^{\textcolor{teal}{+6.3}}$ \\
 &  & Qwen2.5-Coder-14B & 75.7\%$^{\textcolor{teal}{+26.0}}$ & 70.9\%$^{\textcolor{teal}{+1.6}}$ & 70.4\%$^{\textcolor{teal}{+20.7}}$ & 72.5\%$^{\textcolor{teal}{+3.2}}$ & 75.7\%$^{\textcolor{teal}{+26.0}}$ & 74.6\%$^{\textcolor{teal}{+5.3}}$ \\
 &  & Qwen2.5-Coder-0.5B & 3.7\%$^{\textcolor{teal}{+3.7}}$ & 3.2\%$^{\textcolor{teal}{+2.1}}$ & 1.1\%$^{\textcolor{teal}{+1.1}}$ & 4.2\%$^{\textcolor{teal}{+3.1}}$ & 2.6\%$^{\textcolor{teal}{+2.6}}$ & 3.2\%$^{\textcolor{teal}{+2.1}}$ \\
\cline{2-9}
 & \multirow{6}{*}{C} & GPT-4o & 74.1\%$^{\textcolor{red}{-0.5}}$ & 76.2\%$^{\textcolor{teal}{+2.7}}$ & 70.9\%$^{\textcolor{red}{-3.7}}$ & 72.0\%$^{\textcolor{red}{-1.5}}$ & 76.7\%$^{\textcolor{teal}{+2.1}}$ & 77.1\%$^{\textcolor{teal}{+3.6}}$ \\
 &  & o3-mini & 64.6\%$^{\textcolor{red}{-15.3}}$ & 63.0\%$^{\textcolor{red}{-8.4}}$ & 63.5\%$^{\textcolor{red}{-16.4}}$ & 65.6\%$^{\textcolor{red}{-5.8}}$ & 63.0\%$^{\textcolor{red}{-16.9}}$ & 62.9\%$^{\textcolor{red}{-8.5}}$ \\
 &  & DeepSeek-V3.2 & 54.5\%$^{\textcolor{red}{-24.3}}$ & 52.4\%$^{\textcolor{red}{-28.0}}$ & 48.1\%$^{\textcolor{red}{-30.7}}$ & 46.0\%$^{\textcolor{red}{-34.4}}$ & 49.2\%$^{\textcolor{red}{-29.6}}$ & 41.8\%$^{\textcolor{red}{-38.6}}$ \\
\cdashline{3-9}
 &  & Llama-3 & 10.1\%$^{\textcolor{teal}{+0.6}}$ & 11.1\%$^{\textcolor{teal}{+2.6}}$ & 9.5\%$^{\textcolor{teal}{+0.0}}$ & 11.6\%$^{\textcolor{teal}{+3.1}}$ & 13.8\%$^{\textcolor{teal}{+4.3}}$ & 10.1\%$^{\textcolor{teal}{+1.6}}$ \\
 &  & Qwen2.5-Coder-14B & 37.0\%$^{\textcolor{teal}{+18.0}}$ & 36.0\%$^{\textcolor{teal}{+2.7}}$ & 33.3\%$^{\textcolor{teal}{+14.3}}$ & 36.5\%$^{\textcolor{teal}{+3.2}}$ & 37.6\%$^{\textcolor{teal}{+18.6}}$ & 37.0\%$^{\textcolor{teal}{+3.7}}$ \\
 &  & Qwen2.5-Coder-0.5B & 0.0\%$^{\textcolor{teal}{+0.0}}$ & 0.5\%$^{\textcolor{teal}{+0.5}}$ & 0.0\%$^{\textcolor{teal}{+0.0}}$ & 1.1\%$^{\textcolor{teal}{+1.1}}$ & 0.5\%$^{\textcolor{teal}{+0.5}}$ & 0.0\%$^{\textcolor{teal}{+0.0}}$ \\
\cline{2-9}
 & \multirow{6}{*}{D} & GPT-4o & 46.6\%$^{\textcolor{red}{-0.5}}$ & 42.3\%$^{\textcolor{red}{-3.2}}$ & 44.4\%$^{\textcolor{red}{-2.7}}$ & 44.4\%$^{\textcolor{red}{-1.1}}$ & 47.1\%$^{\textcolor{red}{-0.0}}$ & 50.0\%$^{\textcolor{teal}{+4.5}}$ \\
 &  & o3-mini & 47.6\%$^{\textcolor{red}{-15.9}}$ & 46.6\%$^{\textcolor{red}{-5.8}}$ & 48.7\%$^{\textcolor{red}{-14.8}}$ & 46.0\%$^{\textcolor{red}{-6.4}}$ & 51.3\%$^{\textcolor{red}{-12.2}}$ & 53.6\%$^{\textcolor{teal}{+1.2}}$ \\
 &  & DeepSeek-V3.2 & 31.7\%$^{\textcolor{red}{-23.3}}$ & 31.7\%$^{\textcolor{red}{-26.5}}$ & 25.9\%$^{\textcolor{red}{-29.1}}$ & 25.9\%$^{\textcolor{red}{-32.3}}$ & 30.7\%$^{\textcolor{red}{-24.3}}$ & 24.3\%$^{\textcolor{red}{-33.9}}$ \\
\cdashline{3-9}
 &  & Llama-3 & 2.1\%$^{\textcolor{teal}{+2.1}}$ & 2.6\%$^{\textcolor{teal}{+1.0}}$ & 3.7\%$^{\textcolor{teal}{+3.7}}$ & 4.8\%$^{\textcolor{teal}{+3.2}}$ & 3.7\%$^{\textcolor{teal}{+3.7}}$ & 3.2\%$^{\textcolor{teal}{+1.6}}$ \\
 &  & Qwen2.5-Coder-14B & 10.6\%$^{\textcolor{teal}{+5.8}}$ & 12.7\%$^{\textcolor{teal}{+3.2}}$ & 9.0\%$^{\textcolor{teal}{+4.2}}$ & 10.6\%$^{\textcolor{teal}{+1.1}}$ & 12.7\%$^{\textcolor{teal}{+7.9}}$ & 11.6\%$^{\textcolor{teal}{+2.1}}$ \\
 &  & Qwen2.5-Coder-0.5B & 0.0\%$^{\textcolor{teal}{+0.0}}$ & 0.0\%$^{\textcolor{teal}{+0.0}}$ & 0.0\%$^{\textcolor{teal}{+0.0}}$ & 0.0\%$^{\textcolor{teal}{+0.0}}$ & 0.0\%$^{\textcolor{teal}{+0.0}}$ & 0.0\%$^{\textcolor{teal}{+0.0}}$ \\
\hline
\multirow{18}{*}{LeetCode} & \multirow{6}{*}{Easy} & GPT-4o & 93.8\%$^{\textcolor{teal}{+6.2}}$ & 93.8\%$^{\textcolor{teal}{+12.5}}$ & 87.5\%$^{\textcolor{teal}{+0.0}}$ & 93.8\%$^{\textcolor{teal}{+12.5}}$ & 93.8\%$^{\textcolor{teal}{+6.2}}$ & 93.8\%$^{\textcolor{teal}{+12.5}}$ \\
 &  & o3-mini & 93.8\%$^{\textcolor{teal}{+6.2}}$ & 93.8\%$^{\textcolor{teal}{+6.2}}$ & 93.8\%$^{\textcolor{teal}{+6.2}}$ & 93.8\%$^{\textcolor{teal}{+6.2}}$ & 93.8\%$^{\textcolor{teal}{+6.2}}$ & 93.8\%$^{\textcolor{teal}{+6.2}}$ \\
 &  & DeepSeek-V3.2 & 87.5\%$^{\textcolor{teal}{+18.7}}$ & 87.5\%$^{\textcolor{teal}{+0.0}}$ & 93.8\%$^{\textcolor{teal}{+25.0}}$ & 87.5\%$^{\textcolor{teal}{+0.0}}$ & 87.5\%$^{\textcolor{teal}{+18.7}}$ & 87.5\%$^{\textcolor{teal}{+0.0}}$ \\
\cdashline{3-9}
 &  & Llama-3 & 81.2\%$^{\textcolor{teal}{+25.0}}$ & 81.2\%$^{\textcolor{teal}{+18.8}}$ & 68.8\%$^{\textcolor{teal}{+12.5}}$ & 62.5\%$^{\textcolor{teal}{+0.0}}$ & 75.0\%$^{\textcolor{teal}{+18.8}}$ & 75.0\%$^{\textcolor{teal}{+12.5}}$ \\
 &  & Qwen2.5-Coder-14B & 87.5\%$^{\textcolor{teal}{+25.0}}$ & 87.5\%$^{\textcolor{teal}{+6.3}}$ & 81.2\%$^{\textcolor{teal}{+18.8}}$ & 87.5\%$^{\textcolor{teal}{+6.3}}$ & 87.5\%$^{\textcolor{teal}{+25.0}}$ & 87.5\%$^{\textcolor{teal}{+6.3}}$ \\
 &  & Qwen2.5-Coder-0.5B & 18.8\%$^{\textcolor{teal}{+12.6}}$ & 6.2\%$^{\textcolor{red}{-6.2}}$ & 18.8\%$^{\textcolor{teal}{+12.6}}$ & 25.0\%$^{\textcolor{teal}{+12.5}}$ & 18.8\%$^{\textcolor{teal}{+12.6}}$ & 12.5\%$^{\textcolor{teal}{+0.0}}$ \\
\cline{2-9}
 & \multirow{6}{*}{Medium} & GPT-4o & 72.5\%$^{\textcolor{teal}{+5.8}}$ & 76.9\%$^{\textcolor{teal}{+3.8}}$ & 70.6\%$^{\textcolor{teal}{+3.9}}$ & 80.8\%$^{\textcolor{teal}{+7.7}}$ & 74.5\%$^{\textcolor{teal}{+7.8}}$ & 80.8\%$^{\textcolor{teal}{+7.7}}$ \\
 &  & o3-mini & 60.8\%$^{\textcolor{teal}{+2.0}}$ & 82.7\%$^{\textcolor{teal}{+7.7}}$ & 70.6\%$^{\textcolor{teal}{+11.8}}$ & 80.8\%$^{\textcolor{teal}{+5.8}}$ & 66.7\%$^{\textcolor{teal}{+7.9}}$ & 82.7\%$^{\textcolor{teal}{+7.7}}$ \\
 &  & DeepSeek-V3.2 & 84.3\%$^{\textcolor{teal}{+15.7}}$ & 80.8\%$^{\textcolor{red}{-5.7}}$ & 82.4\%$^{\textcolor{teal}{+13.8}}$ & 84.6\%$^{\textcolor{red}{-1.9}}$ & 82.4\%$^{\textcolor{teal}{+13.8}}$ & 80.8\%$^{\textcolor{red}{-5.7}}$ \\
\cdashline{3-9}
 &  & Llama-3 & 41.2\%$^{\textcolor{teal}{+15.7}}$ & 46.2\%$^{\textcolor{teal}{+7.7}}$ & 37.3\%$^{\textcolor{teal}{+11.8}}$ & 50.0\%$^{\textcolor{teal}{+11.5}}$ & 45.1\%$^{\textcolor{teal}{+19.6}}$ & 50.0\%$^{\textcolor{teal}{+11.5}}$ \\
 &  & Qwen2.5-Coder-14B & 78.4\%$^{\textcolor{teal}{+11.7}}$ & 73.1\%$^{\textcolor{red}{-0.0}}$ & 74.5\%$^{\textcolor{teal}{+7.8}}$ & 73.1\%$^{\textcolor{red}{-0.0}}$ & 78.4\%$^{\textcolor{teal}{+11.7}}$ & 75.0\%$^{\textcolor{teal}{+1.9}}$ \\
 &  & Qwen2.5-Coder-0.5B & 7.8\%$^{\textcolor{teal}{+5.8}}$ & 9.6\%$^{\textcolor{teal}{+7.7}}$ & 7.8\%$^{\textcolor{teal}{+5.8}}$ & 7.7\%$^{\textcolor{teal}{+5.8}}$ & 7.8\%$^{\textcolor{teal}{+5.8}}$ & 7.7\%$^{\textcolor{teal}{+5.8}}$ \\
\cline{2-9}
 & \multirow{6}{*}{Hard} & GPT-4o & 68.8\%$^{\textcolor{red}{-12.5}}$ & 81.2\%$^{\textcolor{teal}{+0.0}}$ & 68.8\%$^{\textcolor{red}{-12.5}}$ & 84.4\%$^{\textcolor{teal}{+3.2}}$ & 71.9\%$^{\textcolor{red}{-9.3}}$ & 87.5\%$^{\textcolor{teal}{+6.3}}$ \\
 &  & o3-mini & 62.5\%$^{\textcolor{teal}{+0.0}}$ & 84.4\%$^{\textcolor{red}{-0.0}}$ & 56.2\%$^{\textcolor{red}{-6.2}}$ & 84.4\%$^{\textcolor{red}{-0.0}}$ & 56.2\%$^{\textcolor{red}{-6.2}}$ & 84.4\%$^{\textcolor{red}{-0.0}}$ \\
 &  & DeepSeek-V3.2 & 81.2\%$^{\textcolor{teal}{+9.3}}$ & 84.4\%$^{\textcolor{teal}{+3.2}}$ & 81.2\%$^{\textcolor{teal}{+9.3}}$ & 87.5\%$^{\textcolor{teal}{+6.3}}$ & 84.4\%$^{\textcolor{teal}{+12.5}}$ & 87.5\%$^{\textcolor{teal}{+6.3}}$ \\
\cdashline{3-9}
 &  & Llama-3 & 28.1\%$^{\textcolor{teal}{+6.2}}$ & 40.6\%$^{\textcolor{teal}{+18.7}}$ & 31.2\%$^{\textcolor{teal}{+9.4}}$ & 28.1\%$^{\textcolor{teal}{+6.2}}$ & 31.2\%$^{\textcolor{teal}{+9.4}}$ & 34.4\%$^{\textcolor{teal}{+12.5}}$ \\
 &  & Qwen2.5-Coder-14B & 59.4\%$^{\textcolor{red}{-0.0}}$ & 56.2\%$^{\textcolor{red}{-3.1}}$ & 53.1\%$^{\textcolor{red}{-6.3}}$ & 65.6\%$^{\textcolor{teal}{+6.2}}$ & 62.5\%$^{\textcolor{teal}{+3.1}}$ & 65.6\%$^{\textcolor{teal}{+6.2}}$ \\
 &  & Qwen2.5-Coder-0.5B & 6.2\%$^{\textcolor{teal}{+6.2}}$ & 3.1\%$^{\textcolor{teal}{+0.0}}$ & 6.2\%$^{\textcolor{teal}{+6.2}}$ & 6.2\%$^{\textcolor{teal}{+3.1}}$ & 3.1\%$^{\textcolor{teal}{+3.1}}$ & 6.2\%$^{\textcolor{teal}{+3.1}}$ \\
\hline
\multirow{6}{*}{BigCodeBench} & \multirow{6}{*}{All} & GPT-4o & 28.8\%$^{\textcolor{red}{-28.5}}$ & 31.9\%$^{\textcolor{teal}{+11.8}}$ & 30.9\%$^{\textcolor{red}{-26.4}}$ & 28.5\%$^{\textcolor{teal}{+8.4}}$ & 59.0\%$^{\textcolor{teal}{+1.7}}$ & 56.2\%$^{\textcolor{teal}{+36.1}}$ \\
 &  & o3-mini & 58.2\%$^{\textcolor{teal}{+1.6}}$ & 57.0\%$^{\textcolor{teal}{+6.3}}$ & 59.4\%$^{\textcolor{teal}{+2.8}}$ & 54.9\%$^{\textcolor{teal}{+4.2}}$ & 53.8\%$^{\textcolor{red}{-2.8}}$ & 52.8\%$^{\textcolor{teal}{+2.1}}$ \\
 &  & DeepSeek-V3.2 & 60.6\%$^{\textcolor{teal}{+3.7}}$ & 58.0\%$^{\textcolor{teal}{+2.4}}$ & 59.4\%$^{\textcolor{teal}{+2.5}}$ & 58.0\%$^{\textcolor{teal}{+2.4}}$ & 53.1\%$^{\textcolor{red}{-3.8}}$ & 53.5\%$^{\textcolor{red}{-2.1}}$ \\
\cdashline{3-9}
 &  & Llama-3 & 35.2\%$^{\textcolor{red}{-0.1}}$ & 34.5\%$^{\textcolor{red}{-2.2}}$ & 35.4\%$^{\textcolor{teal}{+0.1}}$ & 34.4\%$^{\textcolor{red}{-2.3}}$ & 17.1\%$^{\textcolor{red}{-18.2}}$ & 12.8\%$^{\textcolor{red}{-23.9}}$ \\
 &  & Qwen2.5-Coder-14B & 54.2\%$^{\textcolor{red}{-0.5}}$ & 55.9\%$^{\textcolor{teal}{+1.2}}$ & 55.2\%$^{\textcolor{teal}{+0.5}}$ & 55.9\%$^{\textcolor{teal}{+1.2}}$ & 50.9\%$^{\textcolor{red}{-3.8}}$ & 48.8\%$^{\textcolor{red}{-5.9}}$ \\
 &  & Qwen2.5-Coder-0.5B & 9.4\%$^{\textcolor{teal}{+1.1}}$ & 10.1\%$^{\textcolor{teal}{+0.7}}$ & 10.8\%$^{\textcolor{teal}{+2.5}}$ & 10.4\%$^{\textcolor{teal}{+1.0}}$ & 9.7\%$^{\textcolor{teal}{+1.4}}$ & 9.8\%$^{\textcolor{teal}{+0.4}}$ \\
\hline
\end{tabular}}
\end{table}

\begin{table}[t]
\caption{Median $\mbox{\sl Accuracy}$ of code generation after translating problem statements into Japanese and $\mbox{\sl Accuracy}$ difference from the original language}
\centering
\label{tab:translation-to-JA}
\scalebox{0.6}{
\begin{tabular}{lllrrrrrr}
\hline
\multicolumn{1}{c}{Problems} & \multicolumn{1}{c}{Difficulty} & \multicolumn{1}{c}{Model} & \multicolumn{2}{c}{Google} & \multicolumn{2}{c}{DeepL} & \multicolumn{2}{c}{GPT} \\
 &  &  & \multicolumn{1}{c}{From English} & \multicolumn{1}{c}{From Chinese} & \multicolumn{1}{c}{From English} & \multicolumn{1}{c}{From Chinese} & \multicolumn{1}{c}{From English} & \multicolumn{1}{c}{From Chinese} \\ \hline
\multirow{24}{*}{AtCoder} & \multirow{6}{*}{A} & GPT-4o & 83.6\%$^{\textcolor{red}{-9.0}}$ & 83.6\%$^{\textcolor{red}{-10.6}}$ & 73.5\%$^{\textcolor{red}{-19.1}}$ & 73.0\%$^{\textcolor{red}{-21.2}}$ & 90.5\%$^{\textcolor{red}{-2.1}}$ & 92.1\%$^{\textcolor{red}{-2.1}}$ \\
 &  & o3-mini & 81.5\%$^{\textcolor{red}{-13.2}}$ & 78.8\%$^{\textcolor{red}{-15.4}}$ & 72.5\%$^{\textcolor{red}{-22.2}}$ & 71.4\%$^{\textcolor{red}{-22.8}}$ & 93.7\%$^{\textcolor{red}{-1.0}}$ & 92.6\%$^{\textcolor{red}{-1.6}}$ \\
 &  & DeepSeek-V3.2 & 75.7\%$^{\textcolor{red}{-19.0}}$ & 79.4\%$^{\textcolor{red}{-14.8}}$ & 70.9\%$^{\textcolor{red}{-23.8}}$ & 72.0\%$^{\textcolor{red}{-22.2}}$ & 75.7\%$^{\textcolor{red}{-19.0}}$ & 74.6\%$^{\textcolor{red}{-19.6}}$ \\
\cdashline{3-9}
 &  & Llama-3 & 51.9\%$^{\textcolor{red}{-15.3}}$ & 44.4\%$^{\textcolor{red}{-14.9}}$ & 45.0\%$^{\textcolor{red}{-22.2}}$ & 46.2\%$^{\textcolor{red}{-13.1}}$ & 50.8\%$^{\textcolor{red}{-16.4}}$ & 55.6\%$^{\textcolor{red}{-3.7}}$ \\
 &  & Qwen2.5-Coder-14B & 54.5\%$^{\textcolor{red}{-33.3}}$ & 36.0\%$^{\textcolor{red}{-47.6}}$ & 55.0\%$^{\textcolor{red}{-32.8}}$ & 54.5\%$^{\textcolor{red}{-29.1}}$ & 81.0\%$^{\textcolor{red}{-6.8}}$ & 82.0\%$^{\textcolor{red}{-1.6}}$ \\
 &  & Qwen2.5-Coder-0.5B & 1.1\%$^{\textcolor{red}{-7.9}}$ & 0.5\%$^{\textcolor{red}{-4.3}}$ & 2.1\%$^{\textcolor{red}{-6.9}}$ & 0.5\%$^{\textcolor{red}{-4.3}}$ & 2.6\%$^{\textcolor{red}{-6.4}}$ & 2.1\%$^{\textcolor{red}{-2.7}}$ \\
\cline{2-9}
 & \multirow{6}{*}{B} & GPT-4o & 81.0\%$^{\textcolor{red}{-7.4}}$ & 82.5\%$^{\textcolor{red}{-4.8}}$ & 70.9\%$^{\textcolor{red}{-17.5}}$ & 68.3\%$^{\textcolor{red}{-19.0}}$ & 87.3\%$^{\textcolor{red}{-1.1}}$ & 87.3\%$^{\textcolor{teal}{+0.0}}$ \\
 &  & o3-mini & 78.8\%$^{\textcolor{red}{-6.4}}$ & 76.7\%$^{\textcolor{red}{-11.1}}$ & 69.3\%$^{\textcolor{red}{-15.9}}$ & 69.3\%$^{\textcolor{red}{-18.5}}$ & 89.4\%$^{\textcolor{teal}{+4.2}}$ & 87.3\%$^{\textcolor{red}{-0.5}}$ \\
 &  & DeepSeek-V3.2 & 67.7\%$^{\textcolor{red}{-24.9}}$ & 65.6\%$^{\textcolor{red}{-25.9}}$ & 57.7\%$^{\textcolor{red}{-34.9}}$ & 60.3\%$^{\textcolor{red}{-31.2}}$ & 65.1\%$^{\textcolor{red}{-27.5}}$ & 57.1\%$^{\textcolor{red}{-34.4}}$ \\
\cdashline{3-9}
 &  & Llama-3 & 31.2\%$^{\textcolor{red}{-15.4}}$ & 25.9\%$^{\textcolor{red}{-13.3}}$ & 34.4\%$^{\textcolor{red}{-12.2}}$ & 28.8\%$^{\textcolor{red}{-10.4}}$ & 32.3\%$^{\textcolor{red}{-14.3}}$ & 30.2\%$^{\textcolor{red}{-9.0}}$ \\
 &  & Qwen2.5-Coder-14B & 28.0\%$^{\textcolor{red}{-48.2}}$ & 15.3\%$^{\textcolor{red}{-54.0}}$ & 37.0\%$^{\textcolor{red}{-39.2}}$ & 38.1\%$^{\textcolor{red}{-31.2}}$ & 63.0\%$^{\textcolor{red}{-13.2}}$ & 63.5\%$^{\textcolor{red}{-5.8}}$ \\
 &  & Qwen2.5-Coder-0.5B & 1.1\%$^{\textcolor{red}{-1.0}}$ & 0.5\%$^{\textcolor{red}{-0.6}}$ & 0.0\%$^{\textcolor{red}{-2.1}}$ & 0.5\%$^{\textcolor{red}{-0.6}}$ & 0.0\%$^{\textcolor{red}{-2.1}}$ & 0.0\%$^{\textcolor{red}{-1.1}}$ \\
\cline{2-9}
 & \multirow{6}{*}{C} & GPT-4o & 72.5\%$^{\textcolor{red}{-5.3}}$ & 72.5\%$^{\textcolor{red}{-1.0}}$ & 67.7\%$^{\textcolor{red}{-10.1}}$ & 66.7\%$^{\textcolor{red}{-6.8}}$ & 79.9\%$^{\textcolor{teal}{+2.1}}$ & 74.1\%$^{\textcolor{teal}{+0.6}}$ \\
 &  & o3-mini & 71.2\%$^{\textcolor{teal}{+3.5}}$ & 69.8\%$^{\textcolor{red}{-1.6}}$ & 69.3\%$^{\textcolor{teal}{+1.6}}$ & 70.9\%$^{\textcolor{red}{-0.5}}$ & 77.2\%$^{\textcolor{teal}{+9.5}}$ & 75.1\%$^{\textcolor{teal}{+3.7}}$ \\
 &  & DeepSeek-V3.2 & 45.0\%$^{\textcolor{red}{-34.9}}$ & 43.9\%$^{\textcolor{red}{-36.5}}$ & 39.7\%$^{\textcolor{red}{-40.2}}$ & 38.6\%$^{\textcolor{red}{-41.8}}$ & 31.2\%$^{\textcolor{red}{-48.7}}$ & 33.9\%$^{\textcolor{red}{-46.5}}$ \\
\cdashline{3-9}
 &  & Llama-3 & 5.8\%$^{\textcolor{red}{-4.3}}$ & 4.8\%$^{\textcolor{red}{-3.7}}$ & 7.4\%$^{\textcolor{red}{-2.7}}$ & 6.1\%$^{\textcolor{red}{-2.4}}$ & 7.9\%$^{\textcolor{red}{-2.2}}$ & 6.9\%$^{\textcolor{red}{-1.6}}$ \\
 &  & Qwen2.5-Coder-14B & 15.3\%$^{\textcolor{red}{-21.2}}$ & 9.0\%$^{\textcolor{red}{-24.3}}$ & 13.8\%$^{\textcolor{red}{-22.7}}$ & 15.9\%$^{\textcolor{red}{-17.4}}$ & 31.2\%$^{\textcolor{red}{-5.3}}$ & 28.6\%$^{\textcolor{red}{-4.7}}$ \\
 &  & Qwen2.5-Coder-0.5B & 0.5\%$^{\textcolor{teal}{+0.5}}$ & 0.5\%$^{\textcolor{teal}{+0.5}}$ & 0.5\%$^{\textcolor{teal}{+0.5}}$ & 0.5\%$^{\textcolor{teal}{+0.5}}$ & 0.5\%$^{\textcolor{teal}{+0.5}}$ & 0.5\%$^{\textcolor{teal}{+0.5}}$ \\
\cline{2-9}
 & \multirow{6}{*}{D} & GPT-4o & 46.0\%$^{\textcolor{red}{-1.1}}$ & 46.0\%$^{\textcolor{teal}{+0.5}}$ & 42.9\%$^{\textcolor{red}{-4.2}}$ & 39.2\%$^{\textcolor{red}{-6.3}}$ & 49.7\%$^{\textcolor{teal}{+2.6}}$ & 48.7\%$^{\textcolor{teal}{+3.2}}$ \\
 &  & o3-mini & 56.3\%$^{\textcolor{teal}{+6.0}}$ & 52.9\%$^{\textcolor{teal}{+0.5}}$ & 56.6\%$^{\textcolor{teal}{+6.3}}$ & 53.4\%$^{\textcolor{teal}{+1.0}}$ & 62.4\%$^{\textcolor{teal}{+12.1}}$ & 56.2\%$^{\textcolor{teal}{+3.8}}$ \\
 &  & DeepSeek-V3.2 & 25.4\%$^{\textcolor{red}{-27.5}}$ & 26.5\%$^{\textcolor{red}{-31.7}}$ & 25.9\%$^{\textcolor{red}{-27.0}}$ & 22.8\%$^{\textcolor{red}{-35.4}}$ & 18.5\%$^{\textcolor{red}{-34.4}}$ & 19.0\%$^{\textcolor{red}{-39.2}}$ \\
\cdashline{3-9}
 &  & Llama-3 & 1.1\%$^{\textcolor{red}{-0.5}}$ & 0.0\%$^{\textcolor{red}{-1.6}}$ & 0.5\%$^{\textcolor{red}{-1.1}}$ & 0.0\%$^{\textcolor{red}{-1.6}}$ & 0.5\%$^{\textcolor{red}{-1.1}}$ & 1.1\%$^{\textcolor{red}{-0.5}}$ \\
 &  & Qwen2.5-Coder-14B & 5.3\%$^{\textcolor{red}{-5.8}}$ & 2.6\%$^{\textcolor{red}{-6.9}}$ & 6.3\%$^{\textcolor{red}{-4.8}}$ & 3.7\%$^{\textcolor{red}{-5.8}}$ & 9.0\%$^{\textcolor{red}{-2.1}}$ & 7.9\%$^{\textcolor{red}{-1.6}}$ \\
 &  & Qwen2.5-Coder-0.5B & 0.0\%$^{\textcolor{teal}{+0.0}}$ & 0.0\%$^{\textcolor{teal}{+0.0}}$ & 0.0\%$^{\textcolor{teal}{+0.0}}$ & 0.0\%$^{\textcolor{teal}{+0.0}}$ & 0.0\%$^{\textcolor{teal}{+0.0}}$ & 0.0\%$^{\textcolor{teal}{+0.0}}$ \\
\hline
\end{tabular}}
\end{table}
\begin{table}[t]
\caption{Median $\mbox{\sl Accuracy}$ of code generation after translating problem statements into Chinese and $\mbox{\sl Accuracy}$ difference from the original language}
\centering
\label{tab:translation-to-ZH}
\scalebox{0.6}{
\begin{tabular}{lllrrrrrr}
\hline
\multicolumn{1}{c}{Problems} & \multicolumn{1}{c}{Difficulty} & \multicolumn{1}{c}{Model} & \multicolumn{2}{c}{Google} & \multicolumn{2}{c}{DeepL} & \multicolumn{2}{c}{GPT} \\
 &  &  & \multicolumn{1}{c}{From English} & \multicolumn{1}{c}{From Japanese} & \multicolumn{1}{c}{From English} & \multicolumn{1}{c}{From Japanese} & \multicolumn{1}{c}{From English} & \multicolumn{1}{c}{From Japanese} \\ \hline
\multirow{18}{*}{LeetCode} & \multirow{6}{*}{Easy} & GPT-4o & 81.2\%$^{\textcolor{teal}{+6.2}}$ & 93.8\%$^{\textcolor{teal}{+6.2}}$ & 93.8\%$^{\textcolor{teal}{+18.8}}$ & 87.5\%$^{\textcolor{teal}{+0.0}}$ & 93.8\%$^{\textcolor{teal}{+18.8}}$ & 93.8\%$^{\textcolor{teal}{+6.2}}$ \\
 &  & o3-mini & 93.8\%$^{\textcolor{teal}{+6.2}}$ & 93.8\%$^{\textcolor{teal}{+6.2}}$ & 87.5\%$^{\textcolor{teal}{+0.0}}$ & 87.5\%$^{\textcolor{teal}{+0.0}}$ & 87.5\%$^{\textcolor{teal}{+0.0}}$ & 93.8\%$^{\textcolor{teal}{+6.2}}$ \\
 &  & DeepSeek-V3.2 & 87.5\%$^{\textcolor{teal}{+0.0}}$ & 87.5\%$^{\textcolor{teal}{+18.7}}$ & 87.5\%$^{\textcolor{teal}{+0.0}}$ & 87.5\%$^{\textcolor{teal}{+18.7}}$ & 87.5\%$^{\textcolor{teal}{+0.0}}$ & 87.5\%$^{\textcolor{teal}{+18.7}}$ \\
\cdashline{3-9}
 &  & Llama-3 & 87.5\%$^{\textcolor{teal}{+31.3}}$ & 75.0\%$^{\textcolor{teal}{+18.8}}$ & 75.0\%$^{\textcolor{teal}{+18.8}}$ & 81.2\%$^{\textcolor{teal}{+25.0}}$ & 81.2\%$^{\textcolor{teal}{+25.0}}$ & 87.5\%$^{\textcolor{teal}{+31.3}}$ \\
 &  & Qwen2.5-Coder-14B & 87.5\%$^{\textcolor{teal}{+12.5}}$ & 81.2\%$^{\textcolor{teal}{+18.8}}$ & 87.5\%$^{\textcolor{teal}{+12.5}}$ & 87.5\%$^{\textcolor{teal}{+25.0}}$ & 87.5\%$^{\textcolor{teal}{+12.5}}$ & 87.5\%$^{\textcolor{teal}{+25.0}}$ \\
 &  & Qwen2.5-Coder-0.5B & 31.2\%$^{\textcolor{teal}{+18.8}}$ & 25.0\%$^{\textcolor{teal}{+18.8}}$ & 31.2\%$^{\textcolor{teal}{+18.8}}$ & 18.8\%$^{\textcolor{teal}{+12.6}}$ & 31.2\%$^{\textcolor{teal}{+18.8}}$ & 25.0\%$^{\textcolor{teal}{+18.8}}$ \\
\cline{2-9}
 & \multirow{6}{*}{Medium} & GPT-4o & 78.8\%$^{\textcolor{teal}{+3.8}}$ & 70.6\%$^{\textcolor{teal}{+3.9}}$ & 80.8\%$^{\textcolor{teal}{+5.8}}$ & 72.5\%$^{\textcolor{teal}{+5.8}}$ & 76.9\%$^{\textcolor{teal}{+1.9}}$ & 70.6\%$^{\textcolor{teal}{+3.9}}$ \\
 &  & o3-mini & 75.0\%$^{\textcolor{teal}{+1.9}}$ & 58.8\%$^{\textcolor{teal}{+0.0}}$ & 73.1\%$^{\textcolor{red}{-0.0}}$ & 58.8\%$^{\textcolor{teal}{+0.0}}$ & 78.8\%$^{\textcolor{teal}{+5.7}}$ & 62.7\%$^{\textcolor{teal}{+3.9}}$ \\
 &  & DeepSeek-V3.2 & 84.6\%$^{\textcolor{teal}{+1.9}}$ & 86.3\%$^{\textcolor{teal}{+17.7}}$ & 84.6\%$^{\textcolor{teal}{+1.9}}$ & 86.3\%$^{\textcolor{teal}{+17.7}}$ & 84.6\%$^{\textcolor{teal}{+1.9}}$ & 86.3\%$^{\textcolor{teal}{+17.7}}$ \\
\cdashline{3-9}
 &  & Llama-3 & 38.5\%$^{\textcolor{red}{-0.0}}$ & 35.3\%$^{\textcolor{teal}{+9.8}}$ & 46.2\%$^{\textcolor{teal}{+7.7}}$ & 35.3\%$^{\textcolor{teal}{+9.8}}$ & 38.5\%$^{\textcolor{red}{-0.0}}$ & 35.3\%$^{\textcolor{teal}{+9.8}}$ \\
 &  & Qwen2.5-Coder-14B & 75.0\%$^{\textcolor{teal}{+5.8}}$ & 78.4\%$^{\textcolor{teal}{+11.7}}$ & 78.8\%$^{\textcolor{teal}{+9.6}}$ & 70.6\%$^{\textcolor{teal}{+3.9}}$ & 76.9\%$^{\textcolor{teal}{+7.7}}$ & 76.5\%$^{\textcolor{teal}{+9.8}}$ \\
 &  & Qwen2.5-Coder-0.5B & 5.8\%$^{\textcolor{red}{-0.0}}$ & 3.9\%$^{\textcolor{teal}{+1.9}}$ & 3.8\%$^{\textcolor{red}{-2.0}}$ & 3.9\%$^{\textcolor{teal}{+1.9}}$ & 7.7\%$^{\textcolor{teal}{+1.9}}$ & 2.0\%$^{\textcolor{red}{-0.0}}$ \\
\cline{2-9}
 & \multirow{6}{*}{Hard} & GPT-4o & 81.2\%$^{\textcolor{red}{-3.2}}$ & 68.8\%$^{\textcolor{red}{-12.5}}$ & 84.4\%$^{\textcolor{red}{-0.0}}$ & 68.8\%$^{\textcolor{red}{-12.5}}$ & 87.5\%$^{\textcolor{teal}{+3.1}}$ & 71.9\%$^{\textcolor{red}{-9.3}}$ \\
 &  & o3-mini & 84.4\%$^{\textcolor{red}{-0.0}}$ & 46.9\%$^{\textcolor{red}{-15.6}}$ & 81.2\%$^{\textcolor{red}{-3.2}}$ & 50.0\%$^{\textcolor{red}{-12.5}}$ & 81.2\%$^{\textcolor{red}{-3.2}}$ & 43.8\%$^{\textcolor{red}{-18.8}}$ \\
 &  & DeepSeek-V3.2 & 84.4\%$^{\textcolor{red}{-3.1}}$ & 87.5\%$^{\textcolor{teal}{+15.6}}$ & 84.4\%$^{\textcolor{red}{-3.1}}$ & 81.2\%$^{\textcolor{teal}{+9.3}}$ & 84.4\%$^{\textcolor{red}{-3.1}}$ & 87.5\%$^{\textcolor{teal}{+15.6}}$ \\
\cdashline{3-9}
 &  & Llama-3 & 31.2\%$^{\textcolor{teal}{+3.1}}$ & 25.0\%$^{\textcolor{teal}{+3.1}}$ & 28.1\%$^{\textcolor{teal}{+0.0}}$ & 21.9\%$^{\textcolor{red}{-0.0}}$ & 28.1\%$^{\textcolor{teal}{+0.0}}$ & 28.1\%$^{\textcolor{teal}{+6.2}}$ \\
 &  & Qwen2.5-Coder-14B & 68.8\%$^{\textcolor{red}{-0.0}}$ & 71.9\%$^{\textcolor{teal}{+12.5}}$ & 71.9\%$^{\textcolor{teal}{+3.1}}$ & 59.4\%$^{\textcolor{red}{-0.0}}$ & 71.9\%$^{\textcolor{teal}{+3.1}}$ & 65.6\%$^{\textcolor{teal}{+6.2}}$ \\
 &  & Qwen2.5-Coder-0.5B & 3.1\%$^{\textcolor{teal}{+3.1}}$ & 0.0\%$^{\textcolor{teal}{+0.0}}$ & 3.1\%$^{\textcolor{teal}{+3.1}}$ & 0.0\%$^{\textcolor{teal}{+0.0}}$ & 3.1\%$^{\textcolor{teal}{+3.1}}$ & 0.0\%$^{\textcolor{teal}{+0.0}}$ \\
\hline
\end{tabular}}
\end{table}
\begin{table}[t]
\caption{Dataset by model-group diff summary: Target language is English}
\centering
\label{tab:english-dataset-model-group-diff-summary}
\scalebox{0.8}{
\begin{tabular}{llrrrrrrr}
\hline
\multicolumn{1}{c}{Problems} & \multicolumn{1}{c}{Model Group} & \multicolumn{1}{c}{Count} & \multicolumn{1}{c}{Mean} & \multicolumn{1}{c}{Median} & \multicolumn{1}{c}{Std} & \multicolumn{1}{c}{Improve} & \multicolumn{1}{c}{Worsen} & \multicolumn{1}{c}{Unchanged} \\
\hline
\multirow{2}{*}{AtCoder} & Closed & 72 & -9.2\% & -5.3\% & 10.5\% & 9.7\% & 90.3\% & 0.0\% \\
 & Open & 72 & 4.6\% & 3.2\% & 5.9\% & 83.3\% & 4.2\% & 12.5\% \\
\cmidrule{2-9}
\multirow{2}{*}{LeetCode} & Closed & 54 & 5.0\% & 6.2\% & 7.5\% & 70.4\% & 20.4\% & 9.3\% \\
 & Open & 54 & 8.5\% & 6.3\% & 7.3\% & 85.2\% & 11.1\% & 3.7\% \\
\cmidrule{2-9}
\multirow{2}{*}{BigCodeBench} & Closed & 18 & 1.2\% & 2.4\% & 13.6\% & 72.2\% & 27.8\% & 0.0\% \\
 & Open & 18 & -2.6\% & 0.2\% & 7.1\% & 55.6\% & 44.4\% & 0.0\% \\
\hline
\end{tabular}}
\end{table}
\begin{table}[t]
\caption{Dataset by model-group diff summary: Target language is Japanese.}
\centering
\label{tab:japanese-dataset-model-group-diff-summary}
\scalebox{0.8}{
\begin{tabular}{llrrrrrrr}
\hline
\multicolumn{1}{c}{Problems} & \multicolumn{1}{c}{Model Group} & \multicolumn{1}{c}{Count} & \multicolumn{1}{c}{Mean} & \multicolumn{1}{c}{Median} & \multicolumn{1}{c}{Std} & \multicolumn{1}{c}{Improve} & \multicolumn{1}{c}{Worsen} & \multicolumn{1}{c}{Unchanged} \\
\hline
\multirow{2}{*}{AtCoder} & Closed & 72 & -13.3\% & -10.3\% & 15.3\% & 23.6\% & 76.4\% & 0.0\% \\
 & Open & 72 & -9.5\% & -4.3\% & 12.7\% & 8.3\% & 83.3\% & 8.3\% \\
\hline
\end{tabular}}
\end{table}
\begin{table}[t]
\caption{Dataset by model-group diff summary: Target language is Chinese}
\centering
\label{tab:chinese-dataset-model-group-diff-summary}
\scalebox{0.8}{
\begin{tabular}{llrrrrrrr}
\hline
\multicolumn{1}{c}{Problems} & \multicolumn{1}{c}{Model Group} & \multicolumn{1}{c}{Count} & \multicolumn{1}{c}{Mean} & \multicolumn{1}{c}{Median} & \multicolumn{1}{c}{Std} & \multicolumn{1}{c}{Improve} & \multicolumn{1}{c}{Worsen} & \multicolumn{1}{c}{Unchanged} \\
\hline
\multirow{2}{*}{LeetCode} & Closed & 54 & 3.2\% & 1.9\% & 9.1\% & 59.3\% & 27.8\% & 13.0\% \\
 & Open & 54 & 9.2\% & 6.9\% & 9.1\% & 79.6\% & 14.8\% & 5.6\% \\
\hline
\end{tabular}}
\end{table}

\begin{table}[t]
\caption{Target-language by Open vs Closed diff summary}
\centering
\label{tab:summary-target-language-open-closed-diff-summary}
\scalebox{0.8}{
\begin{tabular}{llrrrrrrr}
\hline
\multicolumn{1}{c}{Target Language} & \multicolumn{1}{c}{Model Group} & \multicolumn{1}{c}{Count} & \multicolumn{1}{c}{Mean} & \multicolumn{1}{c}{Median} & \multicolumn{1}{c}{Std} & \multicolumn{1}{c}{Improve} & \multicolumn{1}{c}{Worsen} & \multicolumn{1}{c}{Unchanged} \\
\hline
\multirow{2}{*}{Chinese} & Closed & 54 & 3.2\% & 1.9\% & 9.1\% & 59.3\% & 27.8\% & 13.0\% \\
 & Open & 54 & 9.2\% & 6.9\% & 9.1\% & 79.6\% & 14.8\% & 5.6\% \\
\cmidrule{2-9}
\multirow{2}{*}{English} & Closed & 144 & -2.6\% & -1.0\% & 12.0\% & 40.3\% & 56.2\% & 3.5\% \\
 & Open & 144 & 5.2\% & 3.2\% & 7.4\% & 80.6\% & 11.8\% & 7.6\% \\
\cmidrule{2-9}
\multirow{2}{*}{Japanese} & Closed & 72 & -13.3\% & -10.3\% & 15.3\% & 23.6\% & 76.4\% & 0.0\% \\
 & Open & 72 & -9.5\% & -4.3\% & 12.7\% & 8.3\% & 83.3\% & 8.3\% \\
\hline
\end{tabular}}
\end{table}
\begin{table}[t]
\caption{Dataset by model-group diff summary}
\centering
\label{tab:dataset-model-group-diff-summary}
\scalebox{0.8}{
\begin{tabular}{llrrrrrrr}
\hline
\multicolumn{1}{c}{Problems} & \multicolumn{1}{c}{Model Group} & \multicolumn{1}{c}{Count} & \multicolumn{1}{c}{Mean} & \multicolumn{1}{c}{Median} & \multicolumn{1}{c}{Std} & \multicolumn{1}{c}{Improve} & \multicolumn{1}{c}{Worsen} & \multicolumn{1}{c}{Unchanged} \\
\hline
\multirow{2}{*}{AtCoder} & Closed & 144 & -11.2\% & -6.1\% & 13.2\% & 16.7\% & 83.3\% & 0.0\% \\
 & Open & 144 & -2.5\% & 0.0\% & 12.2\% & 45.8\% & 43.8\% & 10.4\% \\
\cmidrule{2-9}
\multirow{2}{*}{LeetCode} & Closed & 108 & 4.1\% & 3.9\% & 8.3\% & 64.8\% & 24.1\% & 11.1\% \\
 & Open & 108 & 8.9\% & 6.3\% & 8.2\% & 82.4\% & 13.0\% & 4.6\% \\
\cmidrule{2-9}
\multirow{2}{*}{BigCodeBench} & Closed & 18 & 1.2\% & 2.4\% & 13.6\% & 72.2\% & 27.8\% & 0.0\% \\
 & Open & 18 & -2.6\% & 0.2\% & 7.1\% & 55.6\% & 44.4\% & 0.0\% \\
\hline
\end{tabular}}
\end{table}
\begin{table}[t]
\caption{Dataset by tool diff summary}
\centering
\label{tab:dataset-tool-diff-summary}
\begin{tabular}{llrrrrrrr}
\hline
\multicolumn{1}{c}{Problems} & \multicolumn{1}{c}{Tool} & \multicolumn{1}{c}{Count} & \multicolumn{1}{c}{Mean} & \multicolumn{1}{c}{Median} & \multicolumn{1}{c}{Std} & \multicolumn{1}{c}{Improve} & \multicolumn{1}{c}{Worsen} & \multicolumn{1}{c}{Unchanged} \\
\hline
\multirow{3}{*}{AtCoder} & Google & 96 & -7.4\% & -3.7\% & 13.6\% & 28.1\% & 66.7\% & 5.2\% \\
 & DeepL & 96 & -9.0\% & -4.3\% & 13.2\% & 25.0\% & 69.8\% & 5.2\% \\
 & GPT & 96 & -4.1\% & -0.9\% & 13.1\% & 40.6\% & 54.2\% & 5.2\% \\
 \cmidrule{2-9}
\multirow{3}{*}{LeetCode} & Google & 72 & 6.4\% & 6.0\% & 8.9\% & 73.6\% & 20.8\% & 5.6\% \\
 & DeepL & 72 & 5.8\% & 5.8\% & 8.2\% & 68.1\% & 20.8\% & 11.1\% \\
 & GPT & 72 & 7.2\% & 6.2\% & 8.8\% & 79.2\% & 13.9\% & 6.9\% \\
 \cmidrule{2-9}
\multirow{3}{*}{BigCodeBench} & Google & 12 & -0.2\% & 1.1\% & 9.6\% & 66.7\% & 33.3\% & 0.0\% \\
 & DeepL & 12 & -0.3\% & 1.8\% & 8.6\% & 83.3\% & 16.7\% & 0.0\% \\
 & GPT & 12 & -1.6\% & -2.5\% & 14.3\% & 41.7\% & 58.3\% & 0.0\% \\
\hline
\end{tabular}
\end{table}

\subsection{Results}
\observation{Depending on the dataset, the $\mbox{\sl Accuracy}$ of the code generated from the translated problem statements may show improvement.}
\label{ob:translation}
Table~\ref{tab:translation-to-EN}, \ref{tab:translation-to-JA}, and \ref{tab:translation-to-ZH} present the $\mbox{\sl Accuracy}$ of code generation after translating AtCoder, LeetCode, and BigCodeBench problems into the target language. 
\red{Table~\ref{tab:translation-to-EN} presents the result for AtCoder, LeetCode, and BigCodeBench problems when Japanese and Chinese were translated into English.}
Table~\ref{tab:translation-to-JA} shows the results for AtCoder problems when English and Chinese were translated into Japanese.
Table~\ref{tab:translation-to-ZH} presents the results for LeetCode problems when English and Japanese were translated into Chinese.
In each table, the superscript next to the $\mbox{\sl Accuracy}$ indicates the $\mbox{\sl Accuracy}$ difference before and after translation.
A positive value (in blue) denotes an improvement, and a negative value (in red) denotes a decline.

\red{Table~\ref{tab:english-dataset-model-group-diff-summary}, \ref{tab:japanese-dataset-model-group-diff-summary}, and \ref{tab:chinese-dataset-model-group-diff-summary} summarize the effect of translation based on the results in Tables~\ref{tab:translation-to-EN}, \ref{tab:translation-to-JA}, and \ref{tab:translation-to-ZH}.
Specifically, for each dataset and model group (closed-source and open-source models), these tables report the mean, median, standard deviation, improvement rate, worsening rate, and unchanged rate of the $\mbox{\sl Accuracy}$ differences before and after translation for each target translation language.
Compared to the $\mbox{\sl Accuracy}$ before translation, the impact of translation varies across datasets.
For AtCoder, translation into English shows contrasting results between model groups. 
Closed-source models exhibit performance degradation, with a median decrease of -5.3\% and only 9.7\% of cases showing improvement. 
In contrast, open-source models benefit from translation into English, achieving a median increase of +3.2\% with an improvement rate of 83.3\%. 
However, translation into Japanese consistently degrades performance for both model groups, 
with median decreases of -10.3\% (closed) and -4.3\% (open), and worsening rates of 76.4\% and 83.3\%, respectively.
For LeetCode, translation generally improves performance. 
When translating into English, both model groups show positive median gains (+6.2\% for closed and +6.3\% for open), 
with high improvement rates of 70.4\% and 85.2\%, respectively. 
Similarly, translation into Chinese also results in performance improvements, 
with median increases of +1.9\% (closed) and +6.9\% (open), and improvement rates of 59.3\% and 79.6\%, respectively.
For BigCodeBench, translation also shows relatively high improvement rates despite smaller median gains. 
Closed-source models achieve an improvement rate of 72.2\% with a median increase of +2.4\%, 
while open-source models show an improvement rate of 55.6\% with a near-zero median change (+0.2\%). 
This suggests that translation often leads to improvements in many cases, even if the magnitude of improvement is limited.}
This indicates that the effect of translation differed across datasets.


\observation{Open-source models benefit from translation into English.}
\red{Table~\ref{tab:summary-target-language-open-closed-diff-summary} shows the impact of translation into each target language, reporting the mean and median differences in $\mbox{\sl Accuracy}$, as well as the proportions of improved, worsened, and unchanged cases, for both closed-source models (GPT-4o, o3-mini, DeepSeek-V3.2) and open-source models (Llama3, Qwen2.5-Coder (0.5B/14B)). 
From this table, we observe that when translating into English, open-source models achieve an improvement rate of 80.6\% with a mean $\mbox{\sl Accuracy}$ increase of 5.2\%. 
This suggests that translating inputs into English is an effective strategy for improving code generation performance in open-source models.
}

\observation{Translation is not always effective for closed-source models.}
\red{Table~\ref{tab:dataset-model-group-diff-summary} summarizes the impact of translation for each dataset, 
reporting the mean and median differences in $\mbox{\sl Accuracy}$, as well as the proportions of improved, worsened, and unchanged cases, for both closed-source models (GPT-4o, o3-mini, DeepSeek-V3.2) and open-source models (Llama3, Qwen2.5-Coder (0.5B/14B)). 
From this table, we observe that in AtCoder, closed-source models experience performance degradation in 83.3\% of the cases, with a mean decrease of -11.2\%. 
This indicates that translation may not provide benefits and can even harm performance for closed-source models, depending on the dataset.}

\observation{The impact of translation does not significantly depend on the choice of translation tool. }
\red{Table~\ref{tab:dataset-tool-diff-summary} shows the summary of the impact of translation into officially supported languages (e.g., English and Japanese for AtCoder) across different translation tools, reporting the mean and median differences in $\mbox{\sl Accuracy}$, as well as the proportions of improved, worsened, and unchanged cases.
Across datasets, different tools (Google, DeepL, and GPT) yield similar performance trends in terms of median improvements and improvement rates.
For example, in LeetCode, the median improvements are similar across tools, with +6.0\% (Google), +5.8\% (DeepL), and +6.2\% (GPT).}

\begin{table}[t]
\centering
\caption{Comparison of BLEU scores across translation tools}
\label{tab:bleu_tools}
\begin{tabular}{lrrrr}
\toprule
\multicolumn{1}{c}{Tool} & \multicolumn{1}{c}{Max} & \multicolumn{1}{c}{Avg} & \multicolumn{1}{c}{Median} & \multicolumn{1}{c}{Min} \\
\midrule
DeepL & 1.000 & 0.501 & 0.484 & 0.072 \\
Google & 0.895 & 0.509 & 0.515 & 0.137 \\
GPT & 1.000 & 0.564 & 0.572 & 0.001 \\
\bottomrule
\end{tabular}
\end{table}

\observation{The quality of translation is comparable across different translation tools. }
\red{
Table~\ref{tab:bleu_tools} shows BLEU scores calculated by using the original target-language prompts as reference texts and the machine-translated prompts as candidate texts.
For example, when translating Chinese prompts into English, we compared the translated English prompts with the original English prompts included in the dataset.
We observe that all translation tools achieve similar BLEU scores across multiple metrics.  
Specifically, the average BLEU scores range from 0.501 to 0.564, and the median BLEU scores range from 0.484 to 0.572, indicating only modest differences among the tools. 
These results suggest that the overall translation quality is comparable across different translation tools.}

\summarybox{Summary of RQ2}{
    Although translating problem statements into languages with high $\mbox{\sl Accuracy}$ can improve code generation $\mbox{\sl Accuracy}$, its effectiveness depends heavily on the dataset and the model type. In our experiments, translation improved $\mbox{\sl Accuracy}$ on LeetCode and BigCodeBench, as well as for open-source models on AtCoder, while it led to a performance drop for closed-source models on AtCoder.
    These results suggest that practitioners should carefully decide whether to apply translation based on the target dataset. Furthermore, researchers should investigate the factors driving these divergent effects across datasets.
}


\section{Additional Analysis}\label{sec:ad}
\subsection{Influence of Problem Statement Narrative}
\label{sec:rq:narrative}
The results of RQ2 showed that the effectiveness of translation varies across datasets and model types. This observation raises the question of why such differences occur. One possible factor is the characteristics of the input prompts, particularly their narrative structure and length. Therefore, we further investigate how the narrative and length aspects of problem statements affect LLM code generation performance.

\subsubsection{Motivation}
\red{Observation~\ref{ob:translation} showed that the impact of translation on code generation $\mbox{\sl Accuracy}$ differs across datasets. Translation often improves $\mbox{\sl Accuracy}$ for LeetCode and BigCodeBench, whereas its effect on AtCoder is more limited and varies depending on the model group and target language. }

\red{Based on these observations, we hypothesize that the problem statements in AtCoder are more challenging for translation-based mitigation than those in LeetCode, as their narrative structure and length may introduce complexities that hinder accurate translation, leading to lower effectiveness of translation and larger language bias.
Although BigCodeBench also showed improvement after translation, we focus on AtCoder and LeetCode in this analysis because they are both programming contest datasets and provide a clearer contrast in narrative structure.
Indeed, we observed that AtCoder problems are more likely to be narrative-style (e.g., containing unnecessary sentences) compared to LeetCode problems. Such narrative aspects may affect the translation accuracy and, in turn, the code generation $\mbox{\sl Accuracy}$. To verify this hypothesis, we investigate the influence of narrative aspects in problem statements on the $\mbox{\sl Accuracy}$ of code generation.}

\subsubsection{Approach}
Since there is no clear definition of narrative-style statements, we define the following three criteria to define narrative-style statements. We view these criteria as capturing narrative elements that contextualize a problem statement, and the removal of such elements can be interpreted as a form of decontextualization and abstraction.
Narrative-style statements are those that satisfy at least one of the following criteria:
\begin{itemize}
    \item \textbf{Subject:}
    The subject mentioned in the problem statement is a noun that describes a specific thing, such as person, place, or organization (\eg, ``Bob''). 
    This noun is used to concretize a problem by adding a specific example, making the problem more understandable. 
    These nouns introduce concrete entities that contextualize the problem. In some cases, the same computational structure can be expressed more abstractly by removing or replacing such entities
    (\eg, ``There'').
    For example, the problem statement \emph{``\underline{Mr. Infinity} has a string S consisting of digits from 1 to 9}''~\cite{AtCoder106_c} can be replaced with ``\underline{There} is a string S consisting of digits from 1 to 9''.
    The underline indicates the subject of the problem statement. 
    \item \textbf{Contextualized Objective:}
    A problem statement has a \emph{contextualized objective} when its goal is expressed using context-specific concepts, such as named entities or domain-specific terms, rather than as a purely abstract computational target (\eg, \emph{``Find the minimum possible sadness of Snuke.''}~\cite{AtCoder102_c}).
    A contextualized objective is a purpose that the problem statement wants to address by an answer program. However, similar to the subject, the problem statement can replace such a contextualized objective with an abstract numerical target.
    For example, the problem statement ``Find the minimum possible sadness of \underline{Snuke}'' can be replaced with ``Find the minimum possible number''. ``Snuke'' and ``sadness'' are concrete nouns, making the objective in this problem statement is not abstract but contextualized.
    \item \textbf{Unrelated Sentences:}
    The problem statement contains sentences that are not necessary for resolving the problem (\eg, flavor text: \emph{``In 2020, AtCoder Inc. with an annual sales of more than one billion yen (the currency of Japan) has started a business in programming education.''}~\cite{AtCoder112_a}). Such a sentence is not necessary for solving the problem and can be omitted.
\end{itemize}


Based on these criteria, the first and second authors manually reviewed 50 sampled problems from the D-level in AtCoder and the complete set of 32 Hard-level problems in LeetCode. Since we manually reviewed the problem statements, we randomly retrieved 50 problems from the D-level in AtCoder instead of all problems.
We labeled whether each problem satisfied each criterion. 
\red{To assess the reliability of the annotation, we measured the inter-rater agreement using Cohen's kappa coefficient~\cite{cohen1960coefficient}, which resulted in 0.941, indicating almost perfect agreement between the annotators~\cite{landis1977measurement}.}
If our labels were inconsistent, we discussed and reached a consensus.
We only considered problems with a difficulty level of D (AtCoder) and Hard (LeetCode) since LLMs generally achieve high $\mbox{\sl Accuracy}$ for lower-difficulty problems regardless of the natural languages, as shown in Table~\ref{tab:all}.

If a problem satisfied at least one of the criteria, we classified it as a \emph{Narrative} problem statement; otherwise, we classified it as an \emph{Expository} problem statement. We then calculated the $\mbox{\sl Accuracy}$ of code generation for \emph{Expository} and \emph{Narrative} problem statements.

The following problem statement is an example of an \emph{Expository} problem statement~\cite{AtCoder101_b}:
\begin{formal}
Let S(n) denote the sum of the digits in the decimal notation of n.
For example, S(101) = 1 + 0 + 1 = 2.
Given an integer N, determine if S(N) divides N.
\end{formal}
The following problem statement is an example of a \emph{Narrative} problem statement~\cite{AtCoder115_b}:
\begin{formal}
In some other world, today is the day before Christmas Eve.
Mr. Takaha is buying $N$ items at a department store. The regular price of the $i$-th item $(1 \leq i \leq N)$ is $p_i$ yen (the currency of Japan).
He has a discount coupon, and can buy one item with the highest price for half the regular price. The remaining $N-1$ items cost their regular prices. What is the total amount he will pay?
\end{formal}
There are five subjects that describe a specific thing: ``today,'' ``Mr. Takaha,'' ``The regular price,'' ``He,'' and ``the $i$-th item'' in the problem statement.
The object is described by ``What is the total amount he will pay?'', which is a contextualized objective. ``In some other world, today is the day before Christmas Eve'' is an unrelated sentence. Hence, this problem statement satisfies all three criteria and is classified as a \emph{Narrative} problem statement.


\begin{table}[t]
\centering
\caption{Number of narrative and expository problems}
\label{tab:narrative-expository}
\begin{tabular}{lcc}
\toprule
Dataset & Narrative & Expository \\
\midrule
AtCoder & 74.0\% (37) & 26.0\% (13) \\
LeetCode & 50.0\% (16) & 50.0\% (16) \\
\bottomrule
\end{tabular}
\end{table}

\begin{table}[t]
\centering
\caption{Percentage and number of problems grouped by the number of narrative criteria they satisfy}
\label{tab:narrative-criteria-count}
\begin{tabular}{lcccc}
\toprule
Dataset & One & Two & Three & Total \\
\midrule
AtCoder & 13.5\% (5) & 70.3\% (26) & 16.2\% (6) & 37 \\
LeetCode & 56.2\% (9) & 37.5\% (6) & 6.2\% (1) & 16 \\
\bottomrule
\end{tabular}
\end{table}

\begin{table}[t]
\caption{\mbox{\sl Accuracy} for D-level problems in AtCoder between narrative and expository problems after translation (closed-source models only)}
\centering
\label{tab:atcoder-narrative-accuracy-translated-by-config-closed}
\begin{tabular}{llccc}
\toprule
\multicolumn{1}{c}{Translation Tool} & \multicolumn{1}{c}{Language} & \multicolumn{1}{c}{Narrative} & \multicolumn{1}{c}{Expository} & \multicolumn{1}{c}{Difference} \\
\midrule
\multirow{2}{*}{Google} & English & 45.9\% & 50.0\% & -4.1\% \\
 & Japanese & 50.0\% & 59.6\% & -9.6\% \\
 \cmidrule{2-5}
\multirow{2}{*}{DeepL} & English & 17.6\% & 30.8\% & -13.2\% \\
 & Japanese & 35.1\% & 19.2\% & 15.9\% \\
 \cmidrule{2-5}
\multirow{2}{*}{GPT} & English & 47.3\% & 46.2\% & 1.1\% \\
 & Japanese & 51.4\% & 65.4\% & -14.0\% \\
\bottomrule
\end{tabular}
\end{table}

\begin{table}[t]
\caption{\mbox{\sl Accuracy} for LeetCode problems between narrative and expository problems after translation (closed-source models only)}
\centering
\label{tab:leetcode-narrative-accuracy-translated-by-config-closed}
\begin{tabular}{llccc}
\toprule
\multicolumn{1}{c}{Translation Tool} & \multicolumn{1}{c}{Language} & \multicolumn{1}{c}{Narrative} & \multicolumn{1}{c}{Expository} & \multicolumn{1}{c}{Difference} \\
\midrule
\multirow{2}{*}{Google} & Chinese & 84.4\% & 81.2\% & 3.1\% \\
 & English & 84.4\% & 81.2\% & 3.1\% \\
 \cmidrule{2-5}
\multirow{2}{*}{DeepL} & Chinese & 81.2\% & 77.5\% & 3.7\% \\
 & English & 87.5\% & 80.6\% & 6.9\% \\
 \cmidrule{2-5}
\multirow{2}{*}{GPT} & Chinese & 84.4\% & 84.4\% & 0.0\% \\
 & English & 87.5\% & 81.2\% & 6.2\% \\
\bottomrule
\end{tabular}
\end{table}

\subsubsection{Results}
\observation{AtCoder has a relatively large proportion of narrative problem statements.}  
Table~\ref{tab:narrative-expository} presents the distribution of narrative and expository problems.
\red{On AtCoder, 74.0\% of problems are narrative and 26.0\% are expository. In contrast, on LeetCode, half of the problems are narrative and the other half are expository.}
This indicates that AtCoder problems are more narrative in style than LeetCode problems. Moreover, narrative problems on AtCoder are more likely to satisfy multiple narrative criteria than those on LeetCode.
Table~\ref{tab:narrative-criteria-count} shows the distribution of problems that are narrative style based on the number of narrative criteria they satisfy.
\red{On AtCoder, 86.5\% of narrative problems satisfy at least two criteria
(70.3\% satisfy two and 16.2\% satisfy three), while on LeetCode, 43.7\%
satisfy at least two criteria (37.5\% satisfy two and 6.2\% satisfy three).}





\observation{Narrative-style problems on AtCoder show lower code generation $\mbox{\sl Accuracy}$, while narrative-style problems on LeetCode show higher $\mbox{\sl Accuracy}$.}
Tables~\ref{tab:atcoder-narrative-accuracy-translated-by-config-closed} and~\ref{tab:leetcode-narrative-accuracy-translated-by-config-closed} show the median $\mbox{\sl Accuracy}$ of code generation for narrative and expository problems on AtCoder and LeetCode, respectively. 
\red{Since Observation~9 showed that translation was less effective for closed-source models on AtCoder, this analysis focuses on closed-source models.}
\red{For AtCoder, narrative problems generally yield lower $\mbox{\sl Accuracy}$ than expository problems after translation. Across translation tools, narrative problems show lower $\mbox{\sl Accuracy}$ than expository problems, with a median difference of -4.1\% for English and -9.6\% for Japanese. Overall, the median difference is -6.9\%, suggesting that narrative-style AtCoder problems remain more difficult even after translation-based mitigation.
In contrast, for LeetCode, narrative problems yield higher $\mbox{\sl Accuracy}$ than expository problems after translation. Across translation tools, narrative problems show higher $\mbox{\sl Accuracy}$ than expository problems, with a median difference of 3.1\% for Chinese and 6.2\% for English. Overall, the median difference is 3.4\%.}

\begin{figure}[t]
	\begin{center}
  \includegraphics[width=\linewidth]{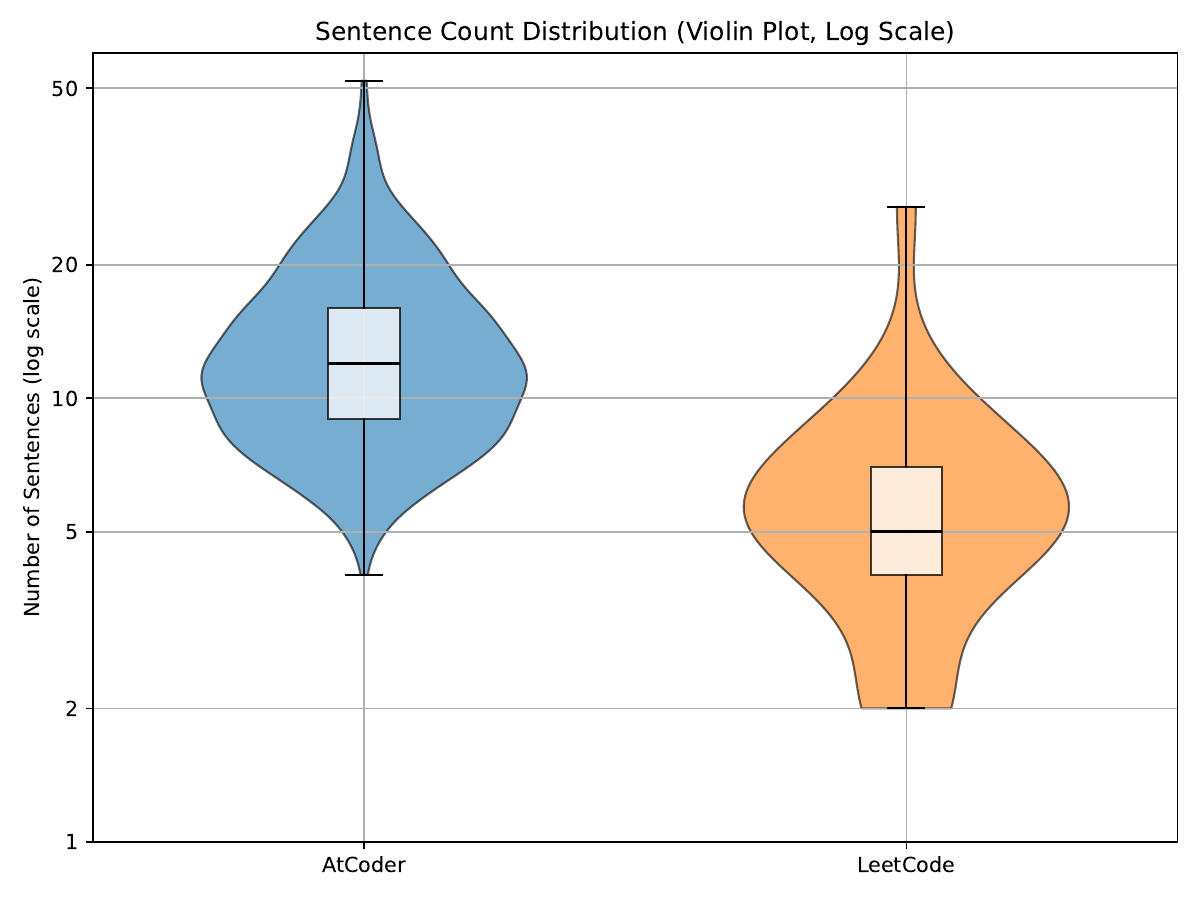}
	\caption{Number of sentences for each problem in AtCoder and LeetCode} 
	\label{fig:ad_boxplot_sentence}
	\end{center}
\end{figure}
\begin{table}[]
    \centering
    \caption{Quartiles of number of sentences in a problem for each dataset}
    \label{tab:add_sentence}
    \begin{tabular}{lrrr}
      \toprule
      \multicolumn{1}{c}{Dataset} & \multicolumn{1}{c}{1st Quartile (Q1)} & \multicolumn{1}{c}{Median (Q2)} & \multicolumn{1}{c}{3rd Quartile (Q3)} \\
      \midrule
      AtCoder   & 9.0 & 12.0 & 16.0 \\
      LeetCode  & 4.0 & 5.0  & 7.0  \\
      \bottomrule
    \end{tabular}
  \end{table}

\observation{Narrative-style long problems are more difficult for LLMs to generate code.} 
To investigate why narrative problems on AtCoder are harder while those on LeetCode are easier, we analyze problem length. Figure~\ref{fig:ad_boxplot_sentence} shows boxplots of the number of sentences in problem statements. 
We clearly observe that AtCoder problems contain more sentences than LeetCode problems. 
Table~\ref{tab:add_sentence} presents quartiles of sentence counts, showing that the median number of sentences on AtCoder is more than twice that on LeetCode. 
These results suggest that narrative-style AtCoder problems generally have longer contexts, which may be related to lower code generation $\mbox{\sl Accuracy}$.
In other words, longer contexts may make it more difficult for LLMs to understand problem statements, whereas shorter narrative-style LeetCode problems may provide more accessible context.
\begin{figure*}[t!]
    \begin{center}
    \includegraphics[width=0.8\linewidth]{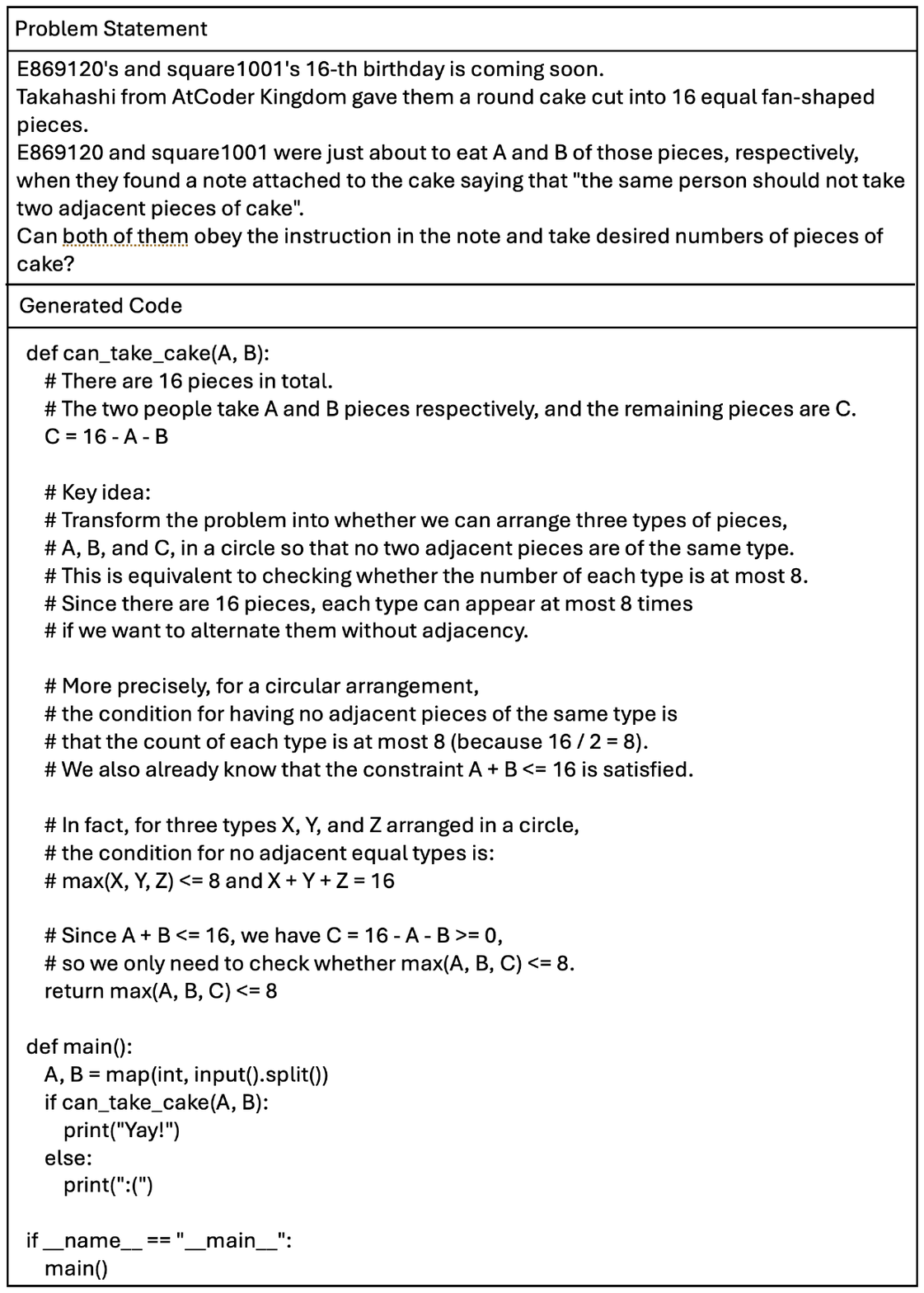}
    \caption{An example of a failure case for narrative-based prompting}
    \label{fig:narrative_failure_case}
    \end{center}
\end{figure*}

\observation{Language bias in code generation is primarily associated with misunderstanding and inappropriate abstraction, rather than logic errors.}
To further examine whether narrative-style problem statements contribute to code generation failures, we qualitatively analyzed failure cases involving narrative-style problems. Specifically, we focused on code generated by DeepSeek, which achieved the best performance in RQ1.
We randomly sampled 50 failure cases from narrative-style problems in the AtCoder dataset.
Two annotators then independently examined whether each failure appeared to involve narrative-related factors (Cohen's kappa = 0.623), and the final labels were determined through discussion and agreement. As a result, 20 out of the 50 cases were caused by narrative-related factors.
Representative examples are shown in Figure~\ref{fig:narrative_failure_case}.

\summarybox{Summary}{
    AtCoder problems are more narrative in style than LeetCode problems.
    Narrative-style longer problem statements decrease LLM code generation $\mbox{\sl Accuracy}$ because LLMs fail to understand extended contexts. In contrast, narrative aspects aid context comprehension when problem statements are short. Hence, the impact of narrative aspects on code generation $\mbox{\sl Accuracy}$ heavily depends on the input prompt context length.
}

\subsection{\red{Applicability and Generalizability}}
\red{In this section, we extend our code generation experiments, which have primarily focused on Python, by incorporating C++ and low-resource natural languages.}

\label{sec:multi_lang}
\subsubsection{Motivation}
\red{From the perspective of programming languages, the existing experiments are limited to Python, and it remains unclear whether the observed findings generalize to other languages. By including C++, which is widely used in competitive programming~\cite{puri2021codenetlargescaleaicode}, we examine whether the observed language bias is specific to Python or consistent across different programming languages.}

\red{From the perspective of natural languages, the main analysis of this study is based on high-resource languages, English, Japanese, and Chinese. However, language bias may be more pronounced in low-resource languages, and therefore it is necessary to evaluate these languages as well.}

\red{Based on the above, we extend the scope of both programming and natural languages to examine the applicability and generalizability of our findings.}



\begin{figure*}[t!]
    \begin{center}
    \includegraphics[width=0.8\linewidth]{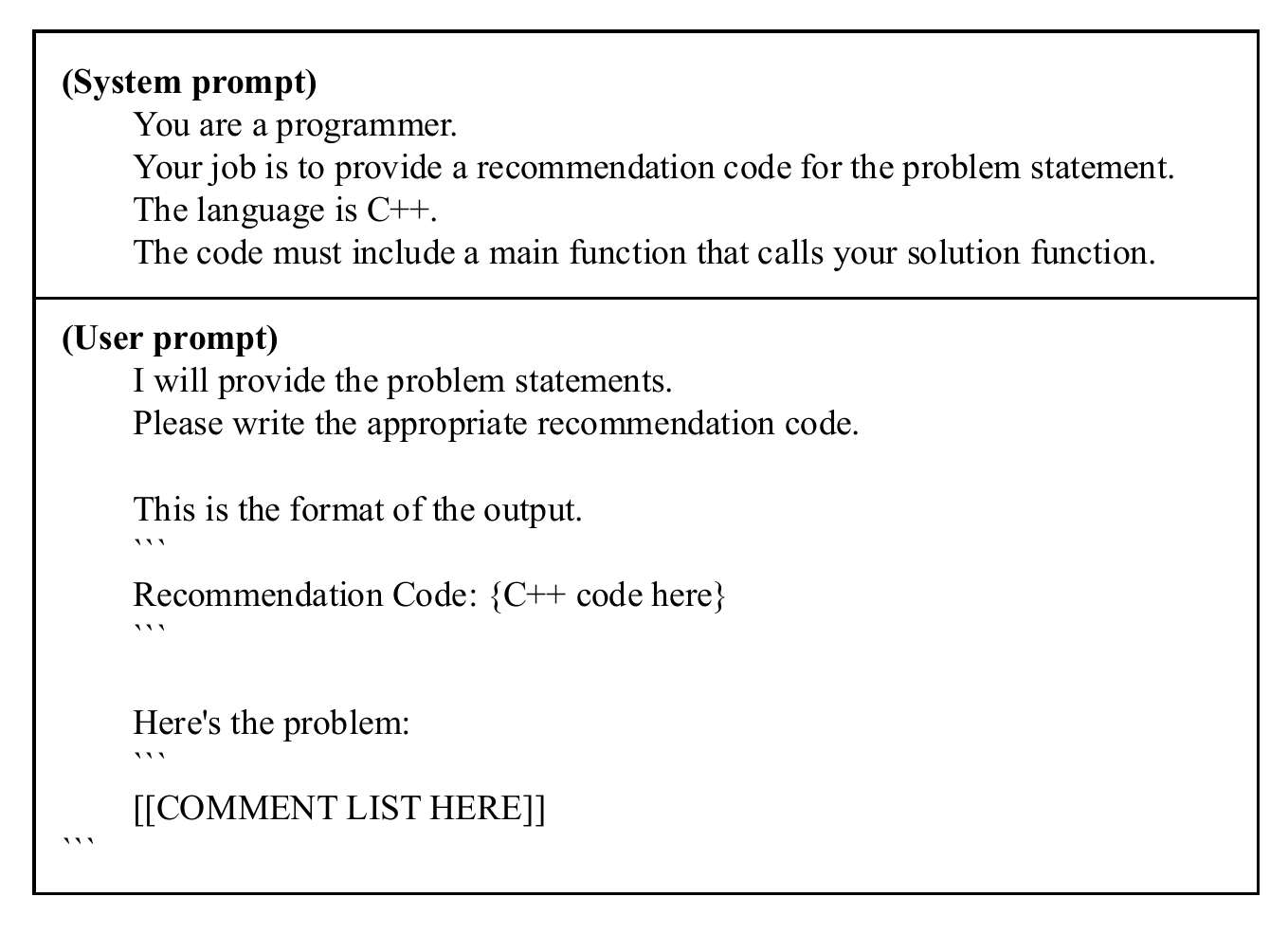}
    \caption{System and user prompts for GPT-4o, o3-mini and DeepSeek}
    \label{fig:prompt-cpp}
    \end{center}
\end{figure*}
\subsubsection{Approach}
\red{
This additional analysis extends the scope of our original experimental setup (Section~\ref{sec:setup}) by incorporating both additional programming languages and natural languages.
First, for the programming language extension, we include C++ in addition to Python. The experimental configuration is kept identical to that of Python, using the same set of problem instances and test cases for code generation and evaluation. Specifically, each problem description is provided as a prompt in a given natural language, and the generated code is evaluated using our unified evaluation pipeline based on test case execution. An example of the prompt used for C++ code generation is shown in Figure~\ref{fig:prompt-cpp}.}

\red{Second, for the natural language extension, we conduct additional experiments on low-resource languages. In addition to the high-resource languages used in the main analysis (English, Japanese, and Chinese), we generate prompts in low-resource languages by automatically translating the original problem descriptions using DeepL. The target languages are selected based on publicly available language distribution data,\footnote{\url{https://commoncrawl.github.io/cc-crawl-statistics/plots/languages}} focusing on those with less than 1.0\% of the data volume of English. Specifically, we consider Thai, Slovak, Finnish, Bulgarian, and Norwegian.} In this study, Norwegian refers to Norwegian Bokm{\aa}l.

\red{All experiments in this additional analysis are conducted using the AtCoder dataset. We do not include LeetCode or BigCodeBench in this extension due to practical constraints. For LeetCode, constructing reliable test cases requires substantial manual effort, making it infeasible to prepare additional multilingual data within our experimental setting. For BigCodeBench, prompts are originally provided in code-centric formats, and adapting them into consistent natural language prompts across multiple languages, as well as preparing corresponding evaluation test cases, is non-trivial.}

\red{For each generated prompt, code is produced using the same set of LLMs as in the main analysis, and performance is evaluated based on test case pass rates. All experimental conditions, including model configurations, prompt formats, and evaluation metrics, are kept consistent with the main experiments, ensuring that any observed differences can be attributed solely to variations in programming or natural languages.}

\red{Through this design, the additional analysis enables us to examine the applicability and generalizability of our findings across both programming languages and natural languages while maintaining the consistency of the experimental framework.
}





\subsubsection{Results}

\begin{table}[t]
\caption{Median $\mbox{\sl Accuracy}$ for each natural language and LLM}
\centering
\label{tab:cpp_all}
\begin{tabular}{lllrrrrl}
\hline
\multicolumn{1}{c}{Problems} & \multicolumn{1}{c}{Difficulty} & \multicolumn{1}{c}{Model} & \multicolumn{1}{c}{English} & \multicolumn{1}{c}{Japanese} & \multicolumn{1}{c}{Chinese} & \multicolumn{1}{c}{Python Best} & \multicolumn{1}{c}{Lang} \\ \hline
\multirow{24}{*}{AtCoder} & \multirow{6}{*}{A} & 4o & \textbf{93.1\%} & 92.1\% & \textbf{93.1\%} & 94.2\% & C \\
 &  & o3 & \textbf{92.6\%} & 88.4\% & \textbf{92.6\%} & 95.8\% & J \\
 &  & deep & \underline{\textbf{95.2\%}} & \underline{92.6\%} & \underline{94.7\%} & 95.2\% & J \\
 &  & llama3 & \textbf{61.9\%} & 55.0\% & 58.2\% & 67.2\% & E \\
 &  & qwen14b & 68.3\% & \textbf{74.6\%} & \textbf{74.6\%} & 87.8\% & E \\
 &  & qwen05b & \textbf{10.1\%} & 3.7\% & 2.6\% & 9.0\% & E \\
\cline{2-8}
 & \multirow{6}{*}{B} & 4o & 85.7\% & \textbf{86.2\%} & 84.7\% & 91.0\% & J \\
 &  & o3 & 90.5\% & 86.8\% & \textbf{91.0\%} & 90.5\% & J \\
 &  & deep & \underline{\textbf{93.1\%}} & \underline{92.6\%} & \underline{92.6\%} & 93.1\% & J \\
 &  & llama3 & 36.0\% & \textbf{37.6\%} & 34.4\% & 46.6\% & E \\
 &  & qwen14b & \textbf{62.4\%} & 51.3\% & 61.4\% & 76.2\% & E \\
 &  & qwen05b & \textbf{3.2\%} & 0.5\% & 2.1\% & 2.1\% & E \\
\cline{2-8}
 & \multirow{6}{*}{C} & 4o & 66.7\% & \textbf{72.5\%} & 66.7\% & 77.8\% & E \\
 &  & o3 & 87.3\% & 85.2\% & \underline{\textbf{88.9\%}} & 79.9\% & J \\
 &  & deep & \underline{\textbf{87.8\%}} & \underline{87.3\%} & \textbf{88.4\%} & 80.4\% & C \\
 &  & llama3 & 6.3\% & \textbf{7.4\%} & 6.9\% & 10.1\% & E \\
 &  & qwen14b & \textbf{30.7\%} & 25.4\% & 24.9\% & 36.5\% & E \\
 &  & qwen05b & \textbf{0.0\%} & \textbf{0.0\%} & \textbf{0.0\%} & 0.0\% & E/J/C \\
\cline{2-8}
 & \multirow{6}{*}{D} & 4o & 39.2\% & \textbf{48.1\%} & 40.2\% & 47.1\% & E/J \\
 &  & o3 & \underline{72.5\%} & \underline{73.0\%} & \underline{\textbf{73.5\%}} & 63.5\% & J \\
 &  & deep & 70.9\% & \underline{\textbf{73.0\%}} & 72.5\% & 58.2\% & C \\
 &  & llama3 & \textbf{3.7\%} & 2.6\% & 2.6\% & 1.6\% & E/C \\
 &  & qwen14b & \textbf{11.6\%} & 9.0\% & 5.8\% & 11.1\% & E \\
 &  & qwen05b & \textbf{0.0\%} & \textbf{0.0\%} & \textbf{0.0\%} & 0.0\% & E/J/C \\
\hline
\end{tabular}
\end{table}

\begin{table}[t]
\caption{Number of highest $\mbox{\sl Accuracy}$ results achieved by each LLM for each language within each difficulty level in C++ code generation}
\centering
\label{tab:highest_accuracy_model_cpp_atcoder}
\setlength{\tabcolsep}{7pt}
\renewcommand{\arraystretch}{1.15}
\begin{tabular}{llrrrrrr}
\hline
\multicolumn{1}{c}{Problem} & \multicolumn{1}{c}{Difficulty} & \multicolumn{1}{c}{GPT-4o} & \multicolumn{1}{c}{o3-mini} & \multicolumn{1}{c}{DeepSeek-V3} & \multicolumn{1}{c}{Llama3} & \multicolumn{1}{c}{Qwen14B} & \multicolumn{1}{c}{Qwen0.5B} \\ \hline
\multirow{4}{*}{AtCoder}
& A & 0 & 0 & 3 & 0 & 0 & 0 \\
& B & 0 & 0 & 3 & 0 & 0 & 0 \\
& C & 0 & 1 & 2 & 0 & 0 & 0 \\
& D & 0 & 3 & 1 & 0 & 0 & 0 \\ \hline
Total & & 0 & 4 & 9 & 0 & 0 & 0 \\ \hline
\end{tabular}
\end{table}

\begin{table}[t]
\caption{Number of LLMs achieving the highest $\mbox{\sl Accuracy}$ for each language}
\centering
\label{tab:highest_accuracy_language_cpp_atcoder}
\setlength{\tabcolsep}{7pt}
\renewcommand{\arraystretch}{1.15}
\begin{tabular}{llrrr}
\hline
\multicolumn{1}{c}{Problem} & \multicolumn{1}{c}{Difficulty} & \multicolumn{1}{c}{English} & \multicolumn{1}{c}{Japanese} & \multicolumn{1}{c}{Chinese} \\ \hline

\multirow{4}{*}{AtCoder}
& A & \textbf{\underline{5}} & \underline{1} & 3 \\
& B & \textbf{\underline{3}} & \underline{2} & 1 \\
& C & \underline{2} & \textbf{\underline{3}} & 3 \\
& D & \textbf{\underline{3}} & \textbf{\underline{3}} & 2 \\ \hdashline

\textbf{Total (All)} & & \underline{13} & \underline{9} & 9 \\ \hline

\end{tabular}
\end{table}
\observation{DeepSeek-V3.2 also achieves the best performance for C++.} 
\red{
Table~\ref{tab:cpp_all} shows the performance of each LLM for C++ code generation.
In this table, the underlined values indicate the highest \mbox{\sl Accuracy} among
the six LLMs for each language within each difficulty level, and the Python Best
column shows the best \mbox{\sl Accuracy} achieved in the corresponding Python
setting.}

\red{
As shown by the underlined values, DeepSeek-V3.2 achieves the highest
\mbox{\sl Accuracy} in many C++ settings, similar to the results for Python.
Table~\ref{tab:highest_accuracy_model_cpp_atcoder} summarizes this tendency by counting the underlined
values in Table~\ref{tab:cpp_all}. DeepSeek-V3.2 achieves the highest count overall,
with 9 cases, and particularly dominates in the lower difficulty levels, A and B.
Although o3-mini achieves the highest \mbox{\sl Accuracy} in several higher-difficulty
settings, especially D, DeepSeek-V3.2 is the most consistently top-performing model
across the C++ results.}

\red{
The bold values in Table~\ref{tab:cpp_all} indicate the highest \mbox{\sl Accuracy} among the three languages. 
In terms of natural languages, English often achieves high performance, while
Japanese and Chinese show comparable performance in many settings.}



\observation{Officially supported languages tend to achieve better performance, with English showing particularly strong results.}
\red{
Table~\ref{tab:highest_accuracy_language_cpp_atcoder} presents the number of LLMs achieving the highest $\mbox{\sl Accuracy}$ for each language.
At the level of individual difficulties, English outperforms the other languages for lower difficulty problems, while Japanese also shows improved performance for higher difficulty problems. This trend is consistent with the observations for Python.}

\red{
Overall, English has the highest count (13), whereas Japanese and Chinese show comparable results, both with a count of 9.
}

\observation{Language bias persists even among low-resource languages.}
Table~\ref{tab:low} shows the median $\mbox{\sl Accuracy}$ for each low-resource language and LLM across AtCoder difficulty levels.
As shown in Table~\ref{tab:low}, across all difficulty levels and models, we observe consistent performance differences among low-resource languages (e.g., Slovak, Finnish, Bulgarian, and Norwegian).
This indicates that language bias is not limited to high-resource languages, but also persists within low-resource settings.
Notably, some languages (e.g., Slovak and Finnish) consistently achieve higher performance than others (e.g., Bulgarian), suggesting that the impact of language varies even within low-resource groups.
A possible explanation is differences in tokenizer-language compatibility. A recent study\cite{petrov2023neurips} suggests that tokenization efficiency can vary across languages, and, for example, Bulgarian may produce more tokens than Slovak or Finnish, which could increase input fragmentation and potentially affect model performance. However, this remains a hypothesis and requires further investigation.

\observation{Model capability does not eliminate language bias in low-resource settings.}
Even for high-performing models (e.g., DeepSeek and o3), performance degradation is observed in certain low-resource languages.
This trend becomes more pronounced as task difficulty increases, where performance variance across languages widens.
These results suggest that improvements in model capability alone are insufficient to eliminate language bias, and that language choice remains a critical factor in code generation performance.

\begin{table}[t]
\caption{Median Accuracy for each natural language and LLM}
\centering
\label{tab:low}
\begin{tabular}{lllrrrrr}
\hline
\multicolumn{1}{c}{Problems} & \multicolumn{1}{c}{Difficulty} & \multicolumn{1}{c}{Model} & \multicolumn{1}{c}{Thai} & \multicolumn{1}{c}{Slovak} & \multicolumn{1}{c}{Finnish} & \multicolumn{1}{c}{Bulgarian} & \multicolumn{1}{c}{Norwegian} \\ \hline
\multirow{24}{*}{AtCoder} & \multirow{6}{*}{A} & 4o & 64.6\% & \textbf{77.2\%} & 66.1\% & 23.3\% & 34.9\% \\
 &  & o3 & \textbf{79.4\%} & 70.4\% & 72.0\% & 70.4\% & 67.2\% \\
 &  & deep & 81.5\% & 81.0\% & 83.6\% & \textbf{93.1\%} & 85.2\% \\
 &  & llama3 & 3.7\% & \textbf{45.5\%} & 20.6\% & 40.2\% & 42.3\% \\
 &  & qwen14b & 53.4\% & \textbf{65.6\%} & 14.8\% & 60.8\% & 61.4\% \\
 &  & qwen05b & 0.0\% & \textbf{0.5\%} & 0.0\% & 0.0\% & 0.0\% \\
\cline{2-8}
 & \multirow{6}{*}{B} & 4o & 63.0\% & \textbf{69.8\%} & 65.1\% & 24.3\% & 32.8\% \\
 &  & o3 & \textbf{76.7\%} & 57.1\% & 63.5\% & 60.8\% & 61.4\% \\
 &  & deep & 76.2\% & 75.1\% & 73.5\% & \textbf{86.8\%} & 83.1\% \\
 &  & llama3 & 1.1\% & \textbf{31.2\%} & 11.6\% & 28.6\% & 27.0\% \\
 &  & qwen14b & \textbf{54.5\%} & 53.4\% & 11.6\% & 42.3\% & 43.4\% \\
 &  & qwen05b & 0.0\% & 0.0\% & 0.0\% & 0.0\% & \textbf{0.5\%} \\
\cline{2-8}
 & \multirow{6}{*}{C} & 4o & 47.6\% & \textbf{67.2\%} & 55.0\% & 16.9\% & 21.2\% \\
 &  & o3 & \textbf{56.1\%} & 43.4\% & 51.9\% & 42.3\% & 51.3\% \\
 &  & deep & 74.6\% & 72.0\% & 70.4\% & \textbf{79.9\%} & 77.2\% \\
 &  & llama3 & 1.6\% & 6.9\% & 4.2\% & \textbf{10.1\%} & 6.3\% \\
 &  & qwen14b & \textbf{28.0\%} & 27.0\% & 5.8\% & 18.5\% & 19.6\% \\
 &  & qwen05b & 0.0\% & 0.0\% & \textbf{0.5\%} & \textbf{0.5\%} & 0.0\% \\
\cline{2-8}
 & \multirow{6}{*}{D} & 4o & 24.3\% & \textbf{40.7\%} & 34.4\% & 9.0\% & 12.7\% \\
 &  & o3 & \textbf{38.6\%} & 27.0\% & 29.6\% & 30.2\% & 34.4\% \\
 &  & deep & 45.5\% & 55.6\% & 47.1\% & \textbf{60.8\%} & 52.4\% \\
 &  & llama3 & 0.0\% & 1.6\% & 0.5\% & \textbf{2.6\%} & 0.5\% \\
 &  & qwen14b & 6.3\% & \textbf{6.9\%} & 0.0\% & 4.2\% & 4.8\% \\
 &  & qwen05b & \textbf{0.0\%} & \textbf{0.0\%} & \textbf{0.0\%} & \textbf{0.0\%} & \textbf{0.0\%} \\
\hline
\end{tabular}
\end{table}

\summarybox{Summary}{
The C++ results are broadly consistent with the Python findings.
DeepSeek-V3.2 achieves the best overall performance, and AtCoder's officially supported languages tend to yield higher Accuracy overall.
However, Chinese, despite not being officially supported by AtCoder, achieves performance comparable to Japanese in some settings.
The analysis of low-resource languages further shows that language bias persists even outside high-resource languages.
These results suggest that natural language choice continues to affect code generation performance, regardless of programming language and model capability.}
\section{Discussion} \label{sec:discussion}
This section discusses the implications of our findings for LLM-based code generation across multiple natural languages and the use of machine translation as a mitigation strategy.

\subsection{Implications to LLM-based Code Generation for Multiple Natural Languages}
\implication{
    LLM-based code generation performance may be influenced by the availability of platform-specific resources in each natural language.
    Even when a language is not officially supported by a dataset, LLMs may still generate accurate code if the input prompt has a simple context.
    }
Observation 2 indicates that code generation performance is higher when the prompt language is officially supported by the corresponding dataset or platform.
This result implies that LLM effectiveness across languages depends on the publicly available resources. In fact, a previous study~\cite{li2024language} has reported similar findings. For example, high-resource languages, defined as natural languages with large volumes of training data, tend to yield better performance. 

Given Table~\ref{tab:all}, for easy problems, LLMs achieved high $\mbox{\sl Accuracy}$ regardless of whether the prompt language was English, Chinese, or Japanese. However, for more difficult problems, prompts in the officially provided language often yielded better results.
One possible reason is that the context of easy problems is simple and easy to understand. In contrast, the context of difficult problems is often more complex and requires a deeper understanding of the problem statement.
Hence, to generate correct code in languages not officially supported by the corresponding dataset or platform, LLMs may benefit from (1) richer platform-specific resources in that language and (2) simpler contexts in the input prompt.

\implication{Context length and narrative aspects are important for code generation when translating the prompt into a natural language with high accuracy.} 
Our additional analysis in Section~\ref{sec:rq:narrative} defines three types of narrative aspects and identifies narrative-style problems. This analysis clearly shows differences between AtCoder and LeetCode problems. Specifically, we found that (1) AtCoder problems contain more narrative aspects than LeetCode problems, (2) narrative-style AtCoder problems result in lower code generation performance than expository-style AtCoder problems, whereas narrative-style LeetCode problems result in higher code generation performance than expository-style LeetCode problems, and (3) the context length of AtCoder problems is longer than that of LeetCode problems. In addition, RQ2 shows that translation improves code generation performance for LeetCode problems but not for AtCoder problems.

These results suggest that context length and narrative aspects are important factors for code generation when using translation to enhance performance. In other words, when focusing on natural language characteristics to improve code generation, context length and narrative aspects are key factors. If context length is long and many narrative aspects are present, translation may not work well.

\subsection{Avenues for Future Research}
This study investigates the impact of multiple natural languages on code generation performance and demonstrates the potential of improving performance by translating prompts into a language with high code generation $\mbox{\sl Accuracy}$. However, we found that translation does not improve code generation performance for prompts with longer context, whereas shorter context prompts benefit from translation. Therefore, it is necessary to develop methods for appropriately handling longer context prompts.

Below, we suggest two directions for future research to improve code generation performance across languages by addressing longer context prompts:
\begin{itemize}
    \item \textbf{Compression:} Natural language sentences can be shortened using synonyms or paraphrases. We suggest that future studies investigate methods to compress context. For example, LLMs could remove \emph{unrelated sentences}, as defined in Section~\ref{sec:rq:narrative}, from prompts, since these sentences are unnecessary for code generation. LLMs could also paraphrase problem statements into shorter versions. These approaches may reduce text length and compress context effectively.
    Previous studies have proposed general-purpose methods for compressing LLM contexts and prompts while preserving task-relevant information~\cite{pan-etal-2024-llmlingua,LinChenghua2023CCtE}. In the context of code generation, Yang et al. showed that docstrings can be compressed while largely retaining their usefulness for generating code~\cite{yang2025less}. Complementarily, Afrin et al.~\cite{Afrin2026Not} found that tokens contribute unequally to LLM-based code summarization, suggesting that future compression approaches for multilingual code-generation prompts should identify and preserve informative content rather than merely reduce prompt length.
    \item \textbf{Organization:}
    Reducing context length can also remove useful information. We suggest that future studies explore ways to include longer context in prompts without losing information. For example, a previous study proposed a recipe pattern~\cite{white2023arxiv}, in which users provide structured information to an LLM and ask it to generate solution steps. Converting a natural language problem into a structured format may enable an LLM to process longer context more efficiently.
\end{itemize}

Additionally, it is important to quantify the impact of context on code generation performance in multilingual environments. Developers can refer to these results to determine the appropriate context for their own code generation tasks, such as the number of statements and narrative aspects in the input prompt. Below, we propose two directions.
\begin{itemize}
    \item \textbf{Investigate the optimal context length for translation efficacy:}
    Translation can both improve and degrade code generation performance depending on the target dataset and model type.
    Context length and narrative aspects in problem statements therefore influence translation effectiveness for code generation. We suggest that future studies quantify the effects of context length and narrative aspects on code generation performance in multilingual environments. 
    \item \textbf{Investigate the necessary context for each language:}
    \red{In this study, we used multiple machine translation tools, including DeepL, Google Translate, and GPT-4o, to directly translate problem statements into the target languages. However, natural languages differ in grammar, vocabulary, and common expressions, and direct translation may omit information critical for code generation.
    Future studies should consider not only direct translation but also language-specific paraphrasing. We recommend using human translators or manual verification to identify the necessary context for each language when translating problem statements.}
\end{itemize}

\section{Threats to Validity} \label{sec:threats}
\red{As with any empirical study, our work is subject to threats to validity.
While we acknowledge ongoing discussions on how validity concerns should be reported in software engineering research~\cite{verdecchia2023threats,lago2024threats,robillard2024communicating}, we use this framing to clearly organize the limitations of our study.}

\subsection{Construct Validity}
When translating problem statements into low-resource languages, we used machine translation without native-speaker validation due to the lack of native speakers among the authors. This may affect the quality of the translated problem statements and should be considered when interpreting the results for low-resource languages.

While any programming language can be used to address problems on AtCoder and LeetCode, we selected Python as the programming language for the code generated by LLMs.
However, the performance of LLM-generated code may vary depending on the programming language~\cite{nguyen2022MSR}.
To partially address this threat, we additionally conducted C++ experiments.
However, due to technical constraints, these experiments were conducted only on AtCoder.
Therefore, future studies would benefit from evaluating the performance of LLMs across a range of programming languages and datasets within this study setting.


\subsection{Internal Validity}
\label{sec:threats:internal}
GPT-4o and o3-mini provided by OpenAI, GitHub Copilot provided by Microsoft, and DeepSeek-V3.2 provided by DeepSeek were used in the experiments. 
These are closed-source models, and the internal structure of the models is not disclosed. Therefore, there is no guarantee that the models used in the experiments are exactly the same. 

\red{Also, the experimental results may change with future updates of LLMs (e.g., the training data includes blogs explaining the problem solutions in a
specific language thereby improving the performance in that language).
The possibility of such data contamination poses a threat to the validity of our results. Widely used platforms such as AtCoder and LeetCode provide a large amount of publicly available solutions, including blogs, repositories, and discussions in multiple languages. Therefore, it is possible that such content is included in the training data of the studied LLMs.
As a result, the observed differences in performance across languages may not solely reflect the models’ intrinsic ability to understand different natural languages, but may also be influenced by the distribution of training data, such as the quantity and quality of available solutions in each language.
Although our translation-based experiments partially mitigate this issue by generating new input texts and controlling the semantic content across languages, this threat cannot be completely eliminated. Therefore, the results of this study should be interpreted with this limitation in mind.}

For LeetCode, the prepared test cases may not fully cover all corner cases because the official hidden test suites are not publicly available; however, we mitigated this threat by generating 15 non-duplicated test cases per problem and validating them one by one using the official LeetCode website.

Problem statements in languages not officially supported by AtCoder, LeetCode, and BigCodeBench were translated from English using machine translation and then checked and corrected by native speakers.
To reduce subjectivity in this process, we prepared a coding book that summarizes the rules for checking and correcting translations.
However, the accuracy of the translation may still be affected by the judgment of the native speakers.
Quantitative evaluation of translation accuracy is needed in future work.

To address the randomness
of the LLMs, we repeated the code generation process five times. This number of repetitions is a parameter in the experiment.

\subsection{External Validity}
\label{sec:threats:external}
In this study, we conducted experiments using AtCoder, LeetCode, and BigCodeBench.
However, these three datasets do not represent the entire landscape of code generation tasks.
Further experiments to collect problems from other datasets will improve the generalizability of the results of this study.

We used seven LLMs in our experiments: GPT-4o, o3-mini, DeepSeek-V3.2, GitHub Copilot, Llama3, qwen2.5-coder-14b, and qwen2.5-coder-0.5b.
Since our goal is to extend the previous study~\cite{koyanagi2023MSR}, we included additional LLMs, such as the state-of-the-art model DeepSeek-V3.2. However, many other LLMs are available, and incorporating more of them into the experiments would provide a more comprehensive evaluation of the results. To encourage further research, we have provided our replication package\footnote{https://doi.org/10.5281/zenodo.20155616}.
\section{Conclusion} \label{sec:conclusion}

In this study, we evaluated the impact of natural language on code generation performance (language bias).
Specifically, we compared English, Japanese, and Chinese prompts across seven recent LLMs (GPT-4o, o3-mini, DeepSeek-V3.2, GitHub Copilot, Llama3, qwen2.5-coder-14b, and qwen2.5-coder-0.5b) using three benchmark datasets, AtCoder, LeetCode, and BigCodeBench.
We quantified the influence of natural language on performance and assessed a mitigation strategy (i.e., translating input prompts into the language that demonstrated the best performance) to reduce the language bias.
Our experiments showed that code generation performance varies by natural language.
Officially supported languages (English and Japanese for AtCoder; English and Chinese for LeetCode; English for BigCodeBench) frequently yielded the highest performance. 
The mitigation strategy demonstrated that translation can improve performance depending on the dataset and model type.
For example, translating Japanese prompts into English improved LeetCode performance, with median changes of +6.2\% and +6.3\% for closed-source and open-source models, respectively. In contrast, for AtCoder, translating into English improved open-source models by +3.2\% but degraded closed-source models by -5.3\%.
For BigCodeBench, translation into English led to small median improvements (+2.4\% for closed-source models and +0.2\% for open-source models).
To explain the language gap differences between the studied datasets, we analyzed narrative aspects and problem statement length, finding that narrative-style statements with long context are more challenging for LLMs and may be associated with larger language bias.


Based on these findings, we suggest:
\begin{itemize}
  \item
  Users of LLMs for code generation should carefully select the prompt language, as it affects performance. We recommend choosing a language for which ample data sources are available. 
  \item
  Translation can enhance code generation performance as well as natural language tasks, depending on the dataset and model type. Therefore, it is important to evaluate the effect of translation in the target dataset and model setting before applying it in practical scenarios. 
  \item
  Narrative-style prompts with long context can hinder code generation, whereas short narrative-style prompts can facilitate it.
\end{itemize}

\red{As future directions, we plan to conduct a more systematic qualitative analysis to better understand the underlying mechanisms behind failure cases and to investigate dialogue-based code generation scenarios, where users iteratively refine prompts through interactions with LLMs in different languages.}

\section*{Acknowledgment}
We gratefully acknowledge the financial support of: (1) JSPS for the KAKENHI grants (JP24K02921, JP25K03100); (2) Japan Science and Technology Agency (JST) as part of Adopting Sustainable Partnerships for Innovative Research Ecosystem (ASPIRE), Grant NumberJPMJAP2415, and (3) the Inamori Research Institute for Science for supporting Yasutaka Kamei via the InaRIS Fellowship.

\section*{Declarations}
\subsection*{Funding and/or Conflicts of interests/Competing interests}
All authors certify that they have no affiliations with or involvement in any organization or entity with any financial
interest or non-financial interest in the subject matter or materials discussed in this manuscript.

\subsection*{Data Availability Statements}
The replication package that supports the findings of this study is available.\footnote{\url{https://doi.org/10.5281/zenodo.20155616}}



\end{document}